\documentclass[%
 reprint,
superscriptaddress,
 amsmath,amssymb,
 aps,
prl,
]{revtex4-2}

\usepackage{graphicx}
\usepackage{dcolumn}
\usepackage{bm}
\usepackage{braket}
\usepackage{hyperref,xcolor,siunitx}
\hypersetup{
	colorlinks,
	linkcolor={magenta},
	citecolor={magenta},
	urlcolor={blue!70!black}
}

\begin{document}


\title{Fermi arc reconstruction in synthetic photonic lattice}
\author{D.-H.-Minh Nguyen}
\email{d.h.minh.ng@gmail.com}
\affiliation{Donostia International Physics Center, 20018 Donostia-San Sebasti\'an, Spain}
\author{Chiara Devescovi}
\affiliation{Donostia International Physics Center, 20018 Donostia-San Sebasti\'an, Spain}
\author{Dung Xuan Nguyen}
\email{dungmuop@gmail.com}
\affiliation{Center for Theoretical Physics of Complex Systems, Institute for Basic Science (IBS), Daejeon, 34126, Republic of Korea}
\author{Hai Son Nguyen}
\email{hai\_son.nguyen@ec-lyon.fr}
\affiliation{Univ Lyon, Ecole Centrale de Lyon, CNRS, INSA Lyon, Universit\'{e} Claude Bernard Lyon 1, CPE Lyon, CNRS, INL, UMR5270, Ecully 69130, France}
\affiliation{Institut Universitaire de France (IUF), F-75231 Paris, France}
\author{Dario Bercioux}
\email{dario.bercioux@dipc.org}
\affiliation{Donostia International Physics Center, 20018 Donostia-San Sebasti\'an, Spain}
\affiliation{IKERBASQUE, Basque Foundation for Science, Euskadi Plaza, 5, 48009 Bilbao, Spain}

\date{\today}

\begin{abstract}
	The chiral surface states of Weyl semimetals have an open Fermi surface called Fermi arc. At the interface between two Weyl semimetals, these Fermi arcs are predicted to hybridize and alter their connectivity. In this letter, we numerically study a one-dimensional (1D) dielectric trilayer grating where the relative displacements between adjacent layers play the role of two synthetic momenta. The lattice emulates 3D crystals without time-reversal symmetry, including Weyl semimetal, nodal line semimetal, and Chern insulator. Besides showing the phase transition between Weyl semimetal and Chern insulator at telecom wavelength, this system allows us to observe the Fermi arc reconstruction between two Weyl semimetals, confirming the theoretical predictions.
\end{abstract}
\maketitle
\date{\today}

\textit{Introduction.|}
Weyl semimetals (WSMs)~\cite{Armitage2018,Burkov2018,Lv2021} have been at the center of intense investigation since their theoretical predictions in 2011~\cite{Wan2011}. They are realized not only in condensed matters but also in photonic~\cite{Dubcek2015,Lu2015,Lin2016,Noh2017,Wang2017,Qin2018,Li2021,Lustig2021,Cheng2020,Han2022,Song2023} and phononic~\cite{Xiao2015,Li2017,Ge2018,Fan2019,Peri2019,Wang2021,Zhang2021,Liu2022} systems with potential applications, such as generation of optical vortex beams~\cite{Cheng2020} and robust transport in the bulk medium~\cite{Peri2019}. One of the well-known signatures of WSM is the appearance of the Fermi arc (FA) surface states~\cite{Wan2011,Xu2011,Burkov2011a}, which are chiral modes propagating unidirectionally on WSM surfaces and have an arc-like Fermi surface. Topologically protected against disorder and defects~\cite{Witten2016}, these FAs give rise to intriguing phenomena such as Weyl orbits~\cite{Potter2014} and magnetic domain walls with electric charge~\cite{Araki2016}.

Recently, some theoretical works~\cite{Dwivedi2018,Ishida2018,Murthy2020,Schroter2020,Abdulla2021,Buccheri2022,Kaushik2022,Bonasera2022} predicted that at the interface between two WSMs, the surface FAs would couple to each other and be reconstructed into new interface states with different spectral shapes and bulk connections. The problem is rich since these works consider distinct junctions of WSMs. For instance, while Dwivedi~\textit{et al.}~\cite{Dwivedi2018} examines two WSMs with the same Weyl point position and chirality, but different FA connectivity, Refs.~\cite{Murthy2020,Abdulla2021,Buccheri2022} study two identical WSMs rotated from each other in their interface plane. The reconstructed FAs are expected to exhibit observable transport signatures, such as unique quantum oscillations~\cite{Kaushik2022} and 3D ``snake states"~\cite{Buccheri2022}. Efforts have been made recently to fabricate a high quality interface between two chiral WSMs~\cite{Mathur2023}. However, for interface states between two 3D crystals, it is challenging to directly observe their spectra using angle-resolved photoemission spectroscopy or scanning tunneling microscopy. On the other hand, the crystal surfaces are usually rough due to defects and disorders, making it difficult to fabricate a clean heterostructure, especially for the interface between rotated WSMs.

In this letter, we propose a simple and versatile photonic lattice as the first platform to directly realize the FA reconstruction. Our system is a 1D trilayer grating where the relative displacements between layers play the role of synthetic momenta. This  1D system can simulate the topological band structure of 3D crystals with broken time-reversal (TR) symmetry, including WSM, nodal line semimetal~\cite{Burkov2011b,Fang2016}, and 3D Chern insulator (CI)~\cite{Yu2010,Vanderbilt2018}. With our trilayer lattice, we can obtain the phase transition WSM-CI just by varying the interlayer distances, or construct a photonic junction to observe the interface states between two WSMs or CIs. We will show that the interface FAs between two WSMs are strongly coupled and deformed, confirming the existence of FA reconstruction.

%
%
\begin{figure}[!t]
	\includegraphics[width=\columnwidth]{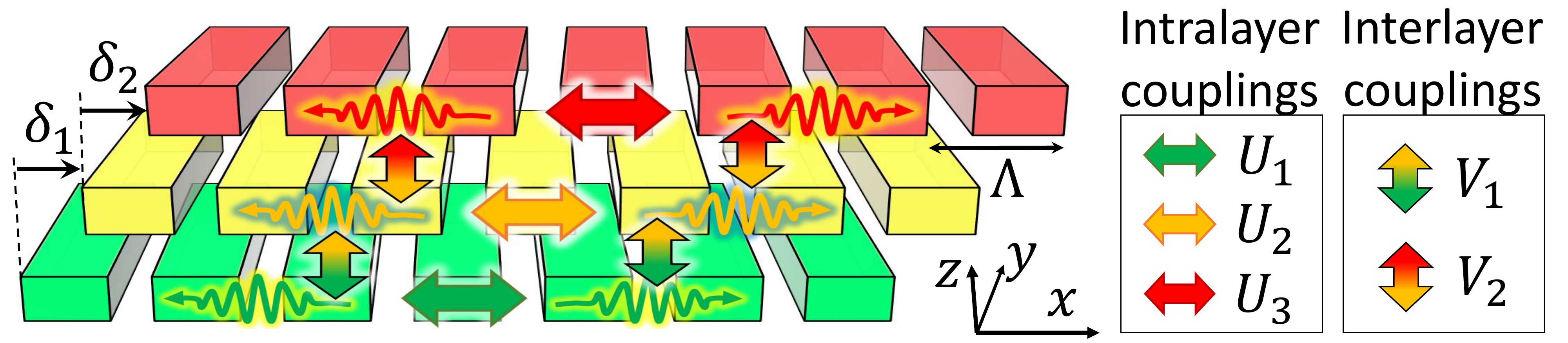}
	\caption{\label{fig:1} Sketch of a 1D trilayer photonic grating with period $\Lambda$. The relative displacements between adjacent layers are denoted by $\delta_1$ and $\delta_2$. The optical guided modes of interest couple with each other via intralayer diffraction and interlayer evanescent field, described by the coupling rates $U_l$ and $V_j$, respectively.}
\end{figure}
%
%
\textit{System and effective Hamiltonian.|} We consider a slab waveguide composed of three 1D dielectric gratings that share the same subwavelength period $\Lambda$ [Fig.~\ref{fig:1}]. Each layer is shifted with respect to its neighbor along the grating direction by $\delta_j$, with $j=1,2$. Other geometrical parameters of the lattice, including the filling fractions, grating thickness and interlayer distances, are also of subwavelength scale. More details on the geometrical parameters for an experimental realization are presented in the Supplemental Material~(SM)~\cite{Supp}

In each grating layer, we consider two counter-propagating guided modes along the $x$-axis, and these modes couple to each other via the grating diffraction described by coefficients $U_l$, $l=1,2,3$. Between adjacent layers, guided modes traveling in the same direction are coupled through the evanescent field with coupling rates $V_j$. Thus, our trilayer lattice has six guided modes described by the effective Hamiltonian~\cite{Son2018,Dung2022,Chau2021}
%
%
\begin{equation}
	H(k,\delta_1,\delta_2) = \begin{pmatrix}\Delta_1 & \Omega_1 & 0\\ \Omega^{\dagger}_1 & \Delta_2 & \Omega_2\\ 0 & \Omega^{\dagger}_2 & \Delta_3 \end{pmatrix},\label{Eq: Hamiltonian}
\end{equation}
%
%
where $\Delta_l=\omega_{0l}+\begin{pmatrix}v_lk & U_l\\ U_l& -v_lk \end{pmatrix}$ represents the guided modes in layer $l$ with their intrlayer coupling, and $\Omega_j=\begin{pmatrix}V_je^{-i\pi\delta_j/\Lambda} & 0\\ 0& V_je^{i\pi\delta_j/\Lambda} \end{pmatrix}$  indicates the interlayer coupling. Here, $v_l$, $\omega_{0l}$, and $k$ are the group velocities, frequency offsets, and wave vector measured from the X-point of the 1D BZ, respectively. The parameters $\omega_{0l}$, $v_l$ and $U_l$ are determined via the filling fraction, whereas the coupling rates $V_j$ are retrieved from the interlayer distances of the practical structure~\cite{Supp}.

The 1D lattice is invariant under the translations $\delta_j\rightarrow\delta_j+\Lambda$, thus, its physical properties vary periodically against $\delta_j$. Consequently, we define the synthetic momenta proportional to the interlayer displacements~\cite{Chau2021,Lee_2022} as
%
%
\begin{equation}
	q_j=2\pi\frac{\delta_j}{\Lambda^2}\label{Eq: synthetic momenta}.
\end{equation}
%
%
Such definitions hold in the presence of any perturbations that preserve the periodicity of our system. The Hamiltonian is then identified in a synthetic 3D momentum space $(k,q_1,q_2)$, i.e., $H(k,\delta_1,\delta_2)\rightarrow H(k,q_1,q_2)$.
\begin{figure}
	\includegraphics[width=\linewidth]{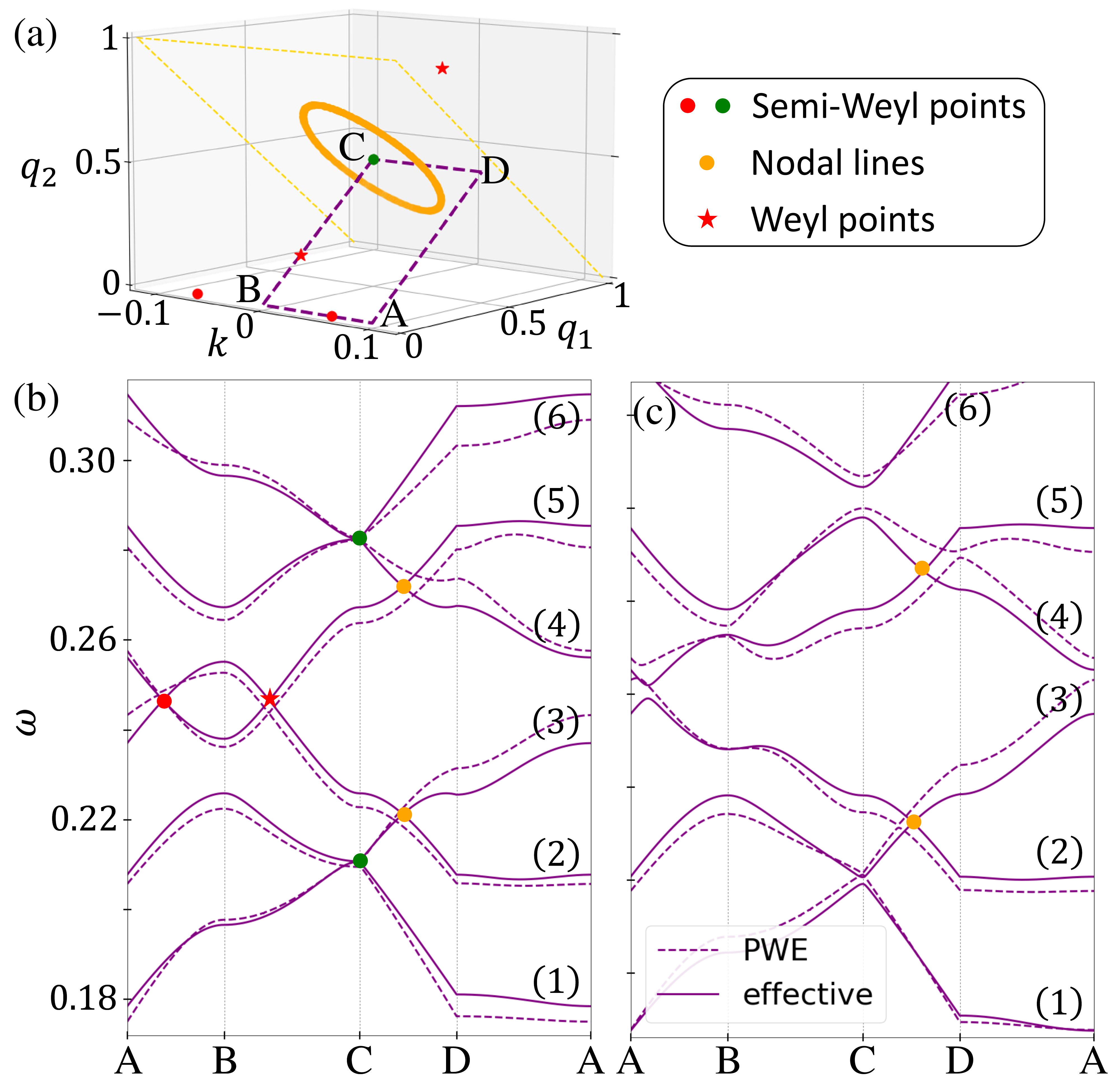}
	\caption{\label{fig:2} (a) Nodal points and lines of the spectrum obtained from the effective model with $U_1=U_2=U_3=V_1=V_2=1$. They are degeneracy points of energy bands shown in (b). The violet dashed line indicates the path for plotting the band structures in (b) and (c). The band structures obtained from the effective model (solid line) and PWE simulation (dashed line) are shown for (b) parameters as in (a), and (c) $U_1=U_3=1$, $U_2=1.2$, $V_1=1.6$ and $V_2=0.6$. The six bands are labeled by numbers.}
\end{figure}
%
The BZ is chosen so that $q_1,q_2\in[0,2\pi/\Lambda)$ [Fig.~\ref{fig:2}(a)]. Hereafter, we express the momenta in units of $2\pi/\Lambda$, frequencies in units of $2\pi c/\Lambda$, and all the coupling rates in units of $\bar{U}=0.0207\,(2\pi c/\Lambda)$. Here, $\bar{U}$ is the intralayer coupling rate of a single grating with filling fraction 0.8~\cite{Supp}. When all coupling rates are equal, the spectrum of this Hamiltonian shows different topological semimetallic phases | see Fig.~\ref{fig:2}(a) and solid lines of Fig.~\ref{fig:2}(b). First, bands (1) and (2), as well as bands (5) and (6), cross each other at a 3D semi-Weyl point~\cite{Murakami2008,He2019,Mohanta2021,Li2022}, which disperses linearly in a 2D plane and quadratically along the direction normal to this plane. Second, bands (2) and (3), as well as bands (4) and (5), form a nodal line~\cite{Burkov2011b,Fang2016} with $\pi$-Berry phase in the plane $q_1+q_2=1$. Finally, bands (3) and (4) touch at two semi-Weyl points aligned along the line $q_1=q_2=0$ and two Weyl points~\cite{Burkov2018,Armitage2018} residing in the plane $k=0$. We can alter these nodal points and lines to obtain other phases by varying the coupling rates. For instance, the semi-Weyl points are gapped out for $U_2>U_1=U_3$, and the Weyl points of bands (3) and (4) are annihilated for an appropriate choice of evanescent coupling rates $V_1$ and $V_2$, as shown by the solid-line band structure of Fig.~\ref{fig:2}(c). More details about topological phases and phase transitions can be found in SM~\cite{Supp}. 

We validate the effective Hamiltonian by comparing its spectrum with the exact solution of Maxwell's equations, which is obtained by the plane wave expansion (PWE) method using the MIT Photonic Bands package~\cite{MPB} | see Figs.~\ref{fig:2}(b) and~\ref{fig:2}(c). In both cases, the two band structures agree well in the overall shape, the positions of nodal points and lines, and their gap opening. The deviations between the two approaches, which are small compared to the bandwidths, result from the coupling mechanisms neglected in our model for simplicity~\cite{Son2018,Letartre2022,Supp}. Hence, the effective model is an efficient tool to examine the trilayer grating.

\textit{Controlling Weyl points.|} The trilayer lattice hosts topological phases owing to the synthetic momenta definition [Eq.~(\ref{Eq: synthetic momenta})]. These momenta are proportional to the interlayer displacements and thus are even under TR, making the Hamiltonian transform as $H(k,q_1,q_2)\xrightarrow{\text{TR}}H(-k,q_1,q_2)$. Consequently, the 1D trilayer grating is equivalent to a 3D crystal with broken TR symmetry. This allows the WSM phase, which requires either inversion or TR symmetry  to be broken~\cite{Armitage2018}. Hereafter, we focus on the two Weyl points at $k=0$ forming between bands (3) and (4), and thus, for simplicity, we gap out the semi-Weyl points of these bands by keeping $U_1=U_3=1$ and $U_2=1.2$. We investigate how the Weyl points move when tuning $V_1$ and $V_2$.

\begin{figure}
	\includegraphics[width=\linewidth]{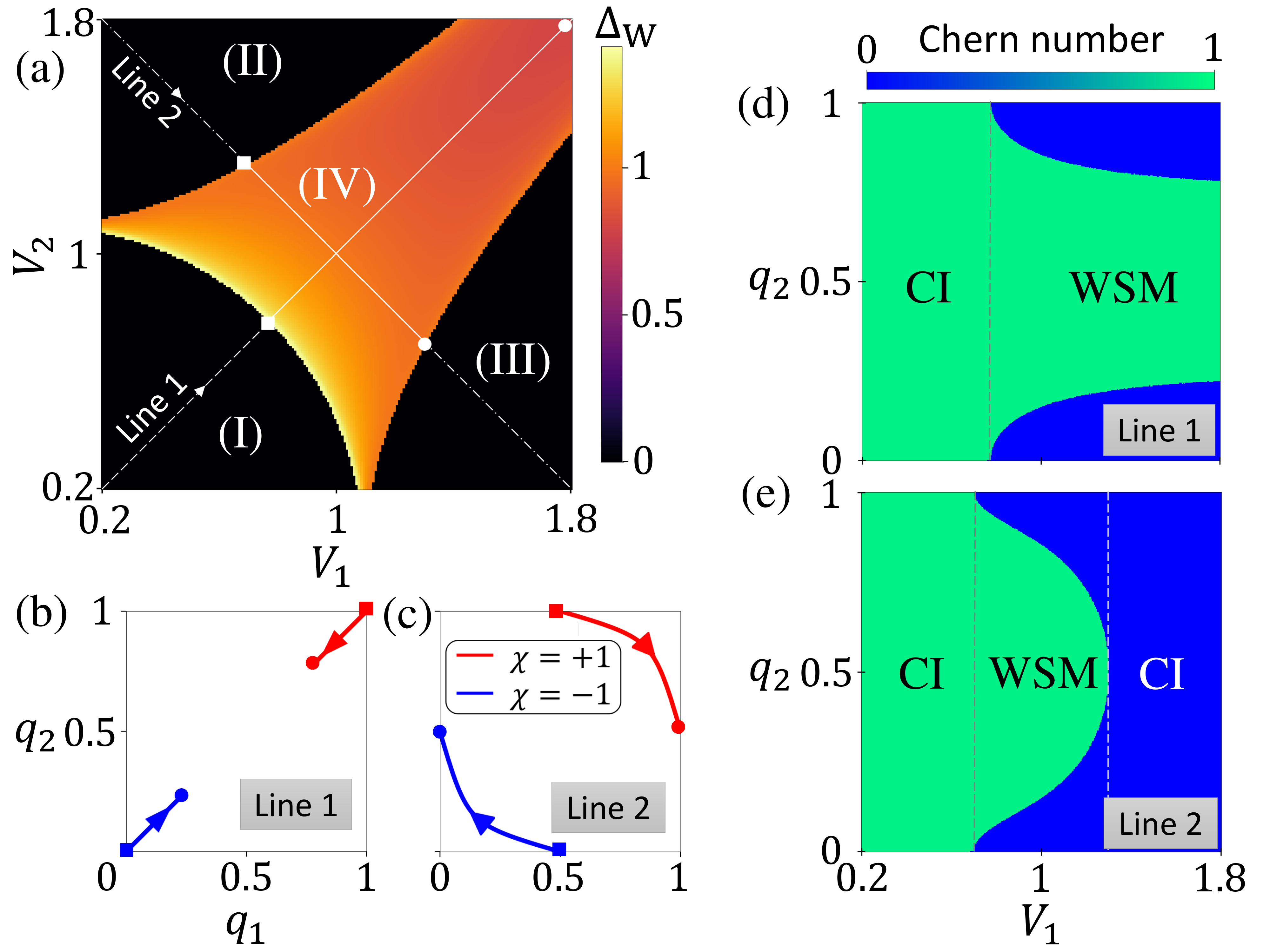}
	\caption{\label{fig:3} (a) Distance $\Delta_\text{W}$ between the two Weyl points at $k=0$ when $U_1=U_3=1$ and $U_2=1.2$. The three domains in black are where bands (3) and (4) separate, and the lattice simulates a 3D CI. (b) and (c) show the Weyl point trajectories in the BZ while (d) and (e) show the Chern number computed in a 2D plane normal to $q_2$ when the interlayer coupling strengths vary along lines~1 and 2 of (a), respectively. In (b) and (c), $\chi$ is the Weyl point chirality.}
\end{figure}

The dependence of the distance between two Weyl points on the evanescent coupling rates is shown in Fig.~\ref{fig:3}(a). We define $\Delta_\text{W}$ as the distance between the Weyl points within the BZ shown in Fig.~\ref{fig:2}(a). Here, the colored domain (IV) is where the WSM phase exists, whereas the three black domains (I), (II), and (III) are where our system spectrum is gapped. The transition between WSM and insulating phases implies the annihilation of Weyl points through merging. The abrupt change in color between the domains indicates that the Weyl points meet at the BZ boundaries. To illustrate how the Weyl points move and merge, we vary the interlayer coupling rates along ``line~1", and ``line~2" of Fig.~\ref{fig:3}(a), and show the corresponding trajectories in Figs.~\ref{fig:3}(b) and \ref{fig:3}(c). Following line~1, bands (3) and (4) are gapped in domain (I) until we enter the WSM domain, where two Weyl points appear at the BZ corner $(0,0)$ and move towards each other along the diagonal $(q_1=q_2)$ as $V_1$ and $V_2$ increase. Regarding line~2, the two Weyl points appear at $(0.5,0)$ when we go from domain (II) to (IV). They rotate about the BZ center and meet each other at $(0,0.5)$ when the system transitions from WSM to insulator. The two Weyl points are annihilated, and bands (3) and (4) separate again. In Figs.~\ref{fig:3}(b) and \ref{fig:3}(c), we distinguish the Weyl points by their chirality $\chi$, which is proportional to the Berry flux threading through a surface enclosing the Weyl point. These trajectories predicted by the effective Hamiltonian~\eqref{Eq: Hamiltonian} are in excellent agreement with PWE simulation~\cite{Supp}.

%
%
%
%
%
%
\begin{figure*}[!th]
	\includegraphics[width=\linewidth]{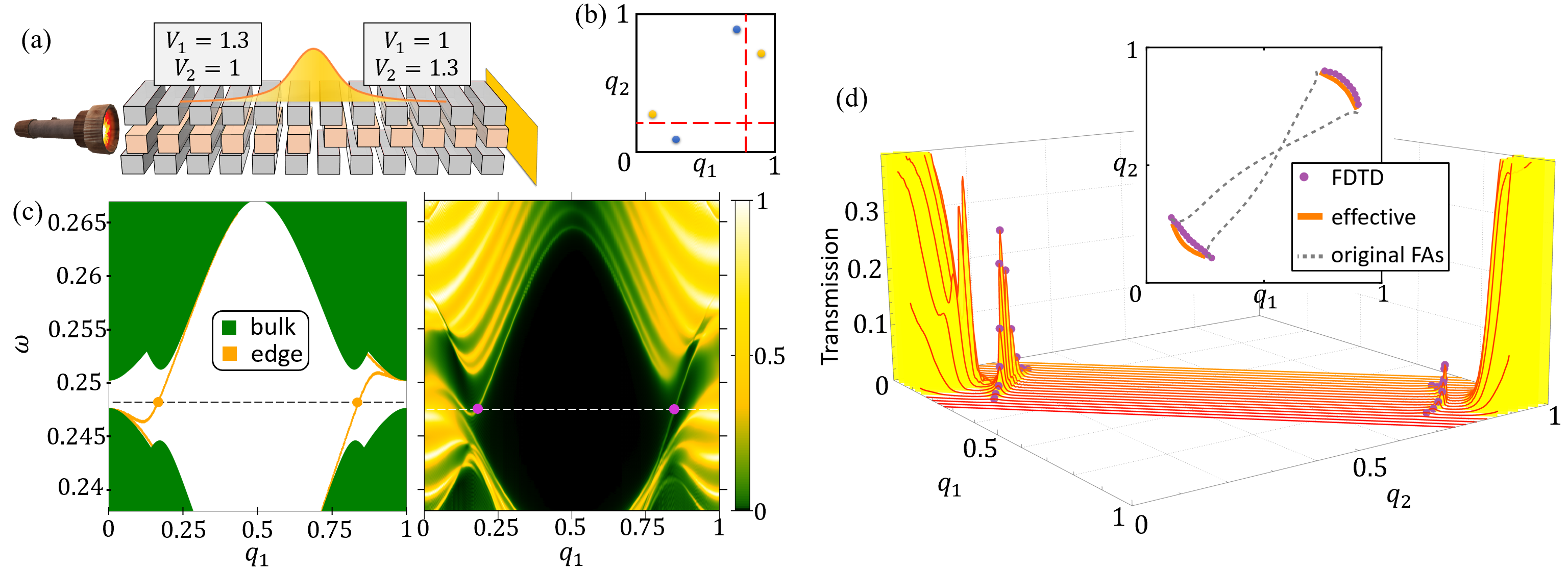}
	\caption{\label{fig:4} Junction of two WSMs. (a) Sketch of the setup for simulation. The intralayer coupling strengths are $U_1=U_3=1$ and $U_2=1.2$. (b) Positions of Weyl points on each side. The yellow dots correspond to $(V_1,V_2)=(1.3,1)$ while the blue ones are for $(V_1,V_2)=(1,1.3)$. (c) shows the energy spectrum of the effective model (left) and the transmission spectrum obtained from FDTD simulation (right) of the systems in the diagonal plane $q_1=q_2$. The dashed lines indicate the frequencies for visualizing the isofrequency transmission over the synthetic BZ in (d). The inset of (d) shows the reconstructed FAs, with the dashed gray lines being the FAs before reconstruction.}
\end{figure*}
%
%
\textit{Topological phases.|} Chern number is the topological invariant characterizing the bulk bands of WSMs. It can be defined for a local energy gap in 2D cross sections of the 3D BZ. Here, we consider 2D planes perpendicular to the $q_2$-axis and compute the Chern number in each plane as a function of $q_2$~\cite{Supp}. The variation of Chern number following lines 1 and 2 of Fig.~\ref{fig:3}(a) are shown in Figs.~\ref{fig:3}(d) and~\ref{fig:3}(e), respectively. Along line~1, domain (I) has a nonzero Chern number over the whole BZ, suggesting a 3D CI phase. When the Weyl points appear and move towards each other, the range of $q_2$ with nonzero Chern number decreases and is limited between these two points. Following line~2, we also realize a CI phase in domain (II) and the WSM phase where the region between two Weyl points has a nonzero Chern number. Remarkably, domain (III) has a vanishing Chern number but is still a CI since the Chern number is now nonzero in all 2D planes normal to $q_1$. Such a WSM-CI transition was theoretically predicted~\cite{Burkov2011a,Yoshimura2016} and observed recently in a photonic lattice for microwave frequencies~\cite{Liu2022v}.

\textit{Fermi arc reconstruction.|} Having a lattice with a controllable WSM phase allows us to emulate the interface states between two WSMs. We consider a photonic junction of two configurations of the trilayer grating | see Fig.~\ref{fig:4}(a). The two WSMs are chosen so that the bulk spectrum of each junction side has two Weyl points, and the straight line connecting them is misaligned with the diagonal plane ($q_1=q_2$). The Weyl points of one side can be transformed into those of the other side via reflection at the diagonal plane~[Fig.~\ref{fig:4}(b)]. Hence, this photonic structure simulates the interface between two WSMs with tilted anisotropy axes. The junction's spectrum in the diagonal plane, obtained by the effective model, is shown in Fig.~\ref{fig:4}(c) (left) together with its transmission spectrum (right) simulated by the finite-difference time-domain (FDTD) method~\cite{FDTD} using MIT Electromagnetic Equation Propagation package~\cite{MEEP}. Comparing with the effective model, which clearly distinguishes the bulk and edge states, we can discern the gapless edge states in the transmission spectrum of our junction.

To see the reconstructed FAs in an isofrequency surface, we choose a frequency close to the bulk Weyl nodes, indicated by the dashed lines in Fig.~\ref{fig:4}(c). The transmission at this frequency over the 2D synthetic BZ is shown in Fig.~\ref{fig:4}(d), where the edge states correspond to the peaks marked by purple dots. The edge-mode transmission is strongest at $q_1=q_2$ and dissipates towards the Weyl points, indicating the nature of FAs to connect with the bulk Weyl nodes. The resultant isofrequency contours are shown in the inset of Fig.~\ref{fig:4}(d), in agreement with the effective model. Besides, we employ the effective model to compute the original FAs of each configuration and demonstrate how they are reconstructed. We see that the two original FAs hybridize with each other and become anticrossing. Whereas the original FAs connect two Weyl points of opposite chirality on the same side, namely the same bulk, the reconstructed ones connect those with the same chirality but on different sides of the junction. This completely agrees with previous theoretical predictions~\cite{Buccheri2022,Ishida2018,Abdulla2021}. The curvature of the reconstructed FAs indicates that the WSMs are strongly coupled~\cite{Ishida2018,Abdulla2021}. Other different cases of photonic junction can be found in the SM~\cite{Supp}.

The reconstructed FAs are robust against perturbations since the bulk Chern number topologically protects them. Specifically, we can find cross sections of the BZ and respectively compute the Chern number on each side so that the difference in Chern number of both sides is nonzero. For instance, the cross sections parallel to $k$ along the red dashed lines of Fig.~\ref{fig:4}(b) have a nonzero difference in Chern number. The interface FAs are thus chiral, as can be seen by the slopes of the edge states in Fig.~\ref{fig:4}(c). To describe their chiral direction, we note that the chiral direction of an original FA is given by $\mathbf{k}_\text{W}\times\mathbf{n}$, where $\mathbf{k}_\text{W}$ is the vector connecting the two Weyl points and $\mathbf{n}$ is a normal vector of the surface that goes towards the trivial media~\cite{Supp}. The vector $\mathbf{k}_\text{W}$ points from the source towards the sink of Berry curvature. The interface FAs are combinations of the original ones, and hence their chiral direction is given by $\Delta\mathbf{k}_\text{W}\times\mathbf{n}$, where $\Delta\mathbf{k}_\text{W}$ is the difference in $\mathbf{k}_\text{W}$ of the two sides.

\textit{Experimental feasibility.|} The gratings in our system can be experimentally realized using standard nanofabrication methods, such as electron beam lithography and ionic dry etching~\cite{Son2018,Cueff2019,Letartre2022}. They are made of silicon (refractive index $n=3.46$) with period $\Lambda=380$ nm for operation in the telecom wavelength range ($\lambda\approx\SI{1.5}{\micro\meter}$). The relative displacement between the three layers can be dynamically tuned via piezoelectric actuators that are combined with goniometer stages for full control of parallelism~\cite{Dufferwiel2014,Li_2019,Geng2020,Vadia2021}. Another possibility is to employ flip-chip bonding~\cite{tang2023onchip} to fabricate rigid structures exhibiting different relative displacements, thus measuring the structures one by one corresponds to probing the eigenmode in the $(q_1,q_2)$-space. The edge states are visualized in far-field spectroscopy by either micro-reflectivity measurements~\cite{Ferrier_2019,Parappurath2020}, or the photoluminescence of emitters that are embedded in the gratings~\cite{Garca_Elcano_2023,Barik2018,JalaliMehrabad:20,Ota2018,Smirnova2020}. In particular, the second scenario can be used to couple edge states to single-photon emitters for quantum effects~\cite{Garca_Elcano_2023, Barik2018,JalaliMehrabad:20}, or to an ensemble of emitters for lasing action~\cite{Ota2018,Smirnova2020}.

\textit{Outlook.|} This letter demonstrates the FA reconstruction between two WSMs in a versatile synthetic photonic lattice. We expect that the confirmation of FA reconstruction presented here would motivate further works about phenomena related to these interface states. Moreover, this photonic system is a stepping stone to investigate not only physics in higher dimensions, such as 4D quantum Hall effect~\cite{Bernevig2013} and 5D Weyl semimetal~\cite{Lian2016} but also non-Hermitian topology, which is a natural question in leaky photonic crystals when operating in the vicinity of the $\Gamma$-point of the 1D BZ~\cite{Letartre2022}.

\emph{Acknowledgments.|}
	We kindly thank Alessandro De Martino for reading the manuscript carefully and giving us valuable comments. We are grateful for the fruitful discussions with Koji Kobayashi, Hai Chau Nguyen, and Aitzol Garcia-Etxarri. The work of D.H.M.N. and D.B. is supported by Ministerio de Ciencia e Innovaci\'on  (MICINN) through Project No. PID2020-120614GB-I00. D.B. thanks the support  of the Transnational Common Laboratory $Quantum-ChemPhys$, the funding from the IKUR Strategy under the collaboration agreement between the Ikerbasque Foundation and DIPC on behalf of the Department of Education of the Basque Government and from the Gipuzkoa Provincial Council within the QUAN-000021-01 project. H.S.N is funded by the French National Research Agency (ANR) under the project POPEYE (ANR-17-CE24-0020) and the IDEXLYON from Université de Lyon, Scientific Breakthrough project TORE within the Programme Investissements d’Avenir (ANR-19-IDEX-0005). D.X.N. is supported by grant IBS-R024-D1. C.D. acknowledges support from the Spanish Ministerio de Ciencia e Innovaci\'on (PID2019-109905GAC2) and Eusko Jaurlaritza (IT1164-19, KK-2019/00101 and KK-2021/00082) through the FPI Ph.D. Fellowship CEX2018-000867-S-19-1.

\bibliography{Minh2023}

\begin{thebibliography}{80}%
\makeatletter
\providecommand \@ifxundefined [1]{%
 \@ifx{#1\undefined}
}%
\providecommand \@ifnum [1]{%
 \ifnum #1\expandafter \@firstoftwo
 \else \expandafter \@secondoftwo
 \fi
}%
\providecommand \@ifx [1]{%
 \ifx #1\expandafter \@firstoftwo
 \else \expandafter \@secondoftwo
 \fi
}%
\providecommand \natexlab [1]{#1}%
\providecommand \enquote  [1]{``#1''}%
\providecommand \bibnamefont  [1]{#1}%
\providecommand \bibfnamefont [1]{#1}%
\providecommand \citenamefont [1]{#1}%
\providecommand \href@noop [0]{\@secondoftwo}%
\providecommand \href [0]{\begingroup \@sanitize@url \@href}%
\providecommand \@href[1]{\@@startlink{#1}\@@href}%
\providecommand \@@href[1]{\endgroup#1\@@endlink}%
\providecommand \@sanitize@url [0]{\catcode `\\12\catcode `\$12\catcode
  `\&12\catcode `\#12\catcode `\^12\catcode `\_12\catcode `\%12\relax}%
\providecommand \@@startlink[1]{}%
\providecommand \@@endlink[0]{}%
\providecommand \url  [0]{\begingroup\@sanitize@url \@url }%
\providecommand \@url [1]{\endgroup\@href {#1}{\urlprefix }}%
\providecommand \urlprefix  [0]{URL }%
\providecommand \Eprint [0]{\href }%
\providecommand \doibase [0]{https://doi.org/}%
\providecommand \selectlanguage [0]{\@gobble}%
\providecommand \bibinfo  [0]{\@secondoftwo}%
\providecommand \bibfield  [0]{\@secondoftwo}%
\providecommand \translation [1]{[#1]}%
\providecommand \BibitemOpen [0]{}%
\providecommand \bibitemStop [0]{}%
\providecommand \bibitemNoStop [0]{.\EOS\space}%
\providecommand \EOS [0]{\spacefactor3000\relax}%
\providecommand \BibitemShut  [1]{\csname bibitem#1\endcsname}%
\let\auto@bib@innerbib\@empty
\bibitem [{\citenamefont {Armitage}\ \emph {et~al.}(2018)\citenamefont
  {Armitage}, \citenamefont {Mele},\ and\ \citenamefont
  {Vishwanath}}]{Armitage2018}%
  \BibitemOpen
  \bibfield  {author} {\bibinfo {author} {\bibfnamefont {N.~P.}\ \bibnamefont
  {Armitage}}, \bibinfo {author} {\bibfnamefont {E.~J.}\ \bibnamefont {Mele}},\
  and\ \bibinfo {author} {\bibfnamefont {A.}~\bibnamefont {Vishwanath}},\
  }\bibfield  {title} {\bibinfo {title} {{Weyl and Dirac semimetals in
  three-dimensional solids}},\ }\href
  {https://doi.org/10.1103/RevModPhys.90.015001} {\bibfield  {journal}
  {\bibinfo  {journal} {Rev. Mod. Phys.}\ }\textbf {\bibinfo {volume} {90}},\
  \bibinfo {pages} {015001} (\bibinfo {year} {2018})}\BibitemShut {NoStop}%
\bibitem [{\citenamefont {Burkov}(2018)}]{Burkov2018}%
  \BibitemOpen
  \bibfield  {author} {\bibinfo {author} {\bibfnamefont {A.}~\bibnamefont
  {Burkov}},\ }\bibfield  {title} {\bibinfo {title} {{Weyl Metals}},\ }\href
  {https://doi.org/10.1146/annurev-conmatphys-033117-054129} {\bibfield
  {journal} {\bibinfo  {journal} {Annu. Rev. Condens. Matter Phys.}\ }\textbf
  {\bibinfo {volume} {9}},\ \bibinfo {pages} {359} (\bibinfo {year}
  {2018})}\BibitemShut {NoStop}%
\bibitem [{\citenamefont {Lv}\ \emph {et~al.}(2021)\citenamefont {Lv},
  \citenamefont {Qian},\ and\ \citenamefont {Ding}}]{Lv2021}%
  \BibitemOpen
  \bibfield  {author} {\bibinfo {author} {\bibfnamefont {B.~Q.}\ \bibnamefont
  {Lv}}, \bibinfo {author} {\bibfnamefont {T.}~\bibnamefont {Qian}},\ and\
  \bibinfo {author} {\bibfnamefont {H.}~\bibnamefont {Ding}},\ }\bibfield
  {title} {\bibinfo {title} {{Experimental perspective on three-dimensional
  topological semimetals}},\ }\href
  {https://doi.org/10.1103/RevModPhys.93.025002} {\bibfield  {journal}
  {\bibinfo  {journal} {Rev. Mod. Phys.}\ }\textbf {\bibinfo {volume} {93}},\
  \bibinfo {pages} {025002} (\bibinfo {year} {2021})}\BibitemShut {NoStop}%
\bibitem [{\citenamefont {Wan}\ \emph {et~al.}(2011)\citenamefont {Wan},
  \citenamefont {Turner}, \citenamefont {Vishwanath},\ and\ \citenamefont
  {Savrasov}}]{Wan2011}%
  \BibitemOpen
  \bibfield  {author} {\bibinfo {author} {\bibfnamefont {X.}~\bibnamefont
  {Wan}}, \bibinfo {author} {\bibfnamefont {A.~M.}\ \bibnamefont {Turner}},
  \bibinfo {author} {\bibfnamefont {A.}~\bibnamefont {Vishwanath}},\ and\
  \bibinfo {author} {\bibfnamefont {S.~Y.}\ \bibnamefont {Savrasov}},\
  }\bibfield  {title} {\bibinfo {title} {{Topological semimetal and Fermi-arc
  surface states in the electronic structure of pyrochlore iridates}},\ }\href
  {https://doi.org/10.1103/PhysRevB.83.205101} {\bibfield  {journal} {\bibinfo
  {journal} {Phys. Rev. B}\ }\textbf {\bibinfo {volume} {83}},\ \bibinfo
  {pages} {205101} (\bibinfo {year} {2011})}\BibitemShut {NoStop}%
\bibitem [{\citenamefont {Dub\ifmmode~\check{c}\else \v{c}\fi{}ek}\ \emph
  {et~al.}(2015)\citenamefont {Dub\ifmmode~\check{c}\else \v{c}\fi{}ek},
  \citenamefont {Kennedy}, \citenamefont {Lu}, \citenamefont {Ketterle},
  \citenamefont {Solja\ifmmode \check{c}\else
  \v{c}\fi{}i\ifmmode~\acute{c}\else \'{c}\fi{}},\ and\ \citenamefont
  {Buljan}}]{Dubcek2015}%
  \BibitemOpen
  \bibfield  {author} {\bibinfo {author} {\bibfnamefont {T.}~\bibnamefont
  {Dub\ifmmode~\check{c}\else \v{c}\fi{}ek}}, \bibinfo {author} {\bibfnamefont
  {C.~J.}\ \bibnamefont {Kennedy}}, \bibinfo {author} {\bibfnamefont
  {L.}~\bibnamefont {Lu}}, \bibinfo {author} {\bibfnamefont {W.}~\bibnamefont
  {Ketterle}}, \bibinfo {author} {\bibfnamefont {M.}~\bibnamefont
  {Solja\ifmmode \check{c}\else \v{c}\fi{}i\ifmmode~\acute{c}\else
  \'{c}\fi{}}},\ and\ \bibinfo {author} {\bibfnamefont {H.}~\bibnamefont
  {Buljan}},\ }\bibfield  {title} {\bibinfo {title} {{Weyl Points in
  Three-Dimensional Optical Lattices: Synthetic Magnetic Monopoles in Momentum
  Space}},\ }\href {https://doi.org/10.1103/PhysRevLett.114.225301} {\bibfield
  {journal} {\bibinfo  {journal} {Phys. Rev. Lett.}\ }\textbf {\bibinfo
  {volume} {114}},\ \bibinfo {pages} {225301} (\bibinfo {year}
  {2015})}\BibitemShut {NoStop}%
\bibitem [{\citenamefont {Lu}\ \emph {et~al.}(2015)\citenamefont {Lu},
  \citenamefont {Wang}, \citenamefont {Ye}, \citenamefont {Ran}, \citenamefont
  {Fu}, \citenamefont {Joannopoulos},\ and\ \citenamefont
  {Solja{\v{c}}i{\'{c}}}}]{Lu2015}%
  \BibitemOpen
  \bibfield  {author} {\bibinfo {author} {\bibfnamefont {L.}~\bibnamefont
  {Lu}}, \bibinfo {author} {\bibfnamefont {Z.}~\bibnamefont {Wang}}, \bibinfo
  {author} {\bibfnamefont {D.}~\bibnamefont {Ye}}, \bibinfo {author}
  {\bibfnamefont {L.}~\bibnamefont {Ran}}, \bibinfo {author} {\bibfnamefont
  {L.}~\bibnamefont {Fu}}, \bibinfo {author} {\bibfnamefont {J.~D.}\
  \bibnamefont {Joannopoulos}},\ and\ \bibinfo {author} {\bibfnamefont
  {M.}~\bibnamefont {Solja{\v{c}}i{\'{c}}}},\ }\bibfield  {title} {\bibinfo
  {title} {{Experimental observation of Weyl points}},\ }\href
  {https://doi.org/10.1126/science.aaa9273} {\bibfield  {journal} {\bibinfo
  {journal} {Science}\ }\textbf {\bibinfo {volume} {349}},\ \bibinfo {pages}
  {622} (\bibinfo {year} {2015})}\BibitemShut {NoStop}%
\bibitem [{\citenamefont {Lin}\ \emph {et~al.}(2016)\citenamefont {Lin},
  \citenamefont {Xiao}, \citenamefont {Yuan},\ and\ \citenamefont
  {Fan}}]{Lin2016}%
  \BibitemOpen
  \bibfield  {author} {\bibinfo {author} {\bibfnamefont {Q.}~\bibnamefont
  {Lin}}, \bibinfo {author} {\bibfnamefont {M.}~\bibnamefont {Xiao}}, \bibinfo
  {author} {\bibfnamefont {L.}~\bibnamefont {Yuan}},\ and\ \bibinfo {author}
  {\bibfnamefont {S.}~\bibnamefont {Fan}},\ }\bibfield  {title} {\bibinfo
  {title} {{Photonic Weyl point in a two-dimensional resonator lattice with a
  synthetic frequency dimension}},\ }\href
  {https://doi.org/10.1038/ncomms13731} {\bibfield  {journal} {\bibinfo
  {journal} {Nat. Commun.}\ }\textbf {\bibinfo {volume} {7}},\ \bibinfo {pages}
  {13731} (\bibinfo {year} {2016})}\BibitemShut {NoStop}%
\bibitem [{\citenamefont {Noh}\ \emph {et~al.}(2017)\citenamefont {Noh},
  \citenamefont {Huang}, \citenamefont {Leykam}, \citenamefont {Chong},
  \citenamefont {Chen},\ and\ \citenamefont {Rechtsman}}]{Noh2017}%
  \BibitemOpen
  \bibfield  {author} {\bibinfo {author} {\bibfnamefont {J.}~\bibnamefont
  {Noh}}, \bibinfo {author} {\bibfnamefont {S.}~\bibnamefont {Huang}}, \bibinfo
  {author} {\bibfnamefont {D.}~\bibnamefont {Leykam}}, \bibinfo {author}
  {\bibfnamefont {Y.~D.}\ \bibnamefont {Chong}}, \bibinfo {author}
  {\bibfnamefont {K.~P.}\ \bibnamefont {Chen}},\ and\ \bibinfo {author}
  {\bibfnamefont {M.~C.}\ \bibnamefont {Rechtsman}},\ }\bibfield  {title}
  {\bibinfo {title} {{Experimental observation of optical Weyl points and Fermi
  arc-like surface states}},\ }\href {https://doi.org/10.1038/nphys4072}
  {\bibfield  {journal} {\bibinfo  {journal} {Nat. Phys.}\ }\textbf {\bibinfo
  {volume} {13}},\ \bibinfo {pages} {611} (\bibinfo {year} {2017})}\BibitemShut
  {NoStop}%
\bibitem [{\citenamefont {Wang}\ \emph {et~al.}(2017)\citenamefont {Wang},
  \citenamefont {Xiao}, \citenamefont {Liu}, \citenamefont {Zhu},\ and\
  \citenamefont {Chan}}]{Wang2017}%
  \BibitemOpen
  \bibfield  {author} {\bibinfo {author} {\bibfnamefont {Q.}~\bibnamefont
  {Wang}}, \bibinfo {author} {\bibfnamefont {M.}~\bibnamefont {Xiao}}, \bibinfo
  {author} {\bibfnamefont {H.}~\bibnamefont {Liu}}, \bibinfo {author}
  {\bibfnamefont {S.}~\bibnamefont {Zhu}},\ and\ \bibinfo {author}
  {\bibfnamefont {C.~T.}\ \bibnamefont {Chan}},\ }\bibfield  {title} {\bibinfo
  {title} {{Optical Interface States Protected by Synthetic Weyl Points}},\
  }\href {https://doi.org/10.1103/PhysRevX.7.031032} {\bibfield  {journal}
  {\bibinfo  {journal} {Phys. Rev. X}\ }\textbf {\bibinfo {volume} {7}},\
  \bibinfo {pages} {031032} (\bibinfo {year} {2017})}\BibitemShut {NoStop}%
\bibitem [{\citenamefont {Qin}\ \emph {et~al.}(2018)\citenamefont {Qin},
  \citenamefont {Liu}, \citenamefont {Wang},\ and\ \citenamefont
  {Lu}}]{Qin2018}%
  \BibitemOpen
  \bibfield  {author} {\bibinfo {author} {\bibfnamefont {C.}~\bibnamefont
  {Qin}}, \bibinfo {author} {\bibfnamefont {Q.}~\bibnamefont {Liu}}, \bibinfo
  {author} {\bibfnamefont {B.}~\bibnamefont {Wang}},\ and\ \bibinfo {author}
  {\bibfnamefont {P.}~\bibnamefont {Lu}},\ }\bibfield  {title} {\bibinfo
  {title} {{Photonic Weyl phase transition in dynamically modulated brick-wall
  waveguide arrays}},\ }\href {https://doi.org/10.1364/oe.26.020929} {\bibfield
   {journal} {\bibinfo  {journal} {Opt. Express}\ }\textbf {\bibinfo {volume}
  {26}},\ \bibinfo {pages} {20929} (\bibinfo {year} {2018})}\BibitemShut
  {NoStop}%
\bibitem [{\citenamefont {Li}\ \emph {et~al.}(2021)\citenamefont {Li},
  \citenamefont {Song},\ and\ \citenamefont {Jiang}}]{Li2021}%
  \BibitemOpen
  \bibfield  {author} {\bibinfo {author} {\bibfnamefont {M.}~\bibnamefont
  {Li}}, \bibinfo {author} {\bibfnamefont {J.}~\bibnamefont {Song}},\ and\
  \bibinfo {author} {\bibfnamefont {Y.}~\bibnamefont {Jiang}},\ }\bibfield
  {title} {\bibinfo {title} {{Photonic topological Weyl degeneracies and ideal
  type-I Weyl points in the gyromagnetic metamaterials}},\ }\href
  {https://doi.org/10.1103/PhysRevB.103.045307} {\bibfield  {journal} {\bibinfo
   {journal} {Phys. Rev. B}\ }\textbf {\bibinfo {volume} {103}},\ \bibinfo
  {pages} {045307} (\bibinfo {year} {2021})}\BibitemShut {NoStop}%
\bibitem [{\citenamefont {Lustig}\ and\ \citenamefont
  {Segev}(2021)}]{Lustig2021}%
  \BibitemOpen
  \bibfield  {author} {\bibinfo {author} {\bibfnamefont {E.}~\bibnamefont
  {Lustig}}\ and\ \bibinfo {author} {\bibfnamefont {M.}~\bibnamefont {Segev}},\
  }\bibfield  {title} {\bibinfo {title} {{Topological photonics in synthetic
  dimensions}},\ }\href {https://doi.org/10.1364/AOP.418074} {\bibfield
  {journal} {\bibinfo  {journal} {Adv. Opt. Photon.}\ }\textbf {\bibinfo
  {volume} {13}},\ \bibinfo {pages} {426} (\bibinfo {year} {2021})}\BibitemShut
  {NoStop}%
\bibitem [{\citenamefont {Cheng}\ \emph {et~al.}(2020)\citenamefont {Cheng},
  \citenamefont {Gao}, \citenamefont {Bi}, \citenamefont {Liu}, \citenamefont
  {Li}, \citenamefont {Guo}, \citenamefont {Yang}, \citenamefont {You},
  \citenamefont {Feng}, \citenamefont {Sun}, \citenamefont {Tian},
  \citenamefont {Chen},\ and\ \citenamefont {Zhang}}]{Cheng2020}%
  \BibitemOpen
  \bibfield  {author} {\bibinfo {author} {\bibfnamefont {H.}~\bibnamefont
  {Cheng}}, \bibinfo {author} {\bibfnamefont {W.}~\bibnamefont {Gao}}, \bibinfo
  {author} {\bibfnamefont {Y.}~\bibnamefont {Bi}}, \bibinfo {author}
  {\bibfnamefont {W.}~\bibnamefont {Liu}}, \bibinfo {author} {\bibfnamefont
  {Z.}~\bibnamefont {Li}}, \bibinfo {author} {\bibfnamefont {Q.}~\bibnamefont
  {Guo}}, \bibinfo {author} {\bibfnamefont {Y.}~\bibnamefont {Yang}}, \bibinfo
  {author} {\bibfnamefont {O.}~\bibnamefont {You}}, \bibinfo {author}
  {\bibfnamefont {J.}~\bibnamefont {Feng}}, \bibinfo {author} {\bibfnamefont
  {H.}~\bibnamefont {Sun}}, \bibinfo {author} {\bibfnamefont {J.}~\bibnamefont
  {Tian}}, \bibinfo {author} {\bibfnamefont {S.}~\bibnamefont {Chen}},\ and\
  \bibinfo {author} {\bibfnamefont {S.}~\bibnamefont {Zhang}},\ }\bibfield
  {title} {\bibinfo {title} {{Vortical Reflection and Spiraling Fermi Arcs with
  Weyl Metamaterials}},\ }\href
  {https://doi.org/10.1103/PhysRevLett.125.093904} {\bibfield  {journal}
  {\bibinfo  {journal} {Phys. Rev. Lett.}\ }\textbf {\bibinfo {volume} {125}},\
  \bibinfo {pages} {093904} (\bibinfo {year} {2020})}\BibitemShut {NoStop}%
\bibitem [{\citenamefont {Han}\ \emph {et~al.}(2022)\citenamefont {Han},
  \citenamefont {Liu}, \citenamefont {Gao}, \citenamefont {Zhou},\ and\
  \citenamefont {Liu}}]{Han2022}%
  \BibitemOpen
  \bibfield  {author} {\bibinfo {author} {\bibfnamefont {N.}~\bibnamefont
  {Han}}, \bibinfo {author} {\bibfnamefont {J.}~\bibnamefont {Liu}}, \bibinfo
  {author} {\bibfnamefont {Y.}~\bibnamefont {Gao}}, \bibinfo {author}
  {\bibfnamefont {K.}~\bibnamefont {Zhou}},\ and\ \bibinfo {author}
  {\bibfnamefont {S.}~\bibnamefont {Liu}},\ }\bibfield  {title} {\bibinfo
  {title} {{Topological phase transitions and Weyl semimetal phases in chiral
  photonic metamaterials}},\ }\href {https://doi.org/10.1088/1367-2630/ac6f78}
  {\bibfield  {journal} {\bibinfo  {journal} {New J. Phys.}\ }\textbf {\bibinfo
  {volume} {24}},\ \bibinfo {pages} {053052} (\bibinfo {year}
  {2022})}\BibitemShut {NoStop}%
\bibitem [{\citenamefont {Song}\ \emph {et~al.}(2023)\citenamefont {Song},
  \citenamefont {Wu}, \citenamefont {Chen}, \citenamefont {Chen}, \citenamefont
  {Huang}, \citenamefont {Yuan}, \citenamefont {Zhu},\ and\ \citenamefont
  {Li}}]{Song2023}%
  \BibitemOpen
  \bibfield  {author} {\bibinfo {author} {\bibfnamefont {W.}~\bibnamefont
  {Song}}, \bibinfo {author} {\bibfnamefont {S.}~\bibnamefont {Wu}}, \bibinfo
  {author} {\bibfnamefont {C.}~\bibnamefont {Chen}}, \bibinfo {author}
  {\bibfnamefont {Y.}~\bibnamefont {Chen}}, \bibinfo {author} {\bibfnamefont
  {C.}~\bibnamefont {Huang}}, \bibinfo {author} {\bibfnamefont
  {L.}~\bibnamefont {Yuan}}, \bibinfo {author} {\bibfnamefont {S.}~\bibnamefont
  {Zhu}},\ and\ \bibinfo {author} {\bibfnamefont {T.}~\bibnamefont {Li}},\
  }\bibfield  {title} {\bibinfo {title} {Observation of weyl interface states
  in non-hermitian synthetic photonic systems},\ }\href
  {https://doi.org/10.1103/PhysRevLett.130.043803} {\bibfield  {journal}
  {\bibinfo  {journal} {Phys. Rev. Lett.}\ }\textbf {\bibinfo {volume} {130}},\
  \bibinfo {pages} {043803} (\bibinfo {year} {2023})}\BibitemShut {NoStop}%
\bibitem [{\citenamefont {Xiao}\ \emph {et~al.}(2015)\citenamefont {Xiao},
  \citenamefont {Chen}, \citenamefont {He},\ and\ \citenamefont
  {Chan}}]{Xiao2015}%
  \BibitemOpen
  \bibfield  {author} {\bibinfo {author} {\bibfnamefont {M.}~\bibnamefont
  {Xiao}}, \bibinfo {author} {\bibfnamefont {W.-J.}\ \bibnamefont {Chen}},
  \bibinfo {author} {\bibfnamefont {W.-Y.}\ \bibnamefont {He}},\ and\ \bibinfo
  {author} {\bibfnamefont {C.~T.}\ \bibnamefont {Chan}},\ }\bibfield  {title}
  {\bibinfo {title} {{Synthetic gauge flux and Weyl points in acoustic
  systems}},\ }\href {https://doi.org/10.1038/nphys3458} {\bibfield  {journal}
  {\bibinfo  {journal} {Nat. Phys.}\ }\textbf {\bibinfo {volume} {11}},\
  \bibinfo {pages} {920} (\bibinfo {year} {2015})}\BibitemShut {NoStop}%
\bibitem [{\citenamefont {Li}\ \emph {et~al.}(2017)\citenamefont {Li},
  \citenamefont {Huang}, \citenamefont {Lu}, \citenamefont {Ma},\ and\
  \citenamefont {Liu}}]{Li2017}%
  \BibitemOpen
  \bibfield  {author} {\bibinfo {author} {\bibfnamefont {F.}~\bibnamefont
  {Li}}, \bibinfo {author} {\bibfnamefont {X.}~\bibnamefont {Huang}}, \bibinfo
  {author} {\bibfnamefont {J.}~\bibnamefont {Lu}}, \bibinfo {author}
  {\bibfnamefont {J.}~\bibnamefont {Ma}},\ and\ \bibinfo {author}
  {\bibfnamefont {Z.}~\bibnamefont {Liu}},\ }\bibfield  {title} {\bibinfo
  {title} {{Weyl points and Fermi arcs in a chiral phononic~crystal}},\ }\href
  {https://doi.org/10.1038/nphys4275} {\bibfield  {journal} {\bibinfo
  {journal} {Nat. Phys.}\ }\textbf {\bibinfo {volume} {14}},\ \bibinfo {pages}
  {30} (\bibinfo {year} {2017})}\BibitemShut {NoStop}%
\bibitem [{\citenamefont {Ge}\ \emph {et~al.}(2018)\citenamefont {Ge},
  \citenamefont {Ni}, \citenamefont {Tian}, \citenamefont {Gupta},
  \citenamefont {Lu}, \citenamefont {Lin}, \citenamefont {Huang}, \citenamefont
  {Chan},\ and\ \citenamefont {Chen}}]{Ge2018}%
  \BibitemOpen
  \bibfield  {author} {\bibinfo {author} {\bibfnamefont {H.}~\bibnamefont
  {Ge}}, \bibinfo {author} {\bibfnamefont {X.}~\bibnamefont {Ni}}, \bibinfo
  {author} {\bibfnamefont {Y.}~\bibnamefont {Tian}}, \bibinfo {author}
  {\bibfnamefont {S.~K.}\ \bibnamefont {Gupta}}, \bibinfo {author}
  {\bibfnamefont {M.-H.}\ \bibnamefont {Lu}}, \bibinfo {author} {\bibfnamefont
  {X.}~\bibnamefont {Lin}}, \bibinfo {author} {\bibfnamefont {W.-D.}\
  \bibnamefont {Huang}}, \bibinfo {author} {\bibfnamefont {C.~T.}\ \bibnamefont
  {Chan}},\ and\ \bibinfo {author} {\bibfnamefont {Y.-F.}\ \bibnamefont
  {Chen}},\ }\bibfield  {title} {\bibinfo {title} {{Experimental Observation of
  Acoustic Weyl Points and Topological Surface States}},\ }\href
  {https://doi.org/10.1103/PhysRevApplied.10.014017} {\bibfield  {journal}
  {\bibinfo  {journal} {Phys. Rev. Applied}\ }\textbf {\bibinfo {volume}
  {10}},\ \bibinfo {pages} {014017} (\bibinfo {year} {2018})}\BibitemShut
  {NoStop}%
\bibitem [{\citenamefont {Fan}\ \emph {et~al.}(2019)\citenamefont {Fan},
  \citenamefont {Qiu}, \citenamefont {Shen}, \citenamefont {He}, \citenamefont
  {Xiao}, \citenamefont {Ke},\ and\ \citenamefont {Liu}}]{Fan2019}%
  \BibitemOpen
  \bibfield  {author} {\bibinfo {author} {\bibfnamefont {X.}~\bibnamefont
  {Fan}}, \bibinfo {author} {\bibfnamefont {C.}~\bibnamefont {Qiu}}, \bibinfo
  {author} {\bibfnamefont {Y.}~\bibnamefont {Shen}}, \bibinfo {author}
  {\bibfnamefont {H.}~\bibnamefont {He}}, \bibinfo {author} {\bibfnamefont
  {M.}~\bibnamefont {Xiao}}, \bibinfo {author} {\bibfnamefont {M.}~\bibnamefont
  {Ke}},\ and\ \bibinfo {author} {\bibfnamefont {Z.}~\bibnamefont {Liu}},\
  }\bibfield  {title} {\bibinfo {title} {{Probing Weyl Physics with
  One-Dimensional Sonic Crystals}},\ }\href
  {https://doi.org/10.1103/PhysRevLett.122.136802} {\bibfield  {journal}
  {\bibinfo  {journal} {Phys. Rev. Lett.}\ }\textbf {\bibinfo {volume} {122}},\
  \bibinfo {pages} {136802} (\bibinfo {year} {2019})}\BibitemShut {NoStop}%
\bibitem [{\citenamefont {Peri}\ \emph {et~al.}(2019)\citenamefont {Peri},
  \citenamefont {Serra-Garcia}, \citenamefont {Ilan},\ and\ \citenamefont
  {Huber}}]{Peri2019}%
  \BibitemOpen
  \bibfield  {author} {\bibinfo {author} {\bibfnamefont {V.}~\bibnamefont
  {Peri}}, \bibinfo {author} {\bibfnamefont {M.}~\bibnamefont {Serra-Garcia}},
  \bibinfo {author} {\bibfnamefont {R.}~\bibnamefont {Ilan}},\ and\ \bibinfo
  {author} {\bibfnamefont {S.~D.}\ \bibnamefont {Huber}},\ }\bibfield  {title}
  {\bibinfo {title} {{Axial-field-induced chiral channels in an acoustic Weyl
  system}},\ }\href {https://doi.org/10.1038/s41567-019-0415-x} {\bibfield
  {journal} {\bibinfo  {journal} {Nat. Phys.}\ }\textbf {\bibinfo {volume}
  {15}},\ \bibinfo {pages} {357} (\bibinfo {year} {2019})}\BibitemShut
  {NoStop}%
\bibitem [{\citenamefont {Wang}\ \emph {et~al.}(2021)\citenamefont {Wang},
  \citenamefont {Wang}, \citenamefont {Li}, \citenamefont {Luo}, \citenamefont
  {Wang}, \citenamefont {Liu},\ and\ \citenamefont {Yang}}]{Wang2021}%
  \BibitemOpen
  \bibfield  {author} {\bibinfo {author} {\bibfnamefont {Z.}~\bibnamefont
  {Wang}}, \bibinfo {author} {\bibfnamefont {Z.}~\bibnamefont {Wang}}, \bibinfo
  {author} {\bibfnamefont {H.}~\bibnamefont {Li}}, \bibinfo {author}
  {\bibfnamefont {J.}~\bibnamefont {Luo}}, \bibinfo {author} {\bibfnamefont
  {X.}~\bibnamefont {Wang}}, \bibinfo {author} {\bibfnamefont {Z.}~\bibnamefont
  {Liu}},\ and\ \bibinfo {author} {\bibfnamefont {H.}~\bibnamefont {Yang}},\
  }\bibfield  {title} {\bibinfo {title} {{Weyl points and nodal lines in
  acoustic synthetic parameter space}},\ }\href
  {https://doi.org/10.35848/1882-0786/ac0c8b} {\bibfield  {journal} {\bibinfo
  {journal} {Appl. Phys. Express}\ ,\ \bibinfo {pages} {077002}} (\bibinfo
  {year} {2021})}\BibitemShut {NoStop}%
\bibitem [{\citenamefont {Zhang}\ \emph {et~al.}(2021)\citenamefont {Zhang},
  \citenamefont {Zhang}, \citenamefont {Liu},\ and\ \citenamefont
  {Liu}}]{Zhang2021}%
  \BibitemOpen
  \bibfield  {author} {\bibinfo {author} {\bibfnamefont {H.}~\bibnamefont
  {Zhang}}, \bibinfo {author} {\bibfnamefont {S.}~\bibnamefont {Zhang}},
  \bibinfo {author} {\bibfnamefont {J.}~\bibnamefont {Liu}},\ and\ \bibinfo
  {author} {\bibfnamefont {B.}~\bibnamefont {Liu}},\ }\bibfield  {title}
  {\bibinfo {title} {{Synthetic Weyl points of the shear horizontal guided
  waves in one-dimensional phononic crystal plates}},\ }\href
  {https://doi.org/10.3390/app12010167} {\bibfield  {journal} {\bibinfo
  {journal} {Appl. Sci. (Basel)}\ }\textbf {\bibinfo {volume} {12}},\ \bibinfo
  {pages} {167} (\bibinfo {year} {2021})}\BibitemShut {NoStop}%
\bibitem [{\citenamefont {Liu}\ \emph {et~al.}(2022{\natexlab{a}})\citenamefont
  {Liu}, \citenamefont {Li}, \citenamefont {Chen}, \citenamefont {Tang},
  \citenamefont {Chen}, \citenamefont {Liang}, \citenamefont {Ma},\ and\
  \citenamefont {Cheng}}]{Liu2022}%
  \BibitemOpen
  \bibfield  {author} {\bibinfo {author} {\bibfnamefont {J.-j.}\ \bibnamefont
  {Liu}}, \bibinfo {author} {\bibfnamefont {Z.-w.}\ \bibnamefont {Li}},
  \bibinfo {author} {\bibfnamefont {Z.-G.}\ \bibnamefont {Chen}}, \bibinfo
  {author} {\bibfnamefont {W.}~\bibnamefont {Tang}}, \bibinfo {author}
  {\bibfnamefont {A.}~\bibnamefont {Chen}}, \bibinfo {author} {\bibfnamefont
  {B.}~\bibnamefont {Liang}}, \bibinfo {author} {\bibfnamefont
  {G.}~\bibnamefont {Ma}},\ and\ \bibinfo {author} {\bibfnamefont {J.-C.}\
  \bibnamefont {Cheng}},\ }\bibfield  {title} {\bibinfo {title} {{Experimental
  Realization of Weyl Exceptional Rings in a Synthetic Three-Dimensional
  Non-Hermitian Phononic Crystal}},\ }\href
  {https://doi.org/10.1103/PhysRevLett.129.084301} {\bibfield  {journal}
  {\bibinfo  {journal} {Phys. Rev. Lett.}\ }\textbf {\bibinfo {volume} {129}},\
  \bibinfo {pages} {084301} (\bibinfo {year} {2022}{\natexlab{a}})}\BibitemShut
  {NoStop}%
\bibitem [{\citenamefont {Xu}\ \emph {et~al.}(2011)\citenamefont {Xu},
  \citenamefont {Weng}, \citenamefont {Wang}, \citenamefont {Dai},\ and\
  \citenamefont {Fang}}]{Xu2011}%
  \BibitemOpen
  \bibfield  {author} {\bibinfo {author} {\bibfnamefont {G.}~\bibnamefont
  {Xu}}, \bibinfo {author} {\bibfnamefont {H.}~\bibnamefont {Weng}}, \bibinfo
  {author} {\bibfnamefont {Z.}~\bibnamefont {Wang}}, \bibinfo {author}
  {\bibfnamefont {X.}~\bibnamefont {Dai}},\ and\ \bibinfo {author}
  {\bibfnamefont {Z.}~\bibnamefont {Fang}},\ }\bibfield  {title} {\bibinfo
  {title} {{Chern Semimetal and the Quantized Anomalous Hall Effect in
  ${\mathrm{HgCr}}_{2}{\mathrm{Se}}_{4}$}},\ }\href
  {https://doi.org/10.1103/PhysRevLett.107.186806} {\bibfield  {journal}
  {\bibinfo  {journal} {Phys. Rev. Lett.}\ }\textbf {\bibinfo {volume} {107}},\
  \bibinfo {pages} {186806} (\bibinfo {year} {2011})}\BibitemShut {NoStop}%
\bibitem [{\citenamefont {Burkov}\ and\ \citenamefont
  {Balents}(2011)}]{Burkov2011a}%
  \BibitemOpen
  \bibfield  {author} {\bibinfo {author} {\bibfnamefont {A.~A.}\ \bibnamefont
  {Burkov}}\ and\ \bibinfo {author} {\bibfnamefont {L.}~\bibnamefont
  {Balents}},\ }\bibfield  {title} {\bibinfo {title} {{Weyl Semimetal in a
  Topological Insulator Multilayer}},\ }\href
  {https://doi.org/10.1103/PhysRevLett.107.127205} {\bibfield  {journal}
  {\bibinfo  {journal} {Phys. Rev. Lett.}\ }\textbf {\bibinfo {volume} {107}},\
  \bibinfo {pages} {127205} (\bibinfo {year} {2011})}\BibitemShut {NoStop}%
\bibitem [{\citenamefont {Witten}(2016)}]{Witten2016}%
  \BibitemOpen
  \bibfield  {author} {\bibinfo {author} {\bibfnamefont {E.}~\bibnamefont
  {Witten}},\ }\bibfield  {title} {\bibinfo {title} {Three lectures on
  topological phases of matter},\ }\href
  {https://doi.org/10.1393/ncr/i2016-10125-3} {\bibfield  {journal} {\bibinfo
  {journal} {Riv. Nuovo Cimento}\ }\textbf {\bibinfo {volume} {39}},\ \bibinfo
  {pages} {313–370} (\bibinfo {year} {2016})}\BibitemShut {NoStop}%
\bibitem [{\citenamefont {Potter}\ \emph {et~al.}(2014)\citenamefont {Potter},
  \citenamefont {Kimchi},\ and\ \citenamefont {Vishwanath}}]{Potter2014}%
  \BibitemOpen
  \bibfield  {author} {\bibinfo {author} {\bibfnamefont {A.~C.}\ \bibnamefont
  {Potter}}, \bibinfo {author} {\bibfnamefont {I.}~\bibnamefont {Kimchi}},\
  and\ \bibinfo {author} {\bibfnamefont {A.}~\bibnamefont {Vishwanath}},\
  }\bibfield  {title} {\bibinfo {title} {{Quantum oscillations from surface
  Fermi arcs in Weyl and Dirac semimetals}},\ }\href
  {https://doi.org/10.1038/ncomms6161} {\bibfield  {journal} {\bibinfo
  {journal} {Nat. Commun.}\ }\textbf {\bibinfo {volume} {5}},\ \bibinfo {pages}
  {5161} (\bibinfo {year} {2014})}\BibitemShut {NoStop}%
\bibitem [{\citenamefont {Araki}\ \emph {et~al.}(2016)\citenamefont {Araki},
  \citenamefont {Yoshida},\ and\ \citenamefont {Nomura}}]{Araki2016}%
  \BibitemOpen
  \bibfield  {author} {\bibinfo {author} {\bibfnamefont {Y.}~\bibnamefont
  {Araki}}, \bibinfo {author} {\bibfnamefont {A.}~\bibnamefont {Yoshida}},\
  and\ \bibinfo {author} {\bibfnamefont {K.}~\bibnamefont {Nomura}},\
  }\bibfield  {title} {\bibinfo {title} {{Universal charge and current on
  magnetic domain walls in Weyl semimetals}},\ }\href
  {https://doi.org/10.1103/PhysRevB.94.115312} {\bibfield  {journal} {\bibinfo
  {journal} {Phys. Rev. B}\ }\textbf {\bibinfo {volume} {94}},\ \bibinfo
  {pages} {115312} (\bibinfo {year} {2016})}\BibitemShut {NoStop}%
\bibitem [{\citenamefont {Dwivedi}(2018)}]{Dwivedi2018}%
  \BibitemOpen
  \bibfield  {author} {\bibinfo {author} {\bibfnamefont {V.}~\bibnamefont
  {Dwivedi}},\ }\bibfield  {title} {\bibinfo {title} {{Fermi arc reconstruction
  at junctions between Weyl semimetals}},\ }\href
  {https://doi.org/10.1103/PhysRevB.97.064201} {\bibfield  {journal} {\bibinfo
  {journal} {Phys. Rev. B}\ }\textbf {\bibinfo {volume} {97}},\ \bibinfo
  {pages} {064201} (\bibinfo {year} {2018})}\BibitemShut {NoStop}%
\bibitem [{\citenamefont {Ishida}\ and\ \citenamefont
  {Liebsch}(2018)}]{Ishida2018}%
  \BibitemOpen
  \bibfield  {author} {\bibinfo {author} {\bibfnamefont {H.}~\bibnamefont
  {Ishida}}\ and\ \bibinfo {author} {\bibfnamefont {A.}~\bibnamefont
  {Liebsch}},\ }\bibfield  {title} {\bibinfo {title} {{Fermi arc engineering at
  the interface between two Weyl semimetals}},\ }\href
  {https://doi.org/10.1103/PhysRevB.98.195426} {\bibfield  {journal} {\bibinfo
  {journal} {Phys. Rev. B}\ }\textbf {\bibinfo {volume} {98}},\ \bibinfo
  {pages} {195426} (\bibinfo {year} {2018})}\BibitemShut {NoStop}%
\bibitem [{\citenamefont {Murthy}\ \emph {et~al.}(2020)\citenamefont {Murthy},
  \citenamefont {Fertig},\ and\ \citenamefont {Shimshoni}}]{Murthy2020}%
  \BibitemOpen
  \bibfield  {author} {\bibinfo {author} {\bibfnamefont {G.}~\bibnamefont
  {Murthy}}, \bibinfo {author} {\bibfnamefont {H.~A.}\ \bibnamefont {Fertig}},\
  and\ \bibinfo {author} {\bibfnamefont {E.}~\bibnamefont {Shimshoni}},\
  }\bibfield  {title} {\bibinfo {title} {{Surface states and arcless angles in
  twisted Weyl semimetals}},\ }\href
  {https://doi.org/10.1103/PhysRevResearch.2.013367} {\bibfield  {journal}
  {\bibinfo  {journal} {Phys. Rev. Research}\ }\textbf {\bibinfo {volume}
  {2}},\ \bibinfo {pages} {013367} (\bibinfo {year} {2020})}\BibitemShut
  {NoStop}%
\bibitem [{\citenamefont {Schr\"{o}ter}\ \emph {et~al.}(2020)\citenamefont
  {Schr\"{o}ter}, \citenamefont {Stolz}, \citenamefont {Manna}, \citenamefont
  {de~Juan}, \citenamefont {Vergniory}, \citenamefont {Krieger}, \citenamefont
  {Pei}, \citenamefont {Schmitt}, \citenamefont {Dudin}, \citenamefont {Kim},
  \citenamefont {Cacho}, \citenamefont {Bradlyn}, \citenamefont {Borrmann},
  \citenamefont {Schmidt}, \citenamefont {Widmer}, \citenamefont {Strocov},\
  and\ \citenamefont {Felser}}]{Schroter2020}%
  \BibitemOpen
  \bibfield  {author} {\bibinfo {author} {\bibfnamefont {N.~B.~M.}\
  \bibnamefont {Schr\"{o}ter}}, \bibinfo {author} {\bibfnamefont
  {S.}~\bibnamefont {Stolz}}, \bibinfo {author} {\bibfnamefont
  {K.}~\bibnamefont {Manna}}, \bibinfo {author} {\bibfnamefont
  {F.}~\bibnamefont {de~Juan}}, \bibinfo {author} {\bibfnamefont {M.~G.}\
  \bibnamefont {Vergniory}}, \bibinfo {author} {\bibfnamefont {J.~A.}\
  \bibnamefont {Krieger}}, \bibinfo {author} {\bibfnamefont {D.}~\bibnamefont
  {Pei}}, \bibinfo {author} {\bibfnamefont {T.}~\bibnamefont {Schmitt}},
  \bibinfo {author} {\bibfnamefont {P.}~\bibnamefont {Dudin}}, \bibinfo
  {author} {\bibfnamefont {T.~K.}\ \bibnamefont {Kim}}, \bibinfo {author}
  {\bibfnamefont {C.}~\bibnamefont {Cacho}}, \bibinfo {author} {\bibfnamefont
  {B.}~\bibnamefont {Bradlyn}}, \bibinfo {author} {\bibfnamefont
  {H.}~\bibnamefont {Borrmann}}, \bibinfo {author} {\bibfnamefont
  {M.}~\bibnamefont {Schmidt}}, \bibinfo {author} {\bibfnamefont
  {R.}~\bibnamefont {Widmer}}, \bibinfo {author} {\bibfnamefont {V.~N.}\
  \bibnamefont {Strocov}},\ and\ \bibinfo {author} {\bibfnamefont
  {C.}~\bibnamefont {Felser}},\ }\bibfield  {title} {\bibinfo {title}
  {{Observation and control of maximal Chern numbers in a chiral topological
  semimetal}},\ }\href {https://doi.org/10.1126/science.aaz3480} {\bibfield
  {journal} {\bibinfo  {journal} {Science}\ }\textbf {\bibinfo {volume}
  {369}},\ \bibinfo {pages} {179} (\bibinfo {year} {2020})}\BibitemShut
  {NoStop}%
\bibitem [{\citenamefont {Abdulla}\ \emph {et~al.}(2021)\citenamefont
  {Abdulla}, \citenamefont {Rao},\ and\ \citenamefont {Murthy}}]{Abdulla2021}%
  \BibitemOpen
  \bibfield  {author} {\bibinfo {author} {\bibfnamefont {F.}~\bibnamefont
  {Abdulla}}, \bibinfo {author} {\bibfnamefont {S.}~\bibnamefont {Rao}},\ and\
  \bibinfo {author} {\bibfnamefont {G.}~\bibnamefont {Murthy}},\ }\bibfield
  {title} {\bibinfo {title} {{Fermi arc reconstruction at the interface of
  twisted Weyl semimetals}},\ }\href
  {https://doi.org/10.1103/PhysRevB.103.235308} {\bibfield  {journal} {\bibinfo
   {journal} {Phys. Rev. B}\ }\textbf {\bibinfo {volume} {103}},\ \bibinfo
  {pages} {235308} (\bibinfo {year} {2021})}\BibitemShut {NoStop}%
\bibitem [{\citenamefont {Buccheri}\ \emph {et~al.}(2022)\citenamefont
  {Buccheri}, \citenamefont {Egger},\ and\ \citenamefont
  {De~Martino}}]{Buccheri2022}%
  \BibitemOpen
  \bibfield  {author} {\bibinfo {author} {\bibfnamefont {F.}~\bibnamefont
  {Buccheri}}, \bibinfo {author} {\bibfnamefont {R.}~\bibnamefont {Egger}},\
  and\ \bibinfo {author} {\bibfnamefont {A.}~\bibnamefont {De~Martino}},\
  }\bibfield  {title} {\bibinfo {title} {{Transport, refraction, and interface
  arcs in junctions of Weyl semimetals}},\ }\href
  {https://doi.org/10.1103/PhysRevB.106.045413} {\bibfield  {journal} {\bibinfo
   {journal} {Phys. Rev. B}\ }\textbf {\bibinfo {volume} {106}},\ \bibinfo
  {pages} {045413} (\bibinfo {year} {2022})}\BibitemShut {NoStop}%
\bibitem [{\citenamefont {Kaushik}\ \emph {et~al.}(2022)\citenamefont
  {Kaushik}, \citenamefont {Robredo}, \citenamefont {Mathur}, \citenamefont
  {Schoop}, \citenamefont {Jin}, \citenamefont {Vergniory},\ and\ \citenamefont
  {Cano}}]{Kaushik2022}%
  \BibitemOpen
  \bibfield  {author} {\bibinfo {author} {\bibfnamefont {S.}~\bibnamefont
  {Kaushik}}, \bibinfo {author} {\bibfnamefont {I.}~\bibnamefont {Robredo}},
  \bibinfo {author} {\bibfnamefont {N.}~\bibnamefont {Mathur}}, \bibinfo
  {author} {\bibfnamefont {L.~M.}\ \bibnamefont {Schoop}}, \bibinfo {author}
  {\bibfnamefont {S.}~\bibnamefont {Jin}}, \bibinfo {author} {\bibfnamefont
  {M.~G.}\ \bibnamefont {Vergniory}},\ and\ \bibinfo {author} {\bibfnamefont
  {J.}~\bibnamefont {Cano}},\ }\href
  {https://doi.org/10.48550/ARXIV.2207.14109} {\bibinfo {title} {{Transport
  signatures of Fermi arcs at twin boundaries in Weyl materials}}} (\bibinfo
  {year} {2022})\BibitemShut {NoStop}%
\bibitem [{\citenamefont {Bonasera}\ \emph {et~al.}(2022)\citenamefont
  {Bonasera}, \citenamefont {Zhang}, \citenamefont {Privitera},\ and\
  \citenamefont {Pellegrino}}]{Bonasera2022}%
  \BibitemOpen
  \bibfield  {author} {\bibinfo {author} {\bibfnamefont {F.}~\bibnamefont
  {Bonasera}}, \bibinfo {author} {\bibfnamefont {S.-B.}\ \bibnamefont {Zhang}},
  \bibinfo {author} {\bibfnamefont {L.}~\bibnamefont {Privitera}},\ and\
  \bibinfo {author} {\bibfnamefont {F.~M.~D.}\ \bibnamefont {Pellegrino}},\
  }\bibfield  {title} {\bibinfo {title} {Tunable interface states between
  floquet-weyl semimetals},\ }\href
  {https://doi.org/10.1103/PhysRevB.106.195115} {\bibfield  {journal} {\bibinfo
   {journal} {Phys. Rev. B}\ }\textbf {\bibinfo {volume} {106}},\ \bibinfo
  {pages} {195115} (\bibinfo {year} {2022})}\BibitemShut {NoStop}%
\bibitem [{\citenamefont {Mathur}\ \emph {et~al.}(2023)\citenamefont {Mathur},
  \citenamefont {Yuan}, \citenamefont {Cheng}, \citenamefont {Kaushik},
  \citenamefont {Robredo}, \citenamefont {Vergniory}, \citenamefont {Cano},
  \citenamefont {Yao}, \citenamefont {Jin},\ and\ \citenamefont
  {Schoop}}]{Mathur2023}%
  \BibitemOpen
  \bibfield  {author} {\bibinfo {author} {\bibfnamefont {N.}~\bibnamefont
  {Mathur}}, \bibinfo {author} {\bibfnamefont {F.}~\bibnamefont {Yuan}},
  \bibinfo {author} {\bibfnamefont {G.}~\bibnamefont {Cheng}}, \bibinfo
  {author} {\bibfnamefont {S.}~\bibnamefont {Kaushik}}, \bibinfo {author}
  {\bibfnamefont {I.}~\bibnamefont {Robredo}}, \bibinfo {author} {\bibfnamefont
  {M.~G.}\ \bibnamefont {Vergniory}}, \bibinfo {author} {\bibfnamefont
  {J.}~\bibnamefont {Cano}}, \bibinfo {author} {\bibfnamefont {N.}~\bibnamefont
  {Yao}}, \bibinfo {author} {\bibfnamefont {S.}~\bibnamefont {Jin}},\ and\
  \bibinfo {author} {\bibfnamefont {L.~M.}\ \bibnamefont {Schoop}},\ }\bibfield
   {title} {\bibinfo {title} {Atomically sharp internal interface in a chiral
  weyl semimetal nanowire},\ }\href
  {https://doi.org/10.1021/acs.nanolett.2c05100} {\bibfield  {journal}
  {\bibinfo  {journal} {Nano Letters}\ }\textbf {\bibinfo {volume} {23}},\
  \bibinfo {pages} {2695} (\bibinfo {year} {2023})}\BibitemShut {NoStop}%
\bibitem [{\citenamefont {Burkov}\ \emph {et~al.}(2011)\citenamefont {Burkov},
  \citenamefont {Hook},\ and\ \citenamefont {Balents}}]{Burkov2011b}%
  \BibitemOpen
  \bibfield  {author} {\bibinfo {author} {\bibfnamefont {A.~A.}\ \bibnamefont
  {Burkov}}, \bibinfo {author} {\bibfnamefont {M.~D.}\ \bibnamefont {Hook}},\
  and\ \bibinfo {author} {\bibfnamefont {L.}~\bibnamefont {Balents}},\
  }\bibfield  {title} {\bibinfo {title} {{Topological nodal semimetals}},\
  }\href {https://doi.org/10.1103/PhysRevB.84.235126} {\bibfield  {journal}
  {\bibinfo  {journal} {Phys. Rev. B}\ }\textbf {\bibinfo {volume} {84}},\
  \bibinfo {pages} {235126} (\bibinfo {year} {2011})}\BibitemShut {NoStop}%
\bibitem [{\citenamefont {Fang}\ \emph {et~al.}(2016)\citenamefont {Fang},
  \citenamefont {Weng}, \citenamefont {Dai},\ and\ \citenamefont
  {Fang}}]{Fang2016}%
  \BibitemOpen
  \bibfield  {author} {\bibinfo {author} {\bibfnamefont {C.}~\bibnamefont
  {Fang}}, \bibinfo {author} {\bibfnamefont {H.}~\bibnamefont {Weng}}, \bibinfo
  {author} {\bibfnamefont {X.}~\bibnamefont {Dai}},\ and\ \bibinfo {author}
  {\bibfnamefont {Z.}~\bibnamefont {Fang}},\ }\bibfield  {title} {\bibinfo
  {title} {{Topological nodal line semimetals}},\ }\href
  {https://doi.org/10.1088/1674-1056/25/11/117106} {\bibfield  {journal}
  {\bibinfo  {journal} {Chin. Phys. B}\ }\textbf {\bibinfo {volume} {25}},\
  \bibinfo {pages} {117106} (\bibinfo {year} {2016})}\BibitemShut {NoStop}%
\bibitem [{\citenamefont {Yu}\ \emph {et~al.}(2010)\citenamefont {Yu},
  \citenamefont {Zhang}, \citenamefont {Zhang}, \citenamefont {Zhang},
  \citenamefont {Dai},\ and\ \citenamefont {Fang}}]{Yu2010}%
  \BibitemOpen
  \bibfield  {author} {\bibinfo {author} {\bibfnamefont {R.}~\bibnamefont
  {Yu}}, \bibinfo {author} {\bibfnamefont {W.}~\bibnamefont {Zhang}}, \bibinfo
  {author} {\bibfnamefont {H.-J.}\ \bibnamefont {Zhang}}, \bibinfo {author}
  {\bibfnamefont {S.-C.}\ \bibnamefont {Zhang}}, \bibinfo {author}
  {\bibfnamefont {X.}~\bibnamefont {Dai}},\ and\ \bibinfo {author}
  {\bibfnamefont {Z.}~\bibnamefont {Fang}},\ }\bibfield  {title} {\bibinfo
  {title} {Quantized anomalous hall effect in magnetic topological
  insulators},\ }\href {https://doi.org/10.1126/science.1187485} {\bibfield
  {journal} {\bibinfo  {journal} {Science}\ }\textbf {\bibinfo {volume}
  {329}},\ \bibinfo {pages} {61} (\bibinfo {year} {2010})}\BibitemShut
  {NoStop}%
\bibitem [{\citenamefont {Vanderbilt}(2018)}]{Vanderbilt2018}%
  \BibitemOpen
  \bibfield  {author} {\bibinfo {author} {\bibfnamefont {D.}~\bibnamefont
  {Vanderbilt}},\ }\href {https://doi.org/10.1017/9781316662205} {\emph
  {\bibinfo {title} {{Berry Phases in Electronic Structure Theory}}}}\
  (\bibinfo  {publisher} {Cambridge University Press},\ \bibinfo {year}
  {2018})\BibitemShut {NoStop}%
\bibitem [{Sup()}]{Supp}%
  \BibitemOpen
  \href@noop {} {}\bibinfo {note} {{S}ee {S}upplemental {M}aterial for
  additional information and data, which includes
  Refs.~\cite{Okamoto2021,Devescovi2021,Devescovi2022,Haldane2004,Halperin1987,Berenger1994,Winkler2003,Lowdin1951}.}\BibitemShut
  {Stop}%
\bibitem [{\citenamefont {Nguyen}\ \emph {et~al.}(2018)\citenamefont {Nguyen},
  \citenamefont {Dubois}, \citenamefont {Deschamps}, \citenamefont {Cueff},
  \citenamefont {Pardon}, \citenamefont {Leclercq}, \citenamefont {Seassal},
  \citenamefont {Letartre},\ and\ \citenamefont {Viktorovitch}}]{Son2018}%
  \BibitemOpen
  \bibfield  {author} {\bibinfo {author} {\bibfnamefont {H.~S.}\ \bibnamefont
  {Nguyen}}, \bibinfo {author} {\bibfnamefont {F.}~\bibnamefont {Dubois}},
  \bibinfo {author} {\bibfnamefont {T.}~\bibnamefont {Deschamps}}, \bibinfo
  {author} {\bibfnamefont {S.}~\bibnamefont {Cueff}}, \bibinfo {author}
  {\bibfnamefont {A.}~\bibnamefont {Pardon}}, \bibinfo {author} {\bibfnamefont
  {J.-L.}\ \bibnamefont {Leclercq}}, \bibinfo {author} {\bibfnamefont
  {C.}~\bibnamefont {Seassal}}, \bibinfo {author} {\bibfnamefont
  {X.}~\bibnamefont {Letartre}},\ and\ \bibinfo {author} {\bibfnamefont
  {P.}~\bibnamefont {Viktorovitch}},\ }\bibfield  {title} {\bibinfo {title}
  {{Symmetry Breaking in Photonic Crystals: On-Demand Dispersion from Flatband
  to Dirac Cones}},\ }\href {https://doi.org/10.1103/PhysRevLett.120.066102}
  {\bibfield  {journal} {\bibinfo  {journal} {Phys. Rev. Lett.}\ }\textbf
  {\bibinfo {volume} {120}},\ \bibinfo {pages} {066102} (\bibinfo {year}
  {2018})}\BibitemShut {NoStop}%
\bibitem [{\citenamefont {Nguyen}\ \emph {et~al.}(2022)\citenamefont {Nguyen},
  \citenamefont {Letartre}, \citenamefont {Drouard}, \citenamefont
  {Viktorovitch}, \citenamefont {Nguyen},\ and\ \citenamefont
  {Nguyen}}]{Dung2022}%
  \BibitemOpen
  \bibfield  {author} {\bibinfo {author} {\bibfnamefont {D.~X.}\ \bibnamefont
  {Nguyen}}, \bibinfo {author} {\bibfnamefont {X.}~\bibnamefont {Letartre}},
  \bibinfo {author} {\bibfnamefont {E.}~\bibnamefont {Drouard}}, \bibinfo
  {author} {\bibfnamefont {P.}~\bibnamefont {Viktorovitch}}, \bibinfo {author}
  {\bibfnamefont {H.~C.}\ \bibnamefont {Nguyen}},\ and\ \bibinfo {author}
  {\bibfnamefont {H.~S.}\ \bibnamefont {Nguyen}},\ }\bibfield  {title}
  {\bibinfo {title} {{Magic configurations in moir\'e superlattice of bilayer
  photonic crystals: Almost-perfect flatbands and unconventional
  localization}},\ }\href {https://doi.org/10.1103/PhysRevResearch.4.L032031}
  {\bibfield  {journal} {\bibinfo  {journal} {Phys. Rev. Research}\ }\textbf
  {\bibinfo {volume} {4}},\ \bibinfo {pages} {L032031} (\bibinfo {year}
  {2022})}\BibitemShut {NoStop}%
\bibitem [{\citenamefont {Nguyen}\ \emph {et~al.}(2021)\citenamefont {Nguyen},
  \citenamefont {Nguyen}, \citenamefont {Louvet}, \citenamefont {Letartre},
  \citenamefont {Viktorovitch},\ and\ \citenamefont {Nguyen}}]{Chau2021}%
  \BibitemOpen
  \bibfield  {author} {\bibinfo {author} {\bibfnamefont {H.~C.}\ \bibnamefont
  {Nguyen}}, \bibinfo {author} {\bibfnamefont {D.~X.}\ \bibnamefont {Nguyen}},
  \bibinfo {author} {\bibfnamefont {T.}~\bibnamefont {Louvet}}, \bibinfo
  {author} {\bibfnamefont {X.}~\bibnamefont {Letartre}}, \bibinfo {author}
  {\bibfnamefont {P.}~\bibnamefont {Viktorovitch}},\ and\ \bibinfo {author}
  {\bibfnamefont {H.~S.}\ \bibnamefont {Nguyen}},\ }\href
  {https://doi.org/10.48550/ARXIV.2111.02843} {\bibinfo {title} {{Topological
  Properties of Photonic Bands with Synthetic Momentum}}} (\bibinfo {year}
  {2021})\BibitemShut {NoStop}%
\bibitem [{\citenamefont {Lee}\ \emph {et~al.}(2022)\citenamefont {Lee},
  \citenamefont {Yoo}, \citenamefont {Cheon}, \citenamefont {Joo},
  \citenamefont {Yoon},\ and\ \citenamefont {Song}}]{Lee_2022}%
  \BibitemOpen
  \bibfield  {author} {\bibinfo {author} {\bibfnamefont {K.~Y.}\ \bibnamefont
  {Lee}}, \bibinfo {author} {\bibfnamefont {K.~W.}\ \bibnamefont {Yoo}},
  \bibinfo {author} {\bibfnamefont {S.}~\bibnamefont {Cheon}}, \bibinfo
  {author} {\bibfnamefont {W.-J.}\ \bibnamefont {Joo}}, \bibinfo {author}
  {\bibfnamefont {J.~W.}\ \bibnamefont {Yoon}},\ and\ \bibinfo {author}
  {\bibfnamefont {S.~H.}\ \bibnamefont {Song}},\ }\bibfield  {title} {\bibinfo
  {title} {{Synthetic Topological Nodal Phase in Bilayer Resonant Gratings}},\
  }\href {https://doi.org/10.1103/PhysRevLett.128.053002} {\bibfield  {journal}
  {\bibinfo  {journal} {Phys. Rev. Lett.}\ }\textbf {\bibinfo {volume} {128}},\
  \bibinfo {pages} {053002} (\bibinfo {year} {2022})}\BibitemShut {NoStop}%
\bibitem [{\citenamefont {Murakami}\ and\ \citenamefont
  {Kuga}(2008)}]{Murakami2008}%
  \BibitemOpen
  \bibfield  {author} {\bibinfo {author} {\bibfnamefont {S.}~\bibnamefont
  {Murakami}}\ and\ \bibinfo {author} {\bibfnamefont {S.-i.}\ \bibnamefont
  {Kuga}},\ }\bibfield  {title} {\bibinfo {title} {{Universal phase diagrams
  for the quantum spin Hall systems}},\ }\href
  {https://doi.org/10.1103/PhysRevB.78.165313} {\bibfield  {journal} {\bibinfo
  {journal} {Phys. Rev. B}\ }\textbf {\bibinfo {volume} {78}},\ \bibinfo
  {pages} {165313} (\bibinfo {year} {2008})}\BibitemShut {NoStop}%
\bibitem [{\citenamefont {He}\ \emph {et~al.}(2019)\citenamefont {He},
  \citenamefont {Yu}, \citenamefont {Wang}, \citenamefont {Ge}, \citenamefont
  {Ruan}, \citenamefont {Zhang}, \citenamefont {Lu},\ and\ \citenamefont
  {Chen}}]{He2019}%
  \BibitemOpen
  \bibfield  {author} {\bibinfo {author} {\bibfnamefont {C.}~\bibnamefont
  {He}}, \bibinfo {author} {\bibfnamefont {S.-Y.}\ \bibnamefont {Yu}}, \bibinfo
  {author} {\bibfnamefont {H.}~\bibnamefont {Wang}}, \bibinfo {author}
  {\bibfnamefont {H.}~\bibnamefont {Ge}}, \bibinfo {author} {\bibfnamefont
  {J.}~\bibnamefont {Ruan}}, \bibinfo {author} {\bibfnamefont {H.}~\bibnamefont
  {Zhang}}, \bibinfo {author} {\bibfnamefont {M.-H.}\ \bibnamefont {Lu}},\ and\
  \bibinfo {author} {\bibfnamefont {Y.-F.}\ \bibnamefont {Chen}},\ }\bibfield
  {title} {\bibinfo {title} {{Hybrid Acoustic Topological Insulator in Three
  Dimensions}},\ }\href {https://doi.org/10.1103/PhysRevLett.123.195503}
  {\bibfield  {journal} {\bibinfo  {journal} {Phys. Rev. Lett.}\ }\textbf
  {\bibinfo {volume} {123}},\ \bibinfo {pages} {195503} (\bibinfo {year}
  {2019})}\BibitemShut {NoStop}%
\bibitem [{\citenamefont {Mohanta}\ \emph {et~al.}(2021)\citenamefont
  {Mohanta}, \citenamefont {Ok}, \citenamefont {Zhang}, \citenamefont {Miao},
  \citenamefont {Dagotto}, \citenamefont {Lee},\ and\ \citenamefont
  {Okamoto}}]{Mohanta2021}%
  \BibitemOpen
  \bibfield  {author} {\bibinfo {author} {\bibfnamefont {N.}~\bibnamefont
  {Mohanta}}, \bibinfo {author} {\bibfnamefont {J.~M.}\ \bibnamefont {Ok}},
  \bibinfo {author} {\bibfnamefont {J.}~\bibnamefont {Zhang}}, \bibinfo
  {author} {\bibfnamefont {H.}~\bibnamefont {Miao}}, \bibinfo {author}
  {\bibfnamefont {E.}~\bibnamefont {Dagotto}}, \bibinfo {author} {\bibfnamefont
  {H.~N.}\ \bibnamefont {Lee}},\ and\ \bibinfo {author} {\bibfnamefont
  {S.}~\bibnamefont {Okamoto}},\ }\bibfield  {title} {\bibinfo {title}
  {{Semi-Dirac and Weyl fermions in transition metal oxides}},\ }\href
  {https://doi.org/10.1103/PhysRevB.104.235121} {\bibfield  {journal} {\bibinfo
   {journal} {Phys. Rev. B}\ }\textbf {\bibinfo {volume} {104}},\ \bibinfo
  {pages} {235121} (\bibinfo {year} {2021})}\BibitemShut {NoStop}%
\bibitem [{\citenamefont {Li}\ \emph {et~al.}(2022)\citenamefont {Li},
  \citenamefont {Li}, \citenamefont {Jia},\ and\ \citenamefont {Liu}}]{Li2022}%
  \BibitemOpen
  \bibfield  {author} {\bibinfo {author} {\bibfnamefont {R.}~\bibnamefont
  {Li}}, \bibinfo {author} {\bibfnamefont {P.}~\bibnamefont {Li}}, \bibinfo
  {author} {\bibfnamefont {Y.}~\bibnamefont {Jia}},\ and\ \bibinfo {author}
  {\bibfnamefont {Y.}~\bibnamefont {Liu}},\ }\bibfield  {title} {\bibinfo
  {title} {{Self-localized topological states in three dimensions}},\ }\href
  {https://doi.org/10.1103/PhysRevB.105.L201111} {\bibfield  {journal}
  {\bibinfo  {journal} {Phys. Rev. B}\ }\textbf {\bibinfo {volume} {105}},\
  \bibinfo {pages} {L201111} (\bibinfo {year} {2022})}\BibitemShut {NoStop}%
\bibitem [{\citenamefont {Johnson}\ and\ \citenamefont
  {Joannopoulos}(2001)}]{MPB}%
  \BibitemOpen
  \bibfield  {author} {\bibinfo {author} {\bibfnamefont {S.~G.}\ \bibnamefont
  {Johnson}}\ and\ \bibinfo {author} {\bibfnamefont {J.~D.}\ \bibnamefont
  {Joannopoulos}},\ }\bibfield  {title} {\bibinfo {title} {{Block-iterative
  frequency-domain methods for Maxwell's equations in a planewave basis}},\
  }\href {https://doi.org/10.1364/OE.8.000173} {\bibfield  {journal} {\bibinfo
  {journal} {Opt. Express}\ }\textbf {\bibinfo {volume} {8}},\ \bibinfo {pages}
  {173} (\bibinfo {year} {2001})}\BibitemShut {NoStop}%
\bibitem [{\citenamefont {Letartre}\ \emph {et~al.}(2022)\citenamefont
  {Letartre}, \citenamefont {Mazauric}, \citenamefont {Cueff}, \citenamefont
  {Benyattou}, \citenamefont {Nguyen},\ and\ \citenamefont
  {Viktorovitch}}]{Letartre2022}%
  \BibitemOpen
  \bibfield  {author} {\bibinfo {author} {\bibfnamefont {X.}~\bibnamefont
  {Letartre}}, \bibinfo {author} {\bibfnamefont {S.}~\bibnamefont {Mazauric}},
  \bibinfo {author} {\bibfnamefont {S.}~\bibnamefont {Cueff}}, \bibinfo
  {author} {\bibfnamefont {T.}~\bibnamefont {Benyattou}}, \bibinfo {author}
  {\bibfnamefont {H.~S.}\ \bibnamefont {Nguyen}},\ and\ \bibinfo {author}
  {\bibfnamefont {P.}~\bibnamefont {Viktorovitch}},\ }\bibfield  {title}
  {\bibinfo {title} {{Analytical non-Hermitian description of photonic crystals
  with arbitrary lateral and transverse symmetry}},\ }\href
  {https://doi.org/10.1103/PhysRevA.106.033510} {\bibfield  {journal} {\bibinfo
   {journal} {Phys. Rev. A}\ }\textbf {\bibinfo {volume} {106}},\ \bibinfo
  {pages} {033510} (\bibinfo {year} {2022})}\BibitemShut {NoStop}%
\bibitem [{\citenamefont {Yoshimura}\ \emph {et~al.}(2016)\citenamefont
  {Yoshimura}, \citenamefont {Onishi}, \citenamefont {Kobayashi}, \citenamefont
  {Ohtsuki},\ and\ \citenamefont {Imura}}]{Yoshimura2016}%
  \BibitemOpen
  \bibfield  {author} {\bibinfo {author} {\bibfnamefont {Y.}~\bibnamefont
  {Yoshimura}}, \bibinfo {author} {\bibfnamefont {W.}~\bibnamefont {Onishi}},
  \bibinfo {author} {\bibfnamefont {K.}~\bibnamefont {Kobayashi}}, \bibinfo
  {author} {\bibfnamefont {T.}~\bibnamefont {Ohtsuki}},\ and\ \bibinfo {author}
  {\bibfnamefont {K.-I.}\ \bibnamefont {Imura}},\ }\bibfield  {title} {\bibinfo
  {title} {{Comparative study of Weyl semimetal and topological/Chern
  insulators: Thin-film point of view}},\ }\href
  {https://doi.org/10.1103/PhysRevB.94.235414} {\bibfield  {journal} {\bibinfo
  {journal} {Phys. Rev. B}\ }\textbf {\bibinfo {volume} {94}},\ \bibinfo
  {pages} {235414} (\bibinfo {year} {2016})}\BibitemShut {NoStop}%
\bibitem [{\citenamefont {Liu}\ \emph {et~al.}(2022{\natexlab{b}})\citenamefont
  {Liu}, \citenamefont {Gao}, \citenamefont {Wang}, \citenamefont {Xi},
  \citenamefont {Hu}, \citenamefont {Wang}, \citenamefont {Liu}, \citenamefont
  {Lin}, \citenamefont {Deng}, \citenamefont {Yang}, \citenamefont {Zhou},
  \citenamefont {Yang}, \citenamefont {Chong},\ and\ \citenamefont
  {Zhang}}]{Liu2022v}%
  \BibitemOpen
  \bibfield  {author} {\bibinfo {author} {\bibfnamefont {G.-G.}\ \bibnamefont
  {Liu}}, \bibinfo {author} {\bibfnamefont {Z.}~\bibnamefont {Gao}}, \bibinfo
  {author} {\bibfnamefont {Q.}~\bibnamefont {Wang}}, \bibinfo {author}
  {\bibfnamefont {X.}~\bibnamefont {Xi}}, \bibinfo {author} {\bibfnamefont
  {Y.-H.}\ \bibnamefont {Hu}}, \bibinfo {author} {\bibfnamefont
  {M.}~\bibnamefont {Wang}}, \bibinfo {author} {\bibfnamefont {C.}~\bibnamefont
  {Liu}}, \bibinfo {author} {\bibfnamefont {X.}~\bibnamefont {Lin}}, \bibinfo
  {author} {\bibfnamefont {L.}~\bibnamefont {Deng}}, \bibinfo {author}
  {\bibfnamefont {S.~A.}\ \bibnamefont {Yang}}, \bibinfo {author}
  {\bibfnamefont {P.}~\bibnamefont {Zhou}}, \bibinfo {author} {\bibfnamefont
  {Y.}~\bibnamefont {Yang}}, \bibinfo {author} {\bibfnamefont {Y.}~\bibnamefont
  {Chong}},\ and\ \bibinfo {author} {\bibfnamefont {B.}~\bibnamefont {Zhang}},\
  }\bibfield  {title} {\bibinfo {title} {{Topological Chern vectors in
  three-dimensional photonic crystals}},\ }\href
  {https://doi.org/10.1038/s41586-022-05077-2} {\bibfield  {journal} {\bibinfo
  {journal} {Nature}\ }\textbf {\bibinfo {volume} {609}},\ \bibinfo {pages}
  {925} (\bibinfo {year} {2022}{\natexlab{b}})}\BibitemShut {NoStop}%
\bibitem [{\citenamefont {{Taflove, Allen and Hagness, Susan}}(2005)}]{FDTD}%
  \BibitemOpen
  \bibfield  {author} {\bibinfo {author} {\bibnamefont {{Taflove, Allen and
  Hagness, Susan}}},\ }\href@noop {} {\emph {\bibinfo {title} {{Computational
  electrodynamics}}}},\ \bibinfo {edition} {3rd}\ ed.,\ Artech House antennas
  and propagation library\ (\bibinfo  {publisher} {Artech House},\ \bibinfo
  {address} {Norwood, MA},\ \bibinfo {year} {2005})\BibitemShut {NoStop}%
\bibitem [{\citenamefont {Oskooi}\ \emph {et~al.}(2010)\citenamefont {Oskooi},
  \citenamefont {Roundy}, \citenamefont {Ibanescu}, \citenamefont {Bermel},
  \citenamefont {Joannopoulos},\ and\ \citenamefont {Johnson}}]{MEEP}%
  \BibitemOpen
  \bibfield  {author} {\bibinfo {author} {\bibfnamefont {A.~F.}\ \bibnamefont
  {Oskooi}}, \bibinfo {author} {\bibfnamefont {D.}~\bibnamefont {Roundy}},
  \bibinfo {author} {\bibfnamefont {M.}~\bibnamefont {Ibanescu}}, \bibinfo
  {author} {\bibfnamefont {P.}~\bibnamefont {Bermel}}, \bibinfo {author}
  {\bibfnamefont {J.}~\bibnamefont {Joannopoulos}},\ and\ \bibinfo {author}
  {\bibfnamefont {S.~G.}\ \bibnamefont {Johnson}},\ }\bibfield  {title}
  {\bibinfo {title} {{Meep: A flexible free-software package for
  electromagnetic simulations by the FDTD method}},\ }\href
  {https://doi.org/https://doi.org/10.1016/j.cpc.2009.11.008} {\bibfield
  {journal} {\bibinfo  {journal} {Comput. Phys. Commun.}\ }\textbf {\bibinfo
  {volume} {181}},\ \bibinfo {pages} {687} (\bibinfo {year}
  {2010})}\BibitemShut {NoStop}%
\bibitem [{\citenamefont {Cueff}\ \emph {et~al.}(2019)\citenamefont {Cueff},
  \citenamefont {Dubois}, \citenamefont {Huang}, \citenamefont {Li},
  \citenamefont {Zia}, \citenamefont {Letartre}, \citenamefont {Viktorovitch},\
  and\ \citenamefont {Nguyen}}]{Cueff2019}%
  \BibitemOpen
  \bibfield  {author} {\bibinfo {author} {\bibfnamefont {S.}~\bibnamefont
  {Cueff}}, \bibinfo {author} {\bibfnamefont {F.}~\bibnamefont {Dubois}},
  \bibinfo {author} {\bibfnamefont {M.~S.~R.}\ \bibnamefont {Huang}}, \bibinfo
  {author} {\bibfnamefont {D.}~\bibnamefont {Li}}, \bibinfo {author}
  {\bibfnamefont {R.}~\bibnamefont {Zia}}, \bibinfo {author} {\bibfnamefont
  {X.}~\bibnamefont {Letartre}}, \bibinfo {author} {\bibfnamefont
  {P.}~\bibnamefont {Viktorovitch}},\ and\ \bibinfo {author} {\bibfnamefont
  {H.~S.}\ \bibnamefont {Nguyen}},\ }\bibfield  {title} {\bibinfo {title}
  {Tailoring the local density of optical states and directionality of light
  emission by symmetry breaking},\ }\href
  {https://doi.org/10.1109/JSTQE.2019.2902915} {\bibfield  {journal} {\bibinfo
  {journal} {IEEE Journal of Selected Topics in Quantum Electronics}\ }\textbf
  {\bibinfo {volume} {25}},\ \bibinfo {pages} {1} (\bibinfo {year}
  {2019})}\BibitemShut {NoStop}%
\bibitem [{\citenamefont {Dufferwiel}\ \emph {et~al.}(2014)\citenamefont
  {Dufferwiel}, \citenamefont {Fras}, \citenamefont {Trichet}, \citenamefont
  {Walker}, \citenamefont {Li}, \citenamefont {Giriunas}, \citenamefont
  {Makhonin}, \citenamefont {Wilson}, \citenamefont {Smith}, \citenamefont
  {Clarke}, \citenamefont {Skolnick},\ and\ \citenamefont
  {Krizhanovskii}}]{Dufferwiel2014}%
  \BibitemOpen
  \bibfield  {author} {\bibinfo {author} {\bibfnamefont {S.}~\bibnamefont
  {Dufferwiel}}, \bibinfo {author} {\bibfnamefont {F.}~\bibnamefont {Fras}},
  \bibinfo {author} {\bibfnamefont {A.}~\bibnamefont {Trichet}}, \bibinfo
  {author} {\bibfnamefont {P.~M.}\ \bibnamefont {Walker}}, \bibinfo {author}
  {\bibfnamefont {F.}~\bibnamefont {Li}}, \bibinfo {author} {\bibfnamefont
  {L.}~\bibnamefont {Giriunas}}, \bibinfo {author} {\bibfnamefont {M.~N.}\
  \bibnamefont {Makhonin}}, \bibinfo {author} {\bibfnamefont {L.~R.}\
  \bibnamefont {Wilson}}, \bibinfo {author} {\bibfnamefont {J.~M.}\
  \bibnamefont {Smith}}, \bibinfo {author} {\bibfnamefont {E.}~\bibnamefont
  {Clarke}}, \bibinfo {author} {\bibfnamefont {M.~S.}\ \bibnamefont
  {Skolnick}},\ and\ \bibinfo {author} {\bibfnamefont {D.~N.}\ \bibnamefont
  {Krizhanovskii}},\ }\bibfield  {title} {\bibinfo {title} {Strong
  exciton-photon coupling in open semiconductor microcavities},\ }\href
  {https://doi.org/10.1063/1.4878504} {\bibfield  {journal} {\bibinfo
  {journal} {Applied Physics Letters}\ }\textbf {\bibinfo {volume} {104}},\
  \bibinfo {pages} {192107} (\bibinfo {year} {2014})}\BibitemShut {NoStop}%
\bibitem [{\citenamefont {Li}\ \emph {et~al.}(2019)\citenamefont {Li},
  \citenamefont {Li}, \citenamefont {Cai}, \citenamefont {Li}, \citenamefont
  {Tang},\ and\ \citenamefont {Zhang}}]{Li_2019}%
  \BibitemOpen
  \bibfield  {author} {\bibinfo {author} {\bibfnamefont {F.}~\bibnamefont
  {Li}}, \bibinfo {author} {\bibfnamefont {Y.}~\bibnamefont {Li}}, \bibinfo
  {author} {\bibfnamefont {Y.}~\bibnamefont {Cai}}, \bibinfo {author}
  {\bibfnamefont {P.}~\bibnamefont {Li}}, \bibinfo {author} {\bibfnamefont
  {H.}~\bibnamefont {Tang}},\ and\ \bibinfo {author} {\bibfnamefont
  {Y.}~\bibnamefont {Zhang}},\ }\bibfield  {title} {\bibinfo {title} {Tunable
  open-access microcavities for solid-state quantum photonics and
  polaritonics},\ }\href {https://doi.org/10.1002/qute.201900060} {\bibfield
  {journal} {\bibinfo  {journal} {Advanced Quantum Technologies}\ }\textbf
  {\bibinfo {volume} {2}},\ \bibinfo {pages} {1900060} (\bibinfo {year}
  {2019})}\BibitemShut {NoStop}%
\bibitem [{\citenamefont {Geng}\ \emph {et~al.}(2020)\citenamefont {Geng},
  \citenamefont {Peters}, \citenamefont {Trichet}, \citenamefont {Malmir},
  \citenamefont {Kolkowski}, \citenamefont {Smith},\ and\ \citenamefont
  {Rodriguez}}]{Geng2020}%
  \BibitemOpen
  \bibfield  {author} {\bibinfo {author} {\bibfnamefont {Z.}~\bibnamefont
  {Geng}}, \bibinfo {author} {\bibfnamefont {K.}~\bibnamefont {Peters}},
  \bibinfo {author} {\bibfnamefont {A.}~\bibnamefont {Trichet}}, \bibinfo
  {author} {\bibfnamefont {K.}~\bibnamefont {Malmir}}, \bibinfo {author}
  {\bibfnamefont {R.}~\bibnamefont {Kolkowski}}, \bibinfo {author}
  {\bibfnamefont {J.}~\bibnamefont {Smith}},\ and\ \bibinfo {author}
  {\bibfnamefont {S.}~\bibnamefont {Rodriguez}},\ }\bibfield  {title} {\bibinfo
  {title} {Universal scaling in the dynamic hysteresis, and non-markovian
  dynamics, of a tunable optical cavity},\ }\bibfield  {journal} {\bibinfo
  {journal} {Physical Review Letters}\ }\textbf {\bibinfo {volume} {124}},\
  \href {https://doi.org/10.1103/physrevlett.124.153603}
  {10.1103/physrevlett.124.153603} (\bibinfo {year} {2020})\BibitemShut
  {NoStop}%
\bibitem [{\citenamefont {Vadia}\ \emph {et~al.}(2021)\citenamefont {Vadia},
  \citenamefont {Scherzer}, \citenamefont {Thierschmann}, \citenamefont
  {Sch\"afermeier}, \citenamefont {Dal~Savio}, \citenamefont {Taniguchi},
  \citenamefont {Watanabe}, \citenamefont {Hunger}, \citenamefont
  {Karra\"{\i}},\ and\ \citenamefont {H\"ogele}}]{Vadia2021}%
  \BibitemOpen
  \bibfield  {author} {\bibinfo {author} {\bibfnamefont {S.}~\bibnamefont
  {Vadia}}, \bibinfo {author} {\bibfnamefont {J.}~\bibnamefont {Scherzer}},
  \bibinfo {author} {\bibfnamefont {H.}~\bibnamefont {Thierschmann}}, \bibinfo
  {author} {\bibfnamefont {C.}~\bibnamefont {Sch\"afermeier}}, \bibinfo
  {author} {\bibfnamefont {C.}~\bibnamefont {Dal~Savio}}, \bibinfo {author}
  {\bibfnamefont {T.}~\bibnamefont {Taniguchi}}, \bibinfo {author}
  {\bibfnamefont {K.}~\bibnamefont {Watanabe}}, \bibinfo {author}
  {\bibfnamefont {D.}~\bibnamefont {Hunger}}, \bibinfo {author} {\bibfnamefont
  {K.}~\bibnamefont {Karra\"{\i}}},\ and\ \bibinfo {author} {\bibfnamefont
  {A.}~\bibnamefont {H\"ogele}},\ }\bibfield  {title} {\bibinfo {title}
  {Open-cavity in closed-cycle cryostat as a quantum optics platform},\ }\href
  {https://doi.org/10.1103/PRXQuantum.2.040318} {\bibfield  {journal} {\bibinfo
   {journal} {PRX Quantum}\ }\textbf {\bibinfo {volume} {2}},\ \bibinfo {pages}
  {040318} (\bibinfo {year} {2021})}\BibitemShut {NoStop}%
\bibitem [{\citenamefont {Tang}\ \emph {et~al.}(2023)\citenamefont {Tang},
  \citenamefont {Lou}, \citenamefont {Du}, \citenamefont {Zhang}, \citenamefont
  {Ni}, \citenamefont {Xu}, \citenamefont {Jin}, \citenamefont {Fan},\ and\
  \citenamefont {Mazur}}]{tang2023onchip}%
  \BibitemOpen
  \bibfield  {author} {\bibinfo {author} {\bibfnamefont {H.}~\bibnamefont
  {Tang}}, \bibinfo {author} {\bibfnamefont {B.}~\bibnamefont {Lou}}, \bibinfo
  {author} {\bibfnamefont {F.}~\bibnamefont {Du}}, \bibinfo {author}
  {\bibfnamefont {M.}~\bibnamefont {Zhang}}, \bibinfo {author} {\bibfnamefont
  {X.}~\bibnamefont {Ni}}, \bibinfo {author} {\bibfnamefont {W.}~\bibnamefont
  {Xu}}, \bibinfo {author} {\bibfnamefont {R.}~\bibnamefont {Jin}}, \bibinfo
  {author} {\bibfnamefont {S.}~\bibnamefont {Fan}},\ and\ \bibinfo {author}
  {\bibfnamefont {E.}~\bibnamefont {Mazur}},\ }\href@noop {} {\bibinfo {title}
  {On-chip optical twisted bilayer photonic crystal}} (\bibinfo {year}
  {2023}),\ \Eprint {https://arxiv.org/abs/2303.02325} {arXiv:2303.02325
  [physics.optics]} \BibitemShut {NoStop}%
\bibitem [{\citenamefont {Ferrier}\ \emph {et~al.}(2019)\citenamefont
  {Ferrier}, \citenamefont {Nguyen}, \citenamefont {Jamois}, \citenamefont
  {Berguiga}, \citenamefont {Symonds}, \citenamefont {Bellessa},\ and\
  \citenamefont {Benyattou}}]{Ferrier_2019}%
  \BibitemOpen
  \bibfield  {author} {\bibinfo {author} {\bibfnamefont {L.}~\bibnamefont
  {Ferrier}}, \bibinfo {author} {\bibfnamefont {H.~S.}\ \bibnamefont {Nguyen}},
  \bibinfo {author} {\bibfnamefont {C.}~\bibnamefont {Jamois}}, \bibinfo
  {author} {\bibfnamefont {L.}~\bibnamefont {Berguiga}}, \bibinfo {author}
  {\bibfnamefont {C.}~\bibnamefont {Symonds}}, \bibinfo {author} {\bibfnamefont
  {J.}~\bibnamefont {Bellessa}},\ and\ \bibinfo {author} {\bibfnamefont
  {T.}~\bibnamefont {Benyattou}},\ }\bibfield  {title} {\bibinfo {title} {Tamm
  plasmon photonic crystals: From bandgap engineering to defect cavity},\
  }\href {https://doi.org/10.1063/1.5104334} {\bibfield  {journal} {\bibinfo
  {journal} {{APL} Photonics}\ }\textbf {\bibinfo {volume} {4}},\ \bibinfo
  {pages} {106101} (\bibinfo {year} {2019})}\BibitemShut {NoStop}%
\bibitem [{\citenamefont {Parappurath}\ \emph {et~al.}(2020)\citenamefont
  {Parappurath}, \citenamefont {Alpeggiani}, \citenamefont {Kuipers},\ and\
  \citenamefont {Verhagen}}]{Parappurath2020}%
  \BibitemOpen
  \bibfield  {author} {\bibinfo {author} {\bibfnamefont {N.}~\bibnamefont
  {Parappurath}}, \bibinfo {author} {\bibfnamefont {F.}~\bibnamefont
  {Alpeggiani}}, \bibinfo {author} {\bibfnamefont {L.}~\bibnamefont
  {Kuipers}},\ and\ \bibinfo {author} {\bibfnamefont {E.}~\bibnamefont
  {Verhagen}},\ }\bibfield  {title} {\bibinfo {title} {Direct observation of
  topological edge states in silicon photonic crystals: Spin, dispersion, and
  chiral routing},\ }\bibfield  {journal} {\bibinfo  {journal} {Science
  Advances}\ }\textbf {\bibinfo {volume} {6}},\ \href
  {https://doi.org/10.1126/sciadv.aaw4137} {10.1126/sciadv.aaw4137} (\bibinfo
  {year} {2020})\BibitemShut {NoStop}%
\bibitem [{\citenamefont {Garc{\'{\i}}a-Elcano}\ \emph
  {et~al.}(2023)\citenamefont {Garc{\'{\i}}a-Elcano}, \citenamefont {Merino},
  \citenamefont {Bravo-Abad},\ and\ \citenamefont
  {Gonz{\'{a}}lez-Tudela}}]{Garca_Elcano_2023}%
  \BibitemOpen
  \bibfield  {author} {\bibinfo {author} {\bibfnamefont {I.}~\bibnamefont
  {Garc{\'{\i}}a-Elcano}}, \bibinfo {author} {\bibfnamefont {J.}~\bibnamefont
  {Merino}}, \bibinfo {author} {\bibfnamefont {J.}~\bibnamefont {Bravo-Abad}},\
  and\ \bibinfo {author} {\bibfnamefont {A.}~\bibnamefont
  {Gonz{\'{a}}lez-Tudela}},\ }\bibfield  {title} {\bibinfo {title} {Probing and
  harnessing photonic fermi arc surface states using light-matter
  interactions},\ }\bibfield  {journal} {\bibinfo  {journal} {Science
  Advances}\ }\textbf {\bibinfo {volume} {9}},\ \href
  {https://doi.org/10.1126/sciadv.adf8257} {10.1126/sciadv.adf8257} (\bibinfo
  {year} {2023})\BibitemShut {NoStop}%
\bibitem [{\citenamefont {Barik}\ \emph {et~al.}(2018)\citenamefont {Barik},
  \citenamefont {Karasahin}, \citenamefont {Flower}, \citenamefont {Cai},
  \citenamefont {Miyake}, \citenamefont {DeGottardi}, \citenamefont {Hafezi},\
  and\ \citenamefont {Waks}}]{Barik2018}%
  \BibitemOpen
  \bibfield  {author} {\bibinfo {author} {\bibfnamefont {S.}~\bibnamefont
  {Barik}}, \bibinfo {author} {\bibfnamefont {A.}~\bibnamefont {Karasahin}},
  \bibinfo {author} {\bibfnamefont {C.}~\bibnamefont {Flower}}, \bibinfo
  {author} {\bibfnamefont {T.}~\bibnamefont {Cai}}, \bibinfo {author}
  {\bibfnamefont {H.}~\bibnamefont {Miyake}}, \bibinfo {author} {\bibfnamefont
  {W.}~\bibnamefont {DeGottardi}}, \bibinfo {author} {\bibfnamefont
  {M.}~\bibnamefont {Hafezi}},\ and\ \bibinfo {author} {\bibfnamefont
  {E.}~\bibnamefont {Waks}},\ }\bibfield  {title} {\bibinfo {title} {A
  topological quantum optics interface},\ }\href
  {https://doi.org/10.1126/science.aaq0327} {\bibfield  {journal} {\bibinfo
  {journal} {Science}\ }\textbf {\bibinfo {volume} {359}},\ \bibinfo {pages}
  {666} (\bibinfo {year} {2018})}\BibitemShut {NoStop}%
\bibitem [{\citenamefont {Mehrabad}\ \emph {et~al.}(2020)\citenamefont
  {Mehrabad}, \citenamefont {Foster}, \citenamefont {Dost}, \citenamefont
  {Clarke}, \citenamefont {Patil}, \citenamefont {Fox}, \citenamefont
  {Skolnick},\ and\ \citenamefont {Wilson}}]{JalaliMehrabad:20}%
  \BibitemOpen
  \bibfield  {author} {\bibinfo {author} {\bibfnamefont {M.~J.}\ \bibnamefont
  {Mehrabad}}, \bibinfo {author} {\bibfnamefont {A.~P.}\ \bibnamefont
  {Foster}}, \bibinfo {author} {\bibfnamefont {R.}~\bibnamefont {Dost}},
  \bibinfo {author} {\bibfnamefont {E.}~\bibnamefont {Clarke}}, \bibinfo
  {author} {\bibfnamefont {P.~K.}\ \bibnamefont {Patil}}, \bibinfo {author}
  {\bibfnamefont {A.~M.}\ \bibnamefont {Fox}}, \bibinfo {author} {\bibfnamefont
  {M.~S.}\ \bibnamefont {Skolnick}},\ and\ \bibinfo {author} {\bibfnamefont
  {L.~R.}\ \bibnamefont {Wilson}},\ }\bibfield  {title} {\bibinfo {title}
  {Chiral topological photonics with an embedded quantum emitter},\ }\href
  {https://doi.org/10.1364/OPTICA.393035} {\bibfield  {journal} {\bibinfo
  {journal} {Optica}\ }\textbf {\bibinfo {volume} {7}},\ \bibinfo {pages}
  {1690} (\bibinfo {year} {2020})}\BibitemShut {NoStop}%
\bibitem [{\citenamefont {Ota}\ \emph {et~al.}(2018)\citenamefont {Ota},
  \citenamefont {Katsumi}, \citenamefont {Watanabe}, \citenamefont {Iwamoto},\
  and\ \citenamefont {Arakawa}}]{Ota2018}%
  \BibitemOpen
  \bibfield  {author} {\bibinfo {author} {\bibfnamefont {Y.}~\bibnamefont
  {Ota}}, \bibinfo {author} {\bibfnamefont {R.}~\bibnamefont {Katsumi}},
  \bibinfo {author} {\bibfnamefont {K.}~\bibnamefont {Watanabe}}, \bibinfo
  {author} {\bibfnamefont {S.}~\bibnamefont {Iwamoto}},\ and\ \bibinfo {author}
  {\bibfnamefont {Y.}~\bibnamefont {Arakawa}},\ }\bibfield  {title} {\bibinfo
  {title} {Topological photonic crystal nanocavity laser},\ }\bibfield
  {journal} {\bibinfo  {journal} {Commun. Phys.}\ }\textbf {\bibinfo {volume}
  {1}},\ \href {https://doi.org/10.1038/s42005-018-0083-7}
  {10.1038/s42005-018-0083-7} (\bibinfo {year} {2018})\BibitemShut {NoStop}%
\bibitem [{\citenamefont {Smirnova}\ \emph {et~al.}(2020)\citenamefont
  {Smirnova}, \citenamefont {Tripathi}, \citenamefont {Kruk}, \citenamefont
  {Hwang}, \citenamefont {Kim}, \citenamefont {Park},\ and\ \citenamefont
  {Kivshar}}]{Smirnova2020}%
  \BibitemOpen
  \bibfield  {author} {\bibinfo {author} {\bibfnamefont {D.}~\bibnamefont
  {Smirnova}}, \bibinfo {author} {\bibfnamefont {A.}~\bibnamefont {Tripathi}},
  \bibinfo {author} {\bibfnamefont {S.}~\bibnamefont {Kruk}}, \bibinfo {author}
  {\bibfnamefont {M.-S.}\ \bibnamefont {Hwang}}, \bibinfo {author}
  {\bibfnamefont {H.-R.}\ \bibnamefont {Kim}}, \bibinfo {author} {\bibfnamefont
  {H.-G.}\ \bibnamefont {Park}},\ and\ \bibinfo {author} {\bibfnamefont
  {Y.}~\bibnamefont {Kivshar}},\ }\bibfield  {title} {\bibinfo {title}
  {{Room-temperature lasing from nanophotonic topological cavities}},\ }\href
  {https://doi.org/10.1038/s41377-020-00350-3} {\bibfield  {journal} {\bibinfo
  {journal} {Light: Science \& Applications}\ }\textbf {\bibinfo {volume}
  {9}},\ \bibinfo {pages} {127} (\bibinfo {year} {2020})}\BibitemShut {NoStop}%
\bibitem [{\citenamefont {Bernevig}\ and\ \citenamefont
  {Hughes}(2013)}]{Bernevig2013}%
  \BibitemOpen
  \bibfield  {author} {\bibinfo {author} {\bibfnamefont {B.~A.}\ \bibnamefont
  {Bernevig}}\ and\ \bibinfo {author} {\bibfnamefont {T.~L.}\ \bibnamefont
  {Hughes}},\ }\href {https://doi.org/10.1515/9781400846733} {\emph {\bibinfo
  {title} {Topological Insulators and Topological Superconductors}}}\ (\bibinfo
   {publisher} {Princeton University Press},\ \bibinfo {year}
  {2013})\BibitemShut {NoStop}%
\bibitem [{\citenamefont {Lian}\ and\ \citenamefont {Zhang}(2016)}]{Lian2016}%
  \BibitemOpen
  \bibfield  {author} {\bibinfo {author} {\bibfnamefont {B.}~\bibnamefont
  {Lian}}\ and\ \bibinfo {author} {\bibfnamefont {S.-C.}\ \bibnamefont
  {Zhang}},\ }\bibfield  {title} {\bibinfo {title} {Five-dimensional
  generalization of the topological weyl semimetal},\ }\href
  {https://doi.org/10.1103/PhysRevB.94.041105} {\bibfield  {journal} {\bibinfo
  {journal} {Phys. Rev. B}\ }\textbf {\bibinfo {volume} {94}},\ \bibinfo
  {pages} {041105} (\bibinfo {year} {2016})}\BibitemShut {NoStop}%
\bibitem [{\citenamefont {Okamoto}(2021)}]{Okamoto2021}%
  \BibitemOpen
  \bibfield  {author} {\bibinfo {author} {\bibfnamefont {K.}~\bibnamefont
  {Okamoto}},\ }\href@noop {} {\emph {\bibinfo {title} {Fundamentals of optical
  waveguides}}},\ \bibinfo {edition} {3rd}\ ed.\ (\bibinfo  {publisher}
  {Academic Press},\ \bibinfo {address} {San Diego, CA},\ \bibinfo {year}
  {2021})\BibitemShut {NoStop}%
\bibitem [{\citenamefont {Devescovi}\ \emph {et~al.}(2021)\citenamefont
  {Devescovi}, \citenamefont {Garc{\'{\i}}a-D{\'{\i}}ez}, \citenamefont
  {Robredo}, \citenamefont {de~Paz}, \citenamefont {Lasa-Alonso}, \citenamefont
  {Bradlyn}, \citenamefont {Ma{\~{n}}es}, \citenamefont {Vergniory},\ and\
  \citenamefont {Garc{\'{\i}}a-Etxarri}}]{Devescovi2021}%
  \BibitemOpen
  \bibfield  {author} {\bibinfo {author} {\bibfnamefont {C.}~\bibnamefont
  {Devescovi}}, \bibinfo {author} {\bibfnamefont {M.}~\bibnamefont
  {Garc{\'{\i}}a-D{\'{\i}}ez}}, \bibinfo {author} {\bibfnamefont
  {I.}~\bibnamefont {Robredo}}, \bibinfo {author} {\bibfnamefont {M.~B.}\
  \bibnamefont {de~Paz}}, \bibinfo {author} {\bibfnamefont {J.}~\bibnamefont
  {Lasa-Alonso}}, \bibinfo {author} {\bibfnamefont {B.}~\bibnamefont
  {Bradlyn}}, \bibinfo {author} {\bibfnamefont {J.~L.}\ \bibnamefont
  {Ma{\~{n}}es}}, \bibinfo {author} {\bibfnamefont {M.~G.}\ \bibnamefont
  {Vergniory}},\ and\ \bibinfo {author} {\bibfnamefont {A.}~\bibnamefont
  {Garc{\'{\i}}a-Etxarri}},\ }\bibfield  {title} {\bibinfo {title} {{Cubic 3D
  Chern photonic insulators with orientable large Chern vectors}},\ }\href
  {https://doi.org/10.1038/s41467-021-27168-w} {\bibfield  {journal} {\bibinfo
  {journal} {Nat. Commun.}\ }\textbf {\bibinfo {volume} {12}},\ \bibinfo
  {pages} {7330} (\bibinfo {year} {2021})}\BibitemShut {NoStop}%
\bibitem [{\citenamefont {Devescovi}\ \emph {et~al.}(2022)\citenamefont
  {Devescovi}, \citenamefont {Garc{\'{\i}}a-D{\'{\i}}ez}, \citenamefont
  {Bradlyn}, \citenamefont {Ma{\~{n}}es}, \citenamefont {Vergniory},\ and\
  \citenamefont {Garc{\'{\i}}a-Etxarri}}]{Devescovi2022}%
  \BibitemOpen
  \bibfield  {author} {\bibinfo {author} {\bibfnamefont {C.}~\bibnamefont
  {Devescovi}}, \bibinfo {author} {\bibfnamefont {M.}~\bibnamefont
  {Garc{\'{\i}}a-D{\'{\i}}ez}}, \bibinfo {author} {\bibfnamefont
  {B.}~\bibnamefont {Bradlyn}}, \bibinfo {author} {\bibfnamefont {J.~L.}\
  \bibnamefont {Ma{\~{n}}es}}, \bibinfo {author} {\bibfnamefont {M.~G.}\
  \bibnamefont {Vergniory}},\ and\ \bibinfo {author} {\bibfnamefont
  {A.}~\bibnamefont {Garc{\'{\i}}a-Etxarri}},\ }\bibfield  {title} {\bibinfo
  {title} {Vectorial bulk-boundary correspondence for 3d photonic chern
  insulators},\ }\href {https://doi.org/10.1002/adom.202200475} {\bibfield
  {journal} {\bibinfo  {journal} {Adv. Opt. Mater.}\ }\textbf {\bibinfo
  {volume} {10}},\ \bibinfo {pages} {2200475} (\bibinfo {year}
  {2022})}\BibitemShut {NoStop}%
\bibitem [{\citenamefont {Haldane}(2004)}]{Haldane2004}%
  \BibitemOpen
  \bibfield  {author} {\bibinfo {author} {\bibfnamefont {F.~D.~M.}\
  \bibnamefont {Haldane}},\ }\bibfield  {title} {\bibinfo {title} {{Berry
  Curvature on the Fermi Surface: Anomalous Hall Effect as a Topological
  Fermi-Liquid Property}},\ }\href
  {https://doi.org/10.1103/PhysRevLett.93.206602} {\bibfield  {journal}
  {\bibinfo  {journal} {Phys. Rev. Lett.}\ }\textbf {\bibinfo {volume} {93}},\
  \bibinfo {pages} {206602} (\bibinfo {year} {2004})}\BibitemShut {NoStop}%
\bibitem [{\citenamefont {Halperin}(1987)}]{Halperin1987}%
  \BibitemOpen
  \bibfield  {author} {\bibinfo {author} {\bibfnamefont {B.~I.}\ \bibnamefont
  {Halperin}},\ }\bibfield  {title} {\bibinfo {title} {Possible states for a
  three-dimensional electron gas in a strong magnetic field},\ }\href
  {https://doi.org/10.7567/jjaps.26s3.1913} {\bibfield  {journal} {\bibinfo
  {journal} {Japanese Journal of Applied Physics}\ }\textbf {\bibinfo {volume}
  {26}},\ \bibinfo {pages} {1913} (\bibinfo {year} {1987})}\BibitemShut
  {NoStop}%
\bibitem [{\citenamefont {Berenger}(1994)}]{Berenger1994}%
  \BibitemOpen
  \bibfield  {author} {\bibinfo {author} {\bibfnamefont {J.-P.}\ \bibnamefont
  {Berenger}},\ }\bibfield  {title} {\bibinfo {title} {A perfectly matched
  layer for the absorption of electromagnetic waves},\ }\href
  {https://doi.org/10.1006/jcph.1994.1159} {\bibfield  {journal} {\bibinfo
  {journal} {J. Comput. Phys.}\ }\textbf {\bibinfo {volume} {114}},\ \bibinfo
  {pages} {185} (\bibinfo {year} {1994})}\BibitemShut {NoStop}%
\bibitem [{\citenamefont {Winkler}(2003)}]{Winkler2003}%
  \BibitemOpen
  \bibfield  {author} {\bibinfo {author} {\bibfnamefont {R.}~\bibnamefont
  {Winkler}},\ }\href {https://doi.org/10.1007/b13586} {\emph {\bibinfo {title}
  {{Spin{\textemdash}Orbit Coupling Effects in Two-Dimensional Electron and
  Hole Systems}}}}\ (\bibinfo  {publisher} {Springer Berlin Heidelberg},\
  \bibinfo {year} {2003})\BibitemShut {NoStop}%
\bibitem [{\citenamefont {{Per-Olov L\"owdin}}(1951)}]{Lowdin1951}%
  \BibitemOpen
  \bibfield  {author} {\bibinfo {author} {\bibnamefont {{Per-Olov L\"owdin}}},\
  }\bibfield  {title} {\bibinfo {title} {{A Note on the Quantum-Mechanical
  Perturbation Theory}},\ }\href {https://doi.org/10.1063/1.1748067} {\bibfield
   {journal} {\bibinfo  {journal} {J. Chem. Phys.}\ }\textbf {\bibinfo {volume}
  {19}},\ \bibinfo {pages} {1396} (\bibinfo {year} {1951})}\BibitemShut
  {NoStop}%
\bibitem [{Note1()}]{Note1}%
  \BibitemOpen
  \bibinfo {note} {Notice that, in PWE simulation, the TE and TM modes are
  defined for 2D photonic lattice and are opposite to our definitions here.
  Technically, we compute the TE mode in MPB by using the command for TM
  mode.}\BibitemShut {Stop}%
\end{thebibliography}%

\onecolumngrid
\pagebreak

\renewcommand{\theequation}{SE\arabic{equation}}
\renewcommand{\thefigure}{SF\arabic{figure}}
\renewcommand{\bibnumfmt}[1]{[S#1]}
\renewcommand{\citenumfont}[1]{S#1}
\setcounter{equation}{0}
\setcounter{figure}{0}
\setcounter{page}{1}

\begin{center}
	\large{\textbf{SUPPLEMENTAL MATERIAL:}}\\
	\Large{\textbf{Fermi arc reconstruction in synthetic photonic lattice}}
\end{center}

\twocolumngrid
\tableofcontents

\onecolumngrid

\section{Effective model}
In this section, we present the detailed derivation of the effective Hamiltonian of our trilayer grating from simple concepts of waveguide mode coupling. We will successively go from the monolayer grating to the bilayer and trilayer ones. Furthermore, we demonstrate the procedure to obtain the parameters for the Hamiltonian by fitting it with PWE simulation. Finally, we show how we compute the topological invariants, including the Berry curvature, Berry phase, and Chern number, with the effective model.
\subsection{Constructing the Hamiltonian\label{Sec: 1}}
\subsubsection{Monolayer grating}
\begin{figure}
	\includegraphics[width=0.65\linewidth]{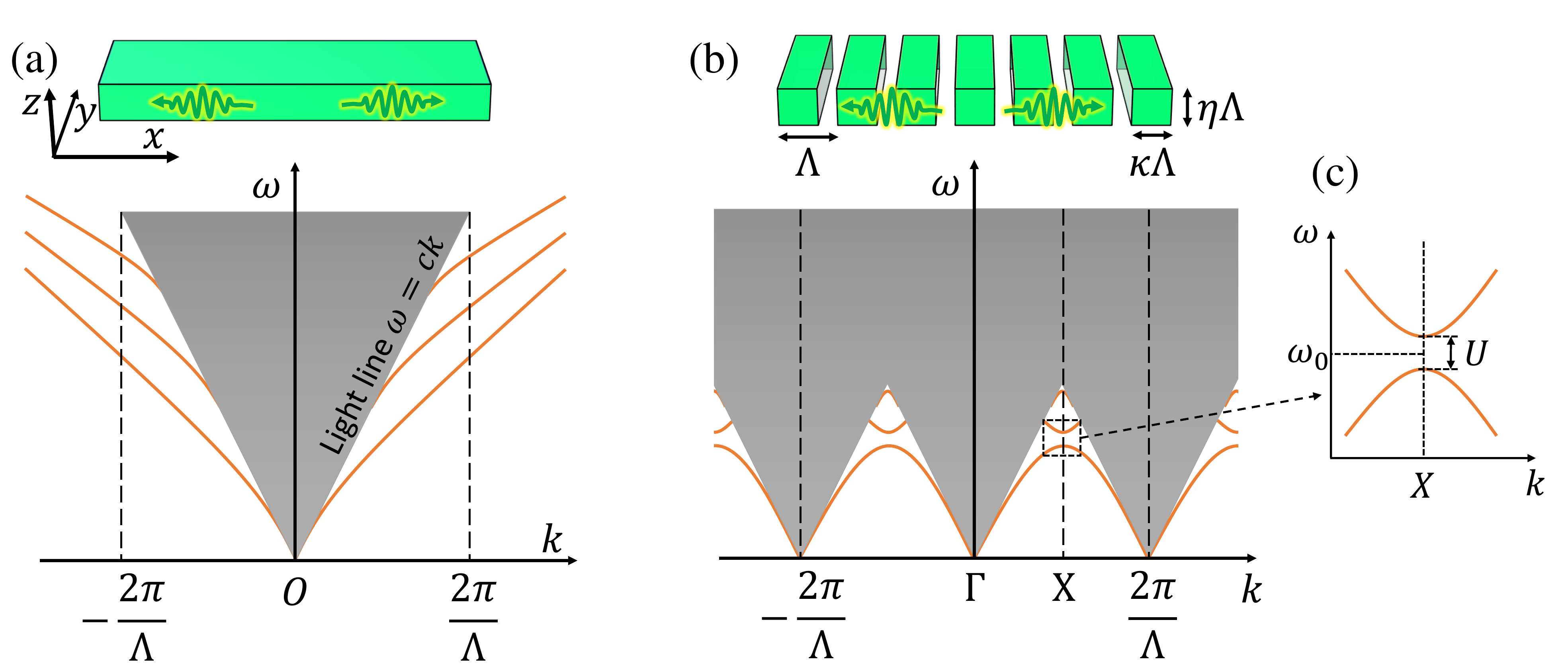}
	\caption{\label{fig:sup1} (a) A slab waveguide with subwavelength thickness and its schematic spectrum. (b) A 1D dielectric grating and its schematic spectrum. The thickness and grating period are of subwavelength scale. (c) The two bands of our interest are in the vicinity of $X$.}
\end{figure}

Light traveling in a homogeneous dielectric has linear dispersion $\omega(\mathbf{k})=v_0|\mathbf{k}|$, where $v_0=c/n$ is the group velocity and $n$ is the refractive index of the dielectric. When this dielectric gets confined in some direction, e.g., along the $z$-axis, with a thickness comparable to the wavelength, the photonic dispersion becomes quantized into discrete subbands, similar to the electronic confinement effect in quantum wells. These subbands disperse isotropically in the $xy$-plane, but we will focus on counter-propagating modes along the $x$ direction hereafter, as shown in Fig.~\ref{fig:sup1}(a) with $k$ being the wave vector along $x$. All guided modes lie below the light line, while those above are radiative modes. The symmetry of this slab waveguide allows the existence of two independent and nondegenerate electromagnetic modes, denoted as TE (transverse electric) mode and TM (transverse magnetic) mode~\cite{Okamoto2021}. In our work, we are interested in the TE mode, which has $E_x=E_z=H_y=0$ \footnote{Notice that, in PWE simulation, the TE and TM modes are defined for 2D photonic lattice and are opposite to our definitions here. Technically, we compute the TE mode in MPB by using the command for TM mode.}.

We create a periodic grating by introducing corrugation to our slab waveguide as shown in Fig~\ref{fig:sup1}(b) with grating period $\Lambda$, filling fraction (or fill factor) $\kappa$ and waveguide thickness $\eta\Lambda$. The corrugation gives rise to two effects in the spectrum. First, similar to condensed matter physics, the momentum space becomes periodic with period $2\pi/\Lambda$, and thus the original dispersion is repeated, or we can say that the bands are folded into the first BZ. With such band folding effect, the counter-propagating modes now cross each other at either $X$-point (guided modes) or $\Gamma$-point (radiative modes). Second, due to the diffractive mechanism, the counter-propagating modes can now couple to each other, depending on their parity. This makes some band crossings due to the band folding effect become anticrossing. We study the band anticrossing around $X$-point of two counter-propagating modes of lowest frequency [Figs.~\ref{fig:sup1}(b) and \ref{fig:sup1}(c)].

Since we only work with the range of $k$ around $X$, we change the variable $k-\pi/\Lambda\rightarrow k$ so that the origin is at $X$ for simplicity. The two bands of interest are described by the Hamiltonian
\begin{equation}
	H_1(k) = \omega_0 + \begin{pmatrix} vk & U \\ U & -vk \end{pmatrix},
\end{equation}
where $\omega_0$ is the midpoint of the photonic gap, $U$ is the diffractive coupling rate, and $v$ is the group velocity of the two guided modes at $X$ before coupling. This effective Hamiltonian is based on the approximation that treats the periodic grating as a perturbation to the homogeneous slab waveguide. The coefficient $U$ can be thought of as a first-order approximation of such a perturbation. As a result, this effective Hamiltonian works better for filling fractions $\kappa$ close to 1.
\subsubsection{Bilayer grating}
\begin{figure}
	\includegraphics[width=\linewidth]{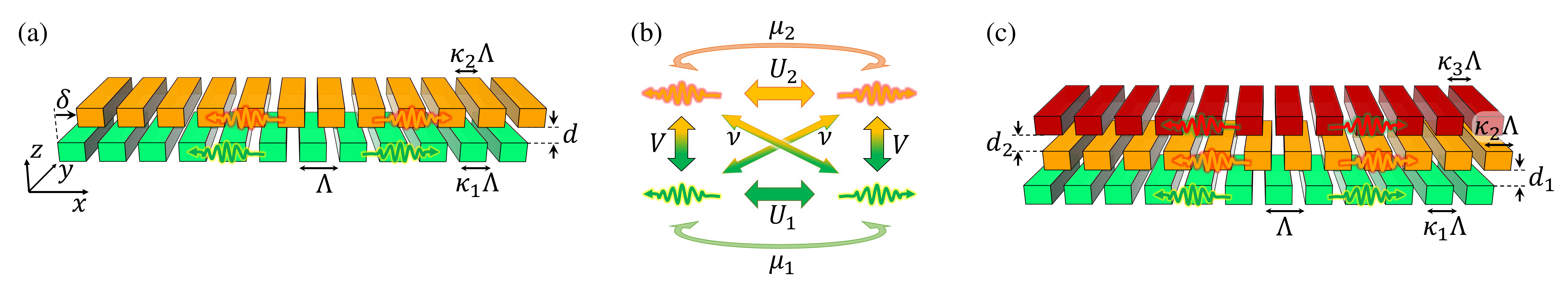}
	\caption{\label{fig:sup2} (a) Structure and geometrical parameters of a bilayer photonic grating. Each layer has two counter-propagating guided modes. (b) The coupling scheme of the guided modes in the bilayer system. (c) Structure and geometrical parameters of a trilayer photonic grating.}
\end{figure}

We consider a bilayer system composed of two 1D dielectric gratings that share the same grating period $\Lambda$ and are shifted from each other along the $x$ direction by $\delta$ [Fig.~\ref{fig:sup2}(a)]. Taking into account only two counter-propagating modes per layer as before, we arrive at four coupling rates for the four modes:
\begin{itemize}
	\item The diffractive coupling rate $U_i$ that couples the intralayer counter-propagating modes due to corrugation of the layer.
	\item The evanescent diffractive coupling rate $\mu_i$ that couples the intralayer counter-propagating modes due to corrugation of the other layer. This is ascribed to the evanescent field of each mode interacting with each other in the other grating.
	\item The evanescent coupling rate $V$ that couples co-propagating guided modes in different layers via the evanescent field.
	\item Diffracto-evanescent coupling rate $\nu$ that couples counter-propagating modes in different layers owing to the corrugation diffraction of the evanescent field. In particular, the evanescent field diffracts in the other grating and couples with a counter-propagating mode there.
\end{itemize}
The coupling scheme is shown in Fig.~\ref{fig:sup2}(b). Except $U_i$, all other coupling rates depend on the interlayer separation. On the other hand, the coupling rates $\mu_i$ and $\nu$ are small compared to $U_i$ and $V$ in general. They become more significant when the layer thickness is small compared to the grating period $\Lambda$, and/or the interlayer separation is small compared to the decay length of evanescent fields. The Hamiltonian describing the four guided modes in the bilayer grating is given by
\begin{equation}
	H_2(k,\delta) = \begin{pmatrix}
		\omega_{01}+v_1k & U_1+\mu_1e^{-i2\pi\frac{\delta}{\Lambda}} & Ve^{-i\pi\frac{\delta}{\Lambda}} & \nu e^{i\pi\frac{\delta}{\Lambda}}\\
		U_1+\mu_1e^{i2\pi\frac{\delta}{\Lambda}} & \omega_{01}-v_1k & \nu e^{-i\pi\frac{\delta}{\Lambda}} & Ve^{i\pi\frac{\delta}{\Lambda}}\\
		Ve^{i\pi\frac{\delta}{\Lambda}} & \nu e^{i\pi\frac{\delta}{\Lambda}} & \omega_{02}+v_2k & U_2+\mu_2e^{i2\pi\frac{\delta}{\Lambda}}\\
		\nu e^{-i\pi\frac{\delta}{\Lambda}} & Ve^{-i\pi\frac{\delta}{\Lambda}} & U_2+\mu_2e^{-i2\pi\frac{\delta}{\Lambda}} & \omega_{02}-v_2k
	\end{pmatrix}.
\end{equation}
The coupling coefficients related to the evanescent field carry a phase factor $\phi=K\delta$ owing to the displacement between two gratings, with $K$ being the wavevector. As we are working in the vicinity of $X$-point, $K=\pi/\Lambda$, and $\phi=\pi\delta/\Lambda$. Consequently, this Hamiltonian is not invariant under the translation $\delta\rightarrow\delta+\Lambda$ but can be transformed into its Bloch form by a unitary operation $H_2(k)\rightarrow\mathcal{U}^{\dagger}H_2(k)\mathcal{U}$ with $\mathcal{U}=\text{diag}(e^{-i\pi\frac{\delta}{\Lambda}}, e^{i\pi\frac{\delta}{\Lambda}}, 1, 1)$. The Bloch Hamiltonian reads
\begin{equation}
	H_2(k,\delta) = \begin{pmatrix}
		\omega_{01}+v_1k & U_1e^{i2\pi\frac{\delta}{\Lambda}}+\mu_1 & V & \nu\\
		U_1e^{-i2\pi\frac{\delta}{\Lambda}}+\mu_1 & \omega_{01}-v_1k & \nu & V\\
		V & \nu & \omega_{02}+v_2k & U_2+\mu_2e^{i2\pi\frac{\delta}{\Lambda}}\\
		\nu & V & U_2+\mu_2e^{-i2\pi\frac{\delta}{\Lambda}} & \omega_{02}-v_2k
	\end{pmatrix}.
\end{equation}
For simplicity, we neglect the coupling rates $\mu_i$ and $\nu$ in our work, i.e., $\mu_1=\mu_2=\nu=0$.

\subsubsection{Trilayer grating}
Finally, based on what we have argued so far, we can straightforwardly construct the Hamiltonian for a trilayer photonic grating. The coupling mechanisms we take into account are shown in Fig.~1. The operator reads
\begin{equation}
	H_3(k,\delta_1,\delta_2) = \begin{pmatrix}
		\omega_{01}+v_1k & U_1 & V_1e^{-i\pi\frac{\delta_1}{\Lambda}} & 0 & 0 & 0\\
		U_1 & \omega_{01}-v_1k & 0 & V_1e^{i\pi\frac{\delta_1}{\Lambda}} & 0 & 0\\
		V_1e^{i\pi\frac{\delta_1}{\Lambda}} & 0 & \omega_{02}+v_2k & U_2 & V_2e^{-i\pi\frac{\delta_2}{\Lambda}} & 0\\
		0 & V_1e^{-i\pi\frac{\delta}{\Lambda}} & U_2 & \omega_{02}-v_2k & 0 & V_2e^{i\pi\frac{\delta_2}{\Lambda}}\\
		0 & 0 & V_2e^{i\pi\frac{\delta_2}{\Lambda}} & 0 & \omega_{03}+v_3k & U_3\\
		0 & 0 & 0 & V_2e^{-i\pi\frac{\delta_2}{\Lambda}} & U_3 & \omega_{03}-v_3k
	\end{pmatrix},
\end{equation}
which is Eq.~(1) in the main text. Similar to the bilayer case, this Hamiltonian is not periodic with respect to $\delta_1$ and $\delta_2$. Hence, we apply a unitary transformation $H_3(k)\rightarrow\mathcal{U}^{\dagger}H_3(k)\mathcal{U}$ with $\mathcal{U}=\text{diag}(e^{-i\pi\frac{\delta}{\Lambda}}, e^{i\pi\frac{\delta}{\Lambda}}, 1, 1, e^{i\pi\frac{\delta}{\Lambda}}, e^{-i\pi\frac{\delta}{\Lambda}})$, which yields
\begin{equation}
	H_3(k,q_1,q_2) = \begin{pmatrix}
		\omega_{01}+v_1k & U_1e^{iq_1\Lambda} & V_1 & 0 & 0 & 0\\
		U_1e^{-iq_1\Lambda} & \omega_{01}-v_1k & 0 & V_1 & 0 & 0\\
		V_1 & 0 & \omega_{02}+v_2k & U_2 & V_2 & 0\\
		0 & V_1 & U_2 & \omega_{02}-v_2k & 0 & V_2\\
		0 & 0 & V_2 & 0 & \omega_{03}+v_3k & U_3e^{-iq_2\Lambda}\\
		0 & 0 & 0 & V_2 & U_3e^{iq_2\Lambda} & \omega_{03}-v_3k
	\end{pmatrix},
\end{equation}
where $q_j=2\pi\dfrac{\delta_j}{\Lambda^2}$ are the synthetic momenta. We make the Hamiltonian periodic in the momentum space in order to guarantee that we can use it later to compute the topological invariants.

\subsection{Obtaining parameters for the Hamiltonian\label{Sec: 1b}}
\begin{figure}
	\includegraphics[width=\linewidth]{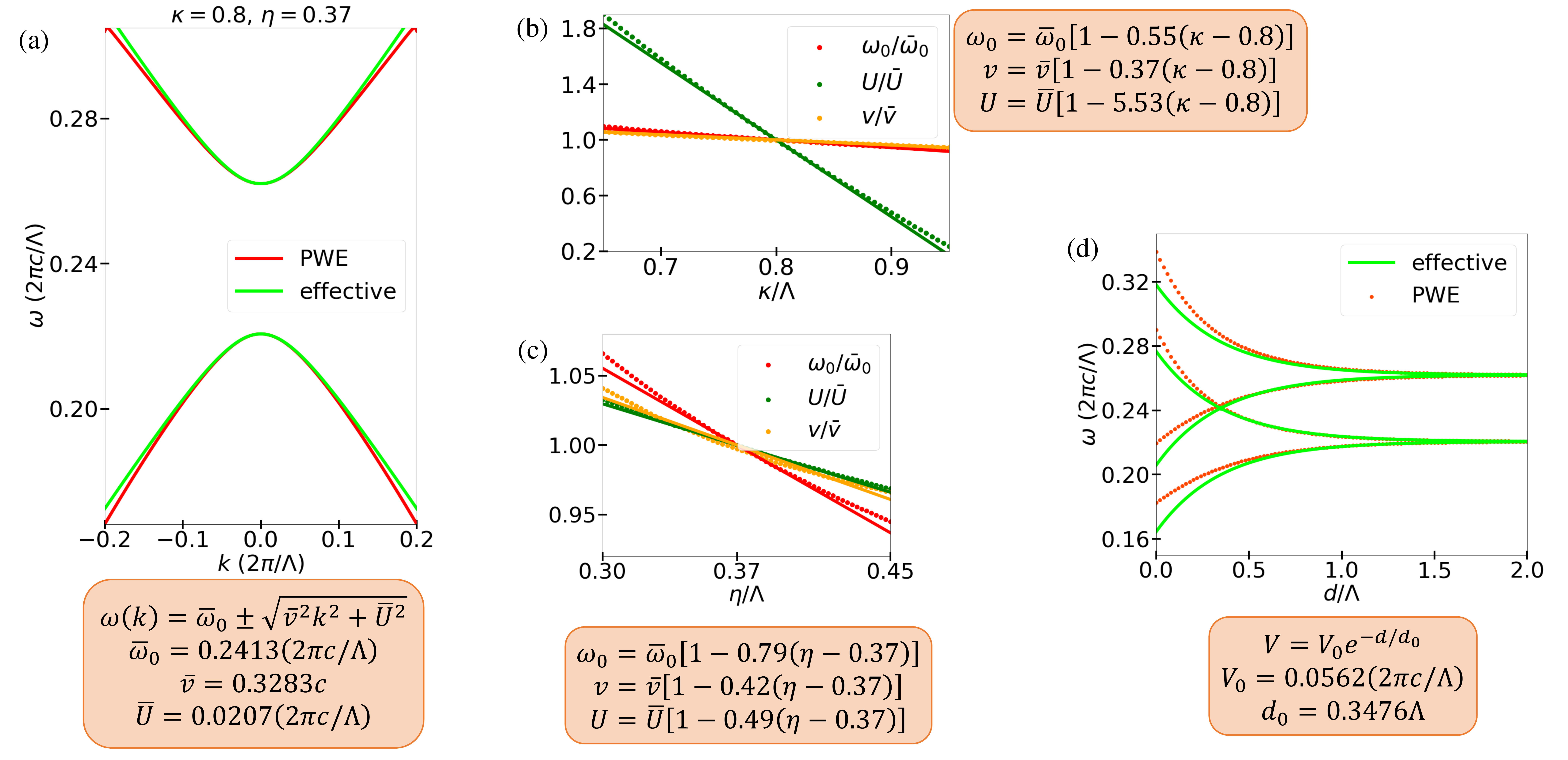}
	\caption{\label{fig:sup3} Parameter retrieval for the effective Hamiltonian from PWE simulations. (a)-(c) Simulations for a monolayer grating are used to obtain (a) $\bar{\omega}_0$, $\bar{U}$, $\bar{v}$, and their dependence on (b) the fill factor and (c) the waveguide thickness. (d) Simulation for aligned and identical gratings of standard geometry ($\kappa=0.8$, $\eta=0.37$) is used to obtain the dependence of $V$ as an exponential function of the interlayer separation $d$.}
\end{figure}

The three parameters $\omega_0$, $v$, and $U$ are obtained by fitting the dispersion $\omega(k)=\omega_0\pm\sqrt{v^2k^2+U^2}$ of a monolayer grating with PWE simulation. To ensure consistency with previous work~\cite{Chau2021}, we choose $\kappa=0.8$ and $\eta=0.37$ to describe a standard geometry of our system. The three parameters corresponding to such geometry are denoted as $\bar{\omega}_0$, $\bar{v}$, and $\bar{U}$. Regarding the PWE simulation with MPB, we construct a 2D supercell of size $(1,10)$ containing a dielectric block of size $(\kappa,\eta)$ with dielectric constant $\varepsilon=12$. The bands are computed around $X$-point with resolution 64 and shown in Fig.~\ref{fig:sup3}(a).

Next, we examine how these parameters vary when the grating deviates from the standard geometry by changing the filling fraction $\kappa$ and thickness $\eta$. The results are shown in Figs.~\ref{fig:sup3}(b) and \ref{fig:sup3}(c), where the round dots are simulated data and the solid lines are linear functions fit with these dots. In Fig.~\ref{fig:sup3}(b), we see that the parameters almost vary linearly with respect to $\kappa$, and the diffractive coupling rate changes more quickly than the frequency offset and group velocity. Hence, we may make a rough approximation to simplify our Hamiltonian by assuming that $\omega_{0l}$ and $v_l$ are the same even when their intralayer coupling rates $U_l$ are different. Besides, without that approximation, we can estimate $\omega_0$ and $v$ from $U$ using the linear functions fit with PWE. Considering Fig.~\ref{fig:sup3}(c), all three parameters vary comparably and slowly against the thickness $\eta$. The dependence of the parameters on $\eta$ is also linearly fit with PWE data.

Now, we consider a bilayer grating in order to parameterize the interlayer interaction. The system now consists of two identical gratings with standard geometry, separated by a distance $d$ [Fig.~\ref{fig:sup2}(a)]. We assume that the evanescent coupling strength of the effective model decays exponentially with respect to $d$, and fit the band edges at $k=0$ and $\delta=0$ with PWE, as shown in Fig.~\ref{fig:sup3}(d). With the relation $V(d)=V_0e^{-d/d_0}$, we can easily determine the interlayer distance given the coupling rate, e.g., $d=0.3472\Lambda$ for $V=1$. All parameters that we obtained so far agree well with those obtained by rigorous coupled-wave analysis (RWCA)~\cite{Chau2021}.

Finally, we consider the trilayer system with the parameters obtained by considering the monolayer and bilayer gratings. The structure of the trilayer lattice is shown again in Fig.~\ref{fig:sup2}(c) with its geometrical parameters, i.e., the filling fractions $\kappa_l$ and interlayer separation $d_j$. All layers share the same lattice constant $\Lambda$ and thickness $\eta\Lambda$. The band structure agrees well with the PWE simulation in shape and nodal point positions. We only make a minor increase in the frequency offset $\bar{\omega}_0=0.2413\rightarrow0.2466(2\pi c/\Lambda)$ so that the semi-Weyl points at C match in energy with the simulation [Fig.~2(b)]. For the PWE simulation, we construct a 2D supercell as before that consists of three dielectric blocks ($\varepsilon=12$) with resolution 32.
\subsection{Computing topological invariants}
\begin{figure}
	\includegraphics[width=\linewidth]{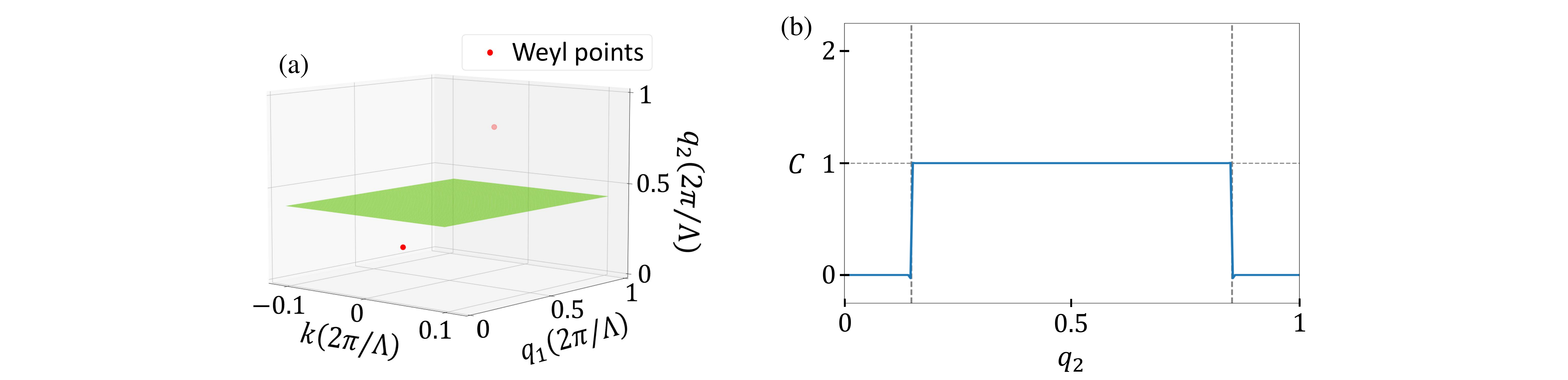}
	\caption{\label{fig:sup4} (a) A 2D cross-section of the BZ to determine the Chern number. As the cross-section is perpendicular to $q_2$ and translates over the BZ, the Chern number becomes a function of $q_2$. (b) The dependence of Chern number on $q_2$ when $U_1=U_3=V_1=V_2=1$ and $U_2=1.2$.}
\end{figure}

In general, the Berry curvature of a band $n$ in a parameter space $\vec{\mathcal{R}}$ is a rank-2 tensor given by~\cite{Bernevig2013}
\begin{align}
	\omega^n_{\mu\nu}(\vec{\mathcal{R}}) &= \partial_{\mathcal{R}_\mu}\mathcal{A}^n_{\nu}(\vec{\mathcal{R}}) - \partial_{\mathcal{R}_\nu}\mathcal{A}^n_{\mu}(\vec{\mathcal{R}})\\
	&= i\sum_{n'\neq n}\frac{\braket{n(\vec{\mathcal{R}})|\partial H/\partial\vec{\mathcal{R}}_\mu|n'(\vec{\mathcal{R}})}\braket{n'(\vec{\mathcal{R}})|\partial H/\partial\vec{\mathcal{R}}_\nu|n(\vec{\mathcal{R}})}-c.c.}{\left[\varepsilon_n(\vec{\mathcal{R}}) - \varepsilon_{n'}(\vec{\mathcal{R}})\right]^2}.
\end{align}
In a 3D parameter space, the Berry curvature is a vector field $\vec{\Omega}_n(\vec{\mathcal{R}})$ whose components are
\begin{equation}
	\Omega^\delta_n(\vec{\mathcal{R}}) = \frac{1}{2}\epsilon^{\delta\mu\nu}\Omega^n_{\mu\nu}(\vec{\mathcal{R}}).
\end{equation}
The Berry flux of band $n$ threading through a closed surface is then given by
\begin{equation}
	\gamma_n = \oint d\vec{S}\cdot\vec{\Omega}_n(\vec{\mathcal{R}}).
\end{equation}
The chiralities of Weyl points of bands (3) and (4) in Figs.~3(b) and 3(c) are the Berry flux of band (3) threading through a cube enclosing the Weyl point divided by $2\pi$. On the other hand, the Chern number in Figs.~3(d) and 3(e) is given by
\begin{equation}
	C(q_2)=\sum_{n=1}^3c_n(q_2)=\frac{1}{2\pi}\sum_{n=1}^3\gamma_n(q_2),
\end{equation}
where $\gamma_n(q_2)$ is the Berry flux of band $n$ threading a 2D cross-section at $q_2$ of the BZ, as schematically shown in Fig.~\ref{fig:sup4}(a). The summation of bands runs from 1 to 3 since there are three bands below the energy gap of our interest. Notice that since we are using an effective model, $k$ is a continuous variable defined in an infinite Euclidean space and thus the 2D cross-section extends to infinity in the $k$-direction.

As an example, we show the Chern number as a function of $q_2$ in Fig.~\ref{fig:sup4}(b). We consider the case of $U_1=U_3=V_1=V_2=1$ and $U_2=1.2$, where the semi-Weyl points are gapped, and only the two Weyl points in the $k=0$ plane remain. Here, the two vertical lines denote the positions of those Weyl points. The Chern number is 1 in the region between those two Weyl points and vanishes elsewhere. By changing also the evanescent coupling rates along line 1 or line 2 [Fig.~3(a)], we obtain the phase diagrams in Figs.~3(d) and 3(e).

In addition, we also calculate the Berry phase of the nodal lines between bands (2) and (3), or bands (4) and (5). The Berry phase of a 1D loop in the BZ is given by
\begin{equation}
	\mathcal{P}(\mathbf{k}) = i\ln\left[\det\left(W_{\mathcal{L}}(\mathbf{k})\right)\right],
\end{equation}
where the discrete Wilson loop matrix is defined as a product of the link matrices $U_n$ of the occupied bands $n$
\begin{equation}
	W_{\mathcal{L}} = \prod_{n=0}^{N-1}\frac{U_n}{|\det(U_n)|}.
\end{equation}
Here, elements of the link matrix are $U_n^{ij}=\braket{u_i(k_n)|u_j(k_{n+1})}$. We find that any closed loop linked with a nodal line in our photonic lattice to form a Hopf link would have $\pi$-Berry phase.
\section{Weyl point trajectories}
\begin{figure}
	\includegraphics[width=\linewidth]{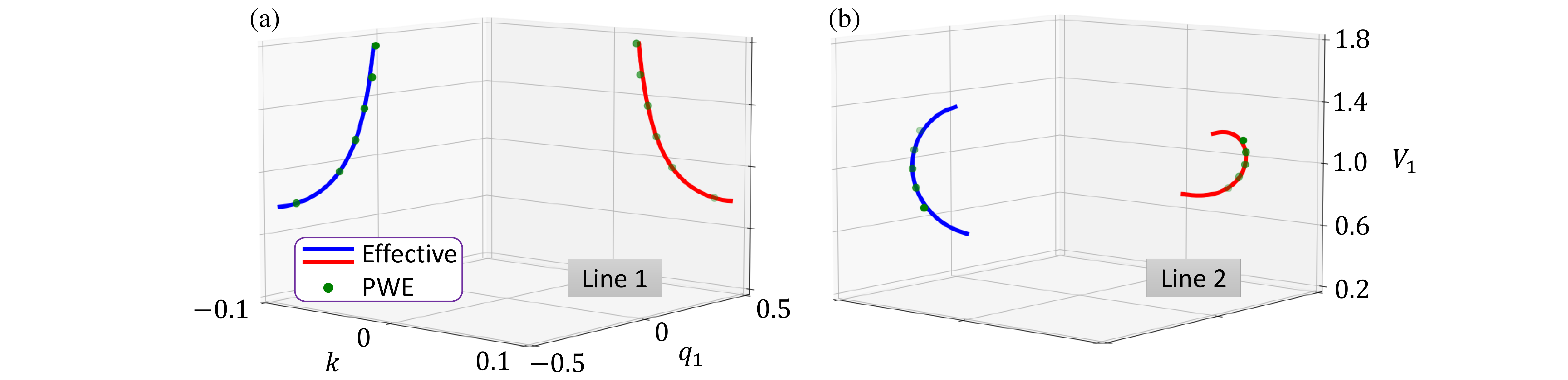}
	\caption{\label{fig:sup8} Weyl point trajectories when the evanescent coupling rates vary along line 1 and line 2 of Fig.~3(a). The solid lines are obtained from the effective Hamiltonian while the round dots are the results of the PWE simulation.}
\end{figure}

We have shown the Weyl point trajectories in Figs.~3(b) and 3(c). Here, in Fig.~\ref{fig:sup8}, we show that these trajectories are in excellent agreement with PWE simulation. The trajectories are consequent on varying the interlayer coupling coefficients. Regarding the effective Hamiltonian, we neglect the dependence of $\omega_{02}$ and $v_2$ on the filling fraction, namely $\omega_{01}=\omega_{02}=\omega_{03}$ and $v_1=v_2=v_3$, yet it still yields surprisingly good results compared with PWE.

\begin{figure}
	\includegraphics[width=\linewidth]{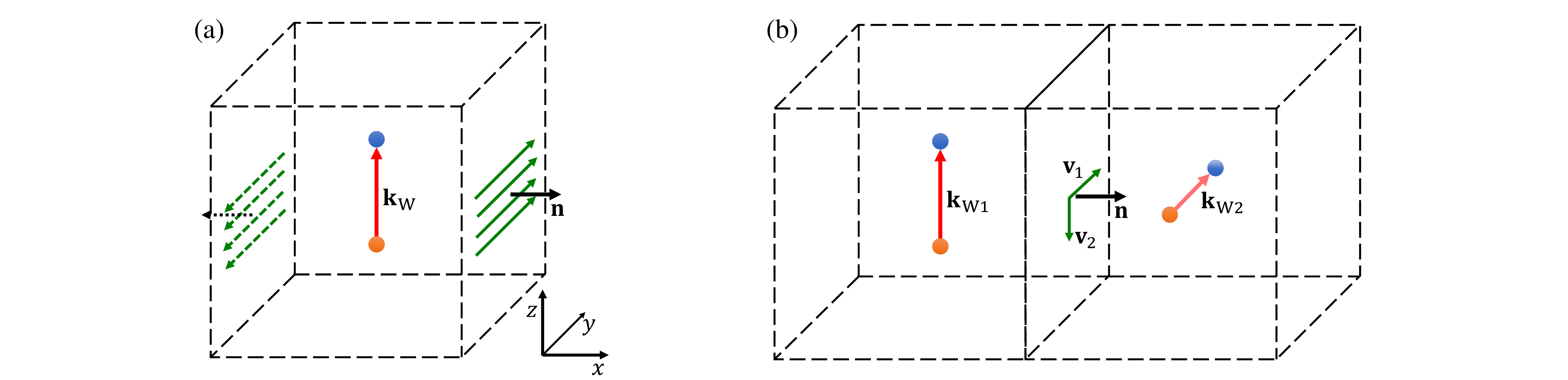}
	\caption{\label{fig:sup15} (a) A WSM with two Weyl nodes aligned along the $z$-axis. Vector $\textbf{k}_\text{W}$ connects the two Weyl nodes and points from the one with positive chirality to the other. The chiral electrons on a surface with normal vector $\textbf{n}$ move towards the direction given by $\textbf{k}_\text{W}\times\textbf{n}$. (b) At the interface between two WSMs, the chiral directions of original edge states are given by vectors $\textbf{v}_1=\textbf{k}_\text{W1}\times\textbf{n}$ and $\textbf{v}_2=\textbf{k}_\text{W2}\times(-\textbf{n})$. The interface Fermi-arc states thus have chirality along the $\textbf{v}_1=(\textbf{k}_\text{W1}-\textbf{k}_\text{W2})\times\textbf{n}$ direction.}
\end{figure}
\section{Chiral direction of the surface states}
Figure~\ref{fig:sup15} schematically illustrates how to determine the chiral direction of an original FA and a reconstructed FA.
\section{Fermi loop reconstruction}
Three-dimensional Chern insulator (CI), also known as 3D quantum anomalous Hall insulator, features chiral surface states whose Fermi surface is a loop traversing the Brillouin zone (BZ), called Fermi loop~\cite{Vanderbilt2018,Devescovi2021,Devescovi2022}. From Fig.~3 of the main text, we see that the trilayer lattice can host three CI phases. To distinguish these phases, we use the Chern vector~\cite{Haldane2004,Vanderbilt2018}, $\mathbf{C} = (C_0, C_1, C_2)$, whose components are the Chern numbers defined in the 2D planes perpendicular to $k$, $q_1$ and $q_2$. In our photonic lattice, these Chern vectors always lie in the $(q_1, q_2)$-plane where the time-reversal symmetry is broken. The Chern vectors of CIs 1, 2 and 3 are $\mathbf{C} = (0, 1, 1)$, $(0, 0, 1)$ and $(0, 1, 0)$, respectively.

Having a lattice that hosts different CI phases allows us to simulate the interface states between two different CIs. We consider a photonic junction of two configurations of the trilayer grating | see Fig.~4(a). The two CIs are chosen to have perpendicular Chern vectors~\cite{Halperin1987,Haldane2004,Vanderbilt2018}, i.e., $\mathbf{C}=(0,1,0)$ and $\mathbf{C}=(0,0,1)$. The junction's spectrum in the diagonal plane, obtained by the effective model, is shown in Fig.~4(b) (left) together with its transmission spectrum (right) simulated by the FDTD method. Comparing with the effective model, which clearly distinguishes the bulk and edge states, we can identify the gapless modes in the transmission spectrum of our junction as the edge modes.

To see the Fermi loops, we choose a frequency close to the middle of the energy gap at $q_1=q_2=0$ indicated by the dashed lines in Fig.~4(b). The resultant isofrequency contours are shown in Fig.~4(c), forming a $(1,1)$ torus knot~\cite{Liu2022v}. Owing to the agreement between the effective model and FDTD simulation, we also employ the effective model to show how the original Fermi loops are reconstructed into this torus knot. When each side of the junction individually interfaces with a trivially non-transparent medium, their corresponding surface Fermi loops can be obtained, which are shown by the dashed lines in Fig.~4(c). We can see that the two original Fermi loops hybridize and become disconnected at their intersection point to form a single loop. The two branches of this loop are greatly pushed away from the BZ center owing to the strong coupling between the two systems at the interface. This torus knot indeed aligns with recent simulation~\cite{Devescovi2022} and experimental observation~\cite{Liu2022v} in other 3D photonic platforms.

\begin{figure}
	\includegraphics[width=\linewidth]{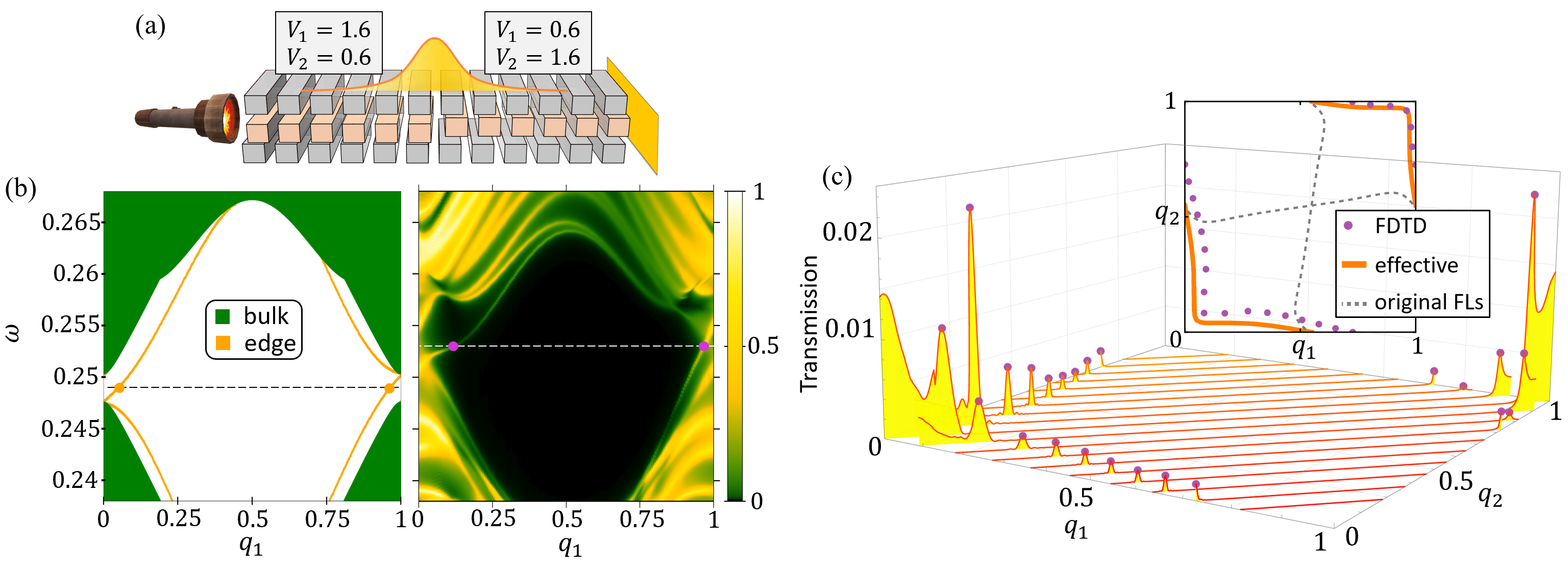}
	\caption{\label{fig:sup9} Junctions of two CIs. (a) Sketch of the setup for simulation. The intralayer coupling strengths are $U_1=U_3=1$ and $U_2=1.2$. (b) shows the energy spectrum of the effective model (left) and the transmission spectrum obtained from FDTD simulation (right) of the systems in the diagonal plane $q_1=q_2$. The dashed lines indicate the frequencies for visualizing the isofrequency transmission over the synthetic BZ in (c). The inset of (c) shows the reconstructed FAs with the dashed gray lines being the FAs before reconstruction.}
\end{figure}
\section{FDTD and PWE simulations\label{Sec: Simulation details}}
In this section, we present the details of our FDTD simulations using MEEP to obtain Fig.~4. To solidify our findings, we also examine the photonic junction with PWE and see the existence of the gapless edge states. Finally, we show two additional cases of photonic interface states, one forms between two WSMs and the other is between two CIs.

\begin{figure}
	\includegraphics[width=\linewidth]{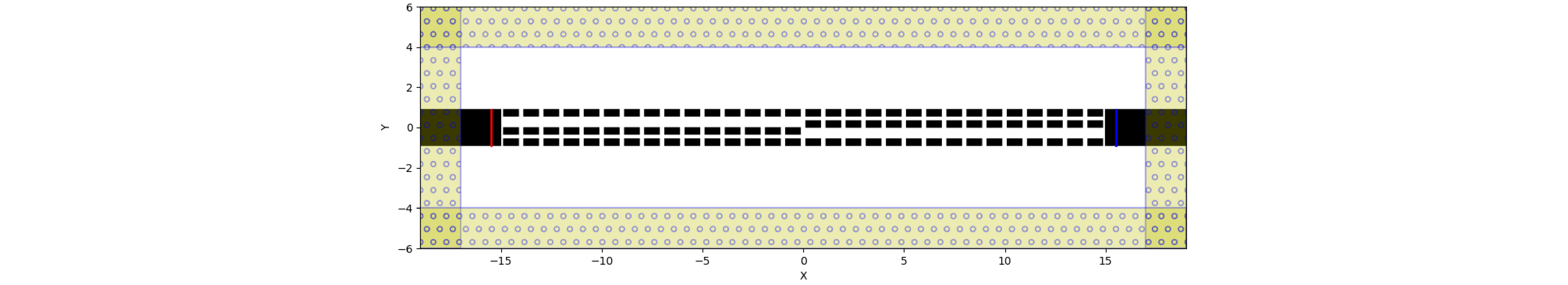}
	\caption{\label{fig:sup11} Computational cell for the FDTD simulation. The red line indicates the light source, while the blue line is where we measure the flux.}
\end{figure}

\subsection{Simulation details}
To construct the computational cell in MEEP for the photonic junction, we first choose the lattice constant $\Lambda$ to be the unit length, namely $\Lambda=1$. The structure is composed of dielectric blocks which are infinite along the $z$-direction and have dielectric constant $\varepsilon=12$ and height $\eta=0.37$. Since we often consider cases with $U_1=U_3=1$ and $U_2=1.2$, the block widths are then $\kappa_1=\kappa_3=0.8$ and $\kappa_2=0.764$ | see Fig.~\ref{fig:sup3}(b) for the relation between $U_l$ and $\kappa_l$. The separations between layers are given by $V=V_0e^{-d/d_0}$ [Fig.~\ref{fig:sup3}(d)] as mentioned in Sec.~\ref{Sec: 1b}. The computational cell is surrounded by four perfectly matched layers (PMLs)~\cite{Berenger1994} of thickness 2, which are supposed to absorb electromagnetic waves without any reflection. The photonic structure is separated from the PMLs by a padding block of dielectric constant 12 and length 2. The setup is shown in Fig.~\ref{fig:sup11}. The resolution per unit length is chosen to be 32.

In order to obtain the transmission spectrum of our system, we put a Gaussian-pulse source in the padding block on the left, as indicated by the red line in Fig.~\ref{fig:sup11}, and measure the flux threading through an area, indicated by the blue line, in the padding block on the right until the electromagnetic wave dissipates. The number of lattice sites is chosen so that we have 15 sites for each side. This number should be sufficiently small for us to observe the interface states that exponentially localize at $x=0$, but it also has to be sufficiently large so that the structure of bulk states is captured. The Gaussian pulse centers at frequency $f_c=0.2525$ ($2\pi c/\Lambda$) with a width $df=0.03$ ($2\pi c/\Lambda$). The fluxes are computed for 500 frequencies centered on $f_c$, from $f_c-df/2$ to $f_c+df/2$. As mentioned before, we are interested in the TE mode, the source is chosen to emit electric current with $E_z$ component.

With the transmission spectrum, we vary the interlayer displacements $\delta_i$ to get the complete spectrum with respect to $q_1$ and $q_2$. For instance, the transmission spectra in Figs.~4(b) and \ref{fig:sup9}(b) are defined by varying $\delta_1$ and $\delta_2$ with the condition $\delta_1=\delta_2$. To produce the isofrequency surface, we compute the spectrum in more planes parallel to the $(\delta_1=\delta_2)$-plane, and then, at the frequency of interest, we detect the peaks that correspond to the edge states. The results are shown in Figs.~4(d) of the main text and \ref{fig:sup9}(c). The frequecy taken in these cases are 0.247720 and 0.252604, respectively.
\subsection{Verifying with PWE}
Besides FDTD simulation using MEEP, we can compute the electromagnetic spectrum of our photonic junctions with PWE using MPB. Here, we consider a junction where both sides have $U_1=U_3=1$, $U_2=1.2$. The coupling rates of the left system are $V_1=0.6$ and $V_2=1.6$ while those of the right system are $V_1=1.6$ and $V_2=0.6$. This case is considered and shown in Figs.~4(a)-(c).

\begin{figure}
	\includegraphics[width=\linewidth]{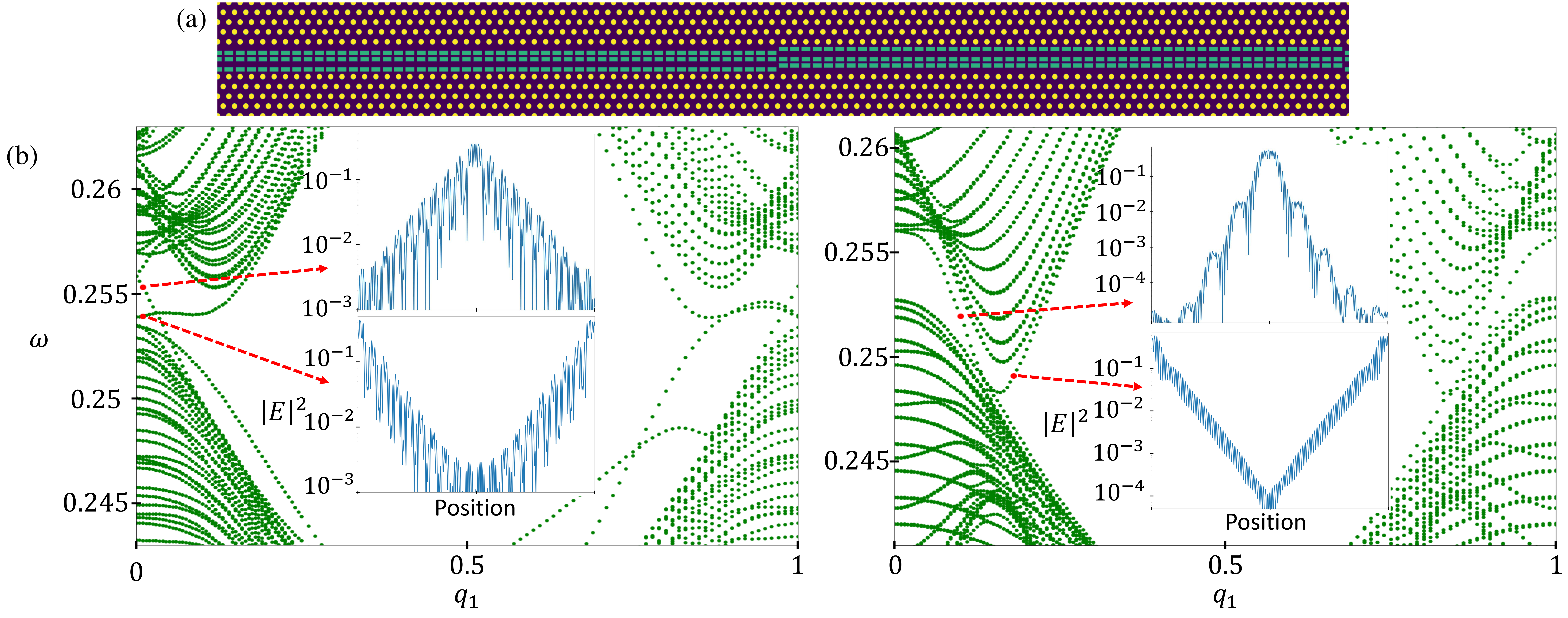}
	\caption{\label{fig:sup12} (a) Computational cell for PWE simulation of a photonic junction. The green rectangles indicate dielectric blocks with $\varepsilon=12$ that constitute our 1D lattice. The yellow circles are dielectric cylinders with $\varepsilon=18$ that form the 2D lattice to prevent the radiative modes from the trilayer lattice. (b) The frequency spectrum of the whole system in the plane $q_1=q_2$. The insets show the electric field distribution of the two red dots in the two gapless bands. The position axis of the insets corresponds to the computational cell structure in (a).}
\end{figure}

Similar to MEEP, we have to define a 2D supercell for our calculation. However, we can now define a large supercell with a large number of lattice sites to reduce the finite size effect, which also requires a significantly greater computational resource. Moreover, we can make the two systems misaligned to see if the edge states withstand a small defect. On the other hand, there are two major disadvantages of using MPB for such a 1D system that makes us work with MEEP instead. Both of them are related to the fact that the supercell is repeated periodically in space. The first disadvantage is that we always have edge states of two different interfaces owing to the periodicity of the supercell along the $x$-direction. Hence, we need to examine the field distribution of those states in order to know where they localize. Besides, the system has to be sufficiently large so that the hybridization of the edge states becomes negligible, otherwise, they will no longer seem gapless. The second one is the confinement of radiative modes in the free-space region due to the periodicity of the supercell along the $y$-direction. To remedy this problem, we replace the free space with an insulating lattice, which, however, would have some deformation effects on the band structure of our 1D system. Such effects can be remedied by replacing the free space with an insulating lattice, which may, however, induce some deformation effects to our 1D system. Such effects can also be regarded as perturbations that should not affect the observation of topological edge states.

We construct the 1D lattice with parameters the same as the FDTD simulation with the number of lattice sites being 50 for each side. In the FDTD simulation, layers 1 and 3 of both sides are aligned, but here we choose only layer 2 to be aligned. The trivial 2D lattice that prevents confinement of radiative modes is chosen to be a triangular lattice of cylindrical rods whose dielectric constant is 18. The computational cell is shown in Fig.~\ref{fig:sup12}(a). The frequency spectrum in the $(q_1=q_2)$-plane is shown in Fig.~\ref{fig:sup12}(b). We can see that, despite some deformation of the bulk bands due to the triangular lattice, the spectrum resembles the transmission spectrum of Fig.~4(b). The main difference between the two methods, as mentioned before, is that here we see two gapless bands connecting the bulk bands due to the existence of two edges. The insets show the field distribution of two states of those two gapless bands, showing clear exponential localization of these states at the edges. Hence, we conclude that the interface states can be observed in PWE simulations, and they persist despite small defects at the interface and external perturbation from the triangular lattice.

\subsection{Additional data}
\begin{figure}
	\includegraphics[width=\linewidth]{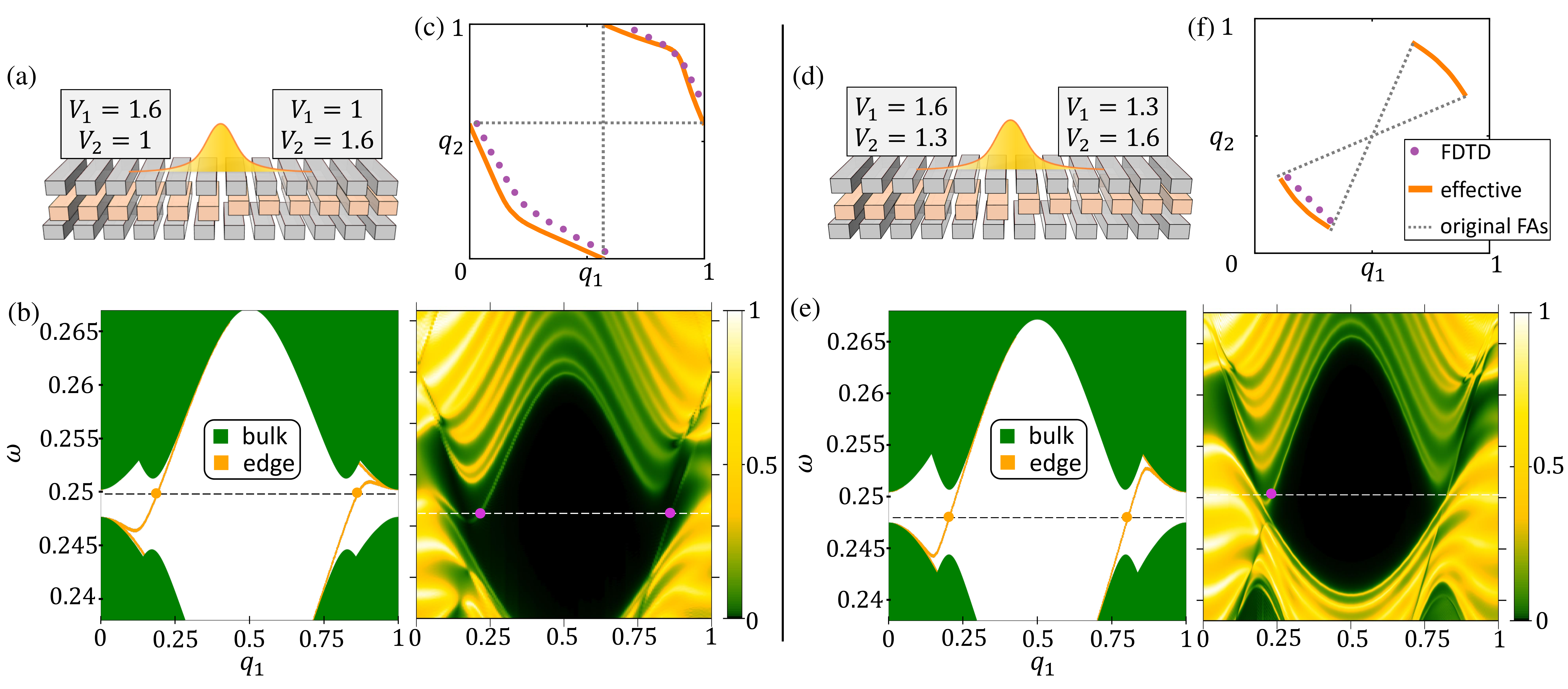}
	\caption{\label{fig:sup13} Junctions of two CIs with perpendicular Chern vectors [(a)-(c)] and of two WSMs with tilted anisotropy axes [(d)-(f)]. (a) and (d) are sketches of the setups. The intralayer coupling strengths are $U_1=U_3=1$ and $U_2=1.2$. (b) and (e) show the energy spectra of the effective model (left) and the transmission spectra obtained from FDTD simulation (right) of the systems in the diagonal plane $q_1=q_2$. The color bar scale indicates the optical transmission. The dashed lines indicate the frequencies for visualizing the isofrequency contours of edge modes in (c) and (f), which are the reconstructed Fermi loops and arcs. The dotted gray lines are schematic plots of the FLs and FAs before reconstruction.}
\end{figure}

We consider two additional photonic junctions to demonstrate FL and FA reconstruction, as shown in Fig.~\ref{fig:sup13}. Here, Figs.~\ref{fig:sup13}(a)-(c) correspond to a junction of two CIs while Figs.~\ref{fig:sup13}(a)-(c) correspond to a junction of two WSMs. In the former case, the mismatch in the energy of the Weyl dome and semi-Weyl dome in the transmission spectrum alters the edge states drastically. Thus, the frequency for the isofrequency surface plot is chosen so that the effective model fits well with FDTD simulation, as indicated by the gray dashed lines in Fig.~\ref{fig:sup13}(b). Considering Figs.~\ref{fig:sup13}(d)-(f), we choose the isofrequency surface to be close to the Weyl point frequency as before, which gives good agreement between the effective model and FDTD simulation. One minor difference is the absence of one reconstructed arc near $(q_1,q_2)=2\pi/\Lambda(1,1)$ because this arc already merges to the bulk states. If we plot at higher frequencies, both arcs of the FDTD simulation will appear and still agree well with the effective model.

\section{Disorders and defects}
The definition of the synthetic momenta is based on our system being invariant under the translations $\delta_j\rightarrow\delta_j+\Lambda$. A question then naturally arises: what is the influence of disorders and defects onto the spectrum? This question is important since structural errors are certainly present in the fabrication of samples.
\subsection{Structural disorders}
In this section, we examine this question, focusing on the structural errors due to the fabrication of the silicon gratings. The error is assumed to occur randomly following a normal distribution with standard deviation $\sigma_{\text{error}}=2$~nm or 5 nm. These values are taken based on realistic fabrication of silicon gratings. In particular, the gratings are composed of silicon rods whose width and height are $w=\kappa\Lambda$ and $h=\eta\Lambda$, respectively | see Fig.~\ref{fig:sup1}(b). These values are perturbed independently by the random error $\Delta d$, i.e., $w+\Delta d$ and $h+\Delta d$. As we have proposed two experimental schemes | one is to fabricate several rigid samples and the other is to dynamically tune one sample, we consider the influence of disorders in these two cases.

\begin{figure}
	\includegraphics[width=\linewidth]{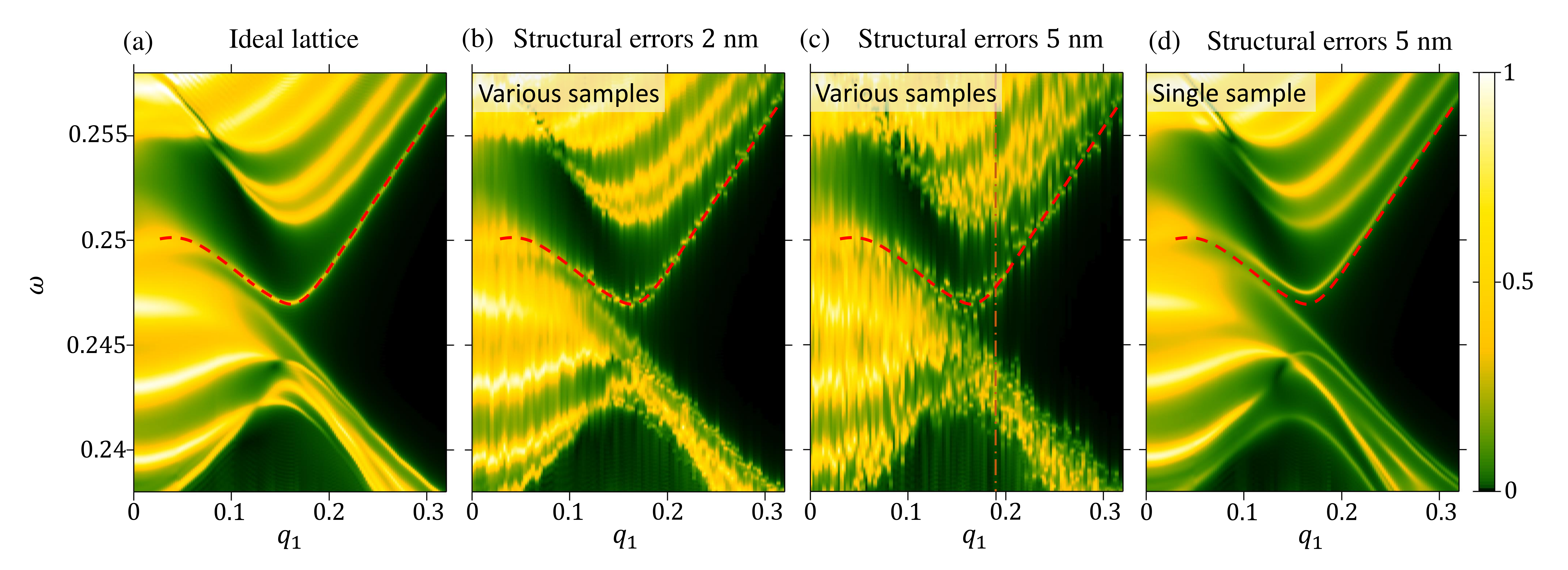}
	\caption{\label{fig:disorders} Transmission spectra at $q_1=q_2$ surface in the presence of structural disorders. (a) an ideal lattice without defects or disorders; (b) and (c) show two lattices with structural disorders whose distribution has a standard deviation of 2 nm and 5 nm, respectively. The disorder configuration varies with respect to the synthetic momenta; (d) a lattice with structural disorders whose distribution has a standard deviation of 5 nm. The disorder configuration remains unchanged with respect to the synthetic momenta.}
\end{figure}

The results are shown in Fig.~\ref{fig:disorders} where we compute the transmission spectra along the diagonal line $q_1=q_2$. Here, Fig.~\ref{fig:disorders}(b) and \ref{fig:disorders}(c) are associated with the experimental scheme of fabricating several rigid bodies. Each value of $q$ corresponds to a sample, and the disorder configurations of all samples are different from each other. We see that the edge states are randomly scatter around the unperturbed value (notated by the red dashed line) and the distribution width of these points depends on $\sigma_{\text{error}}$. Nevertheless, the shape of the gapless edge state is still recognizable for $\sigma_{\text{error}}=5$~nm. On the other hand, Fig.~\ref{fig:disorders}(d) is associated with the experimental scheme of dynamically tuning sample, i.e., all values of $q$ have the same disorder configuration. We see that the spectrum is slightly altered and certainly shares the same physics as an ideal lattice.

\begin{figure}
	\includegraphics[width=\linewidth]{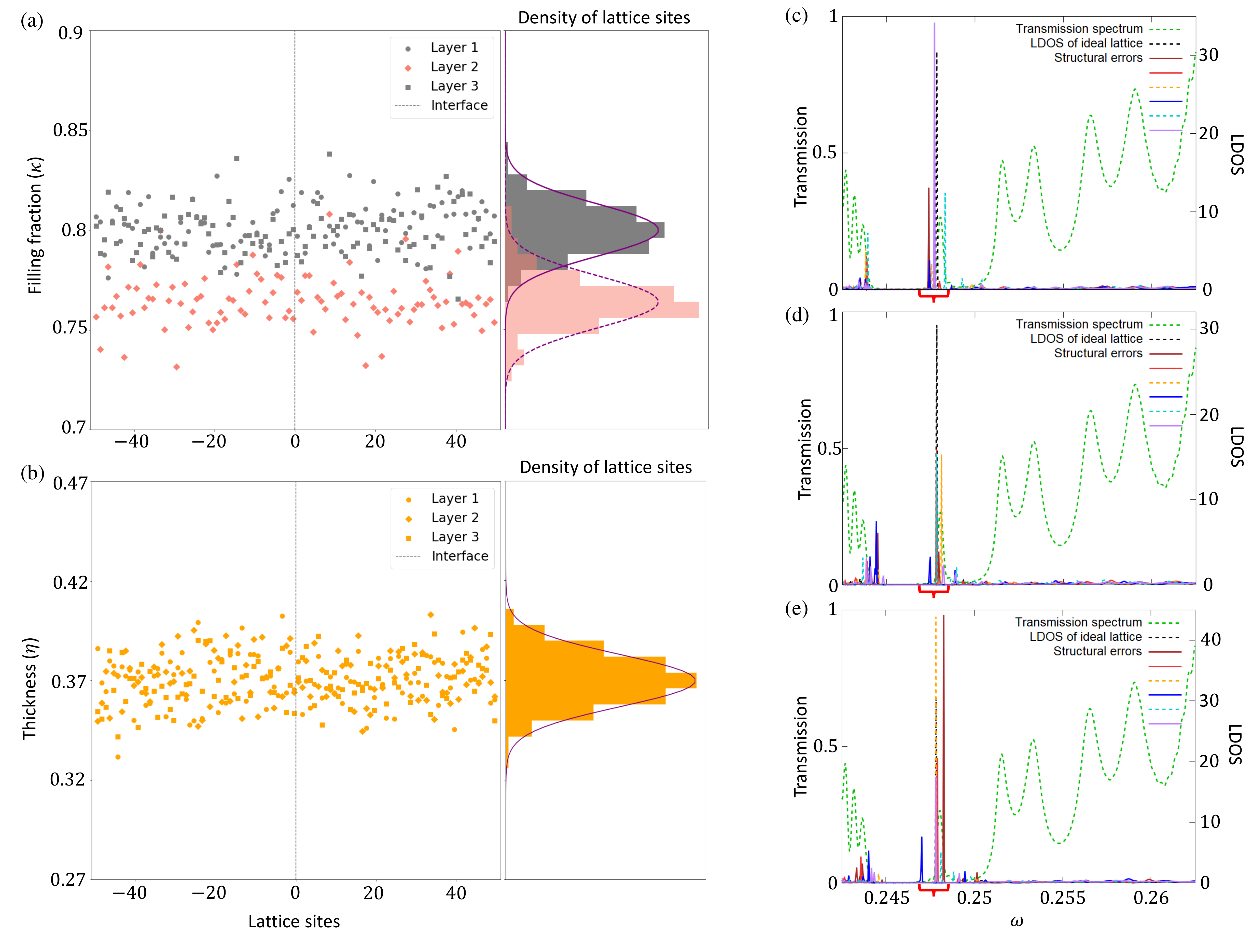}
	\caption{\label{fig:ldos} Local density of states (LDOS) at $q_1=q_2=0.19$ in the presence of structural disorders. (a) and (b) show an exemplary configuration of 5 nm structural disorders. While (a) represents the filling fraction of each grating, i.e., each lattice site, (b) shows their thicknesses. (c), (d) and (e) show the LDOS for several disorder configurations: (c) only filling fraction fluctuates, (d) only thickness fluctuates, and (e) both filling fraction and thickness are disordered. The green dashed line denotes the transmission spectrum, the black dashed line shows the LDOS of an ideal lattice, and the other lines are the LDOS for different disorder configurations.}
\end{figure}

As mentioned in the simulation details (Sec.~\ref{Sec: Simulation details}), the number of lattice sites must be sufficiently small so that we can see the edge state in the transmission spectrum. This makes transmission spectrum not an entirely ideal simulation scheme for examining the structural errors. To see how the edge state changes in larger samples, such as a junction with 100 lattice sites, we compute the local density of states (LDOS) at the junction interface. Since LDOS is more computational demanding, we consider the lattice at $q_1=q_2=0.19$ for several disorder configurations. The results are shown in Fig.~\ref{fig:ldos} for 5~nm structural errors. The simulated structure is similar to Fig.~\ref{fig:sup11} but there are 50 lattice sites per side and there are no padding blocks - the periodic lattice is merged directly into the PML. Figs.~\ref{fig:ldos}(a) and \ref{fig:ldos}(b) show an example of disorder configuration - the thickness (height) and filling fraction (width) of the grating rods are randomly distributed around a mean value from the ideal lattice. Specifically, the mean values of filling fraction are 0.8 and 0.764 while the mean value of thickness is $0.37\Lambda$. Figs.~\ref{fig:ldos}(c)-(e) show the LDOS for three different cases of adding disorders, in each case we consider six disorder configurations. Overall, we see that the LDOS of an ideal lattice concurs with the transmission spectrum, and the edge states are scattered around the unperturbed value - as observed in Fig.~\ref{fig:disorders}(c). All edge state peaks of these cases lie within an energy window shown in the figures. Hence, we conclude that although structural errors have a great influence on the spectrum, the reconstructed edge modes are still experimentally observable.
\subsection{Edge defects}
\begin{figure}
	\includegraphics[width=\linewidth]{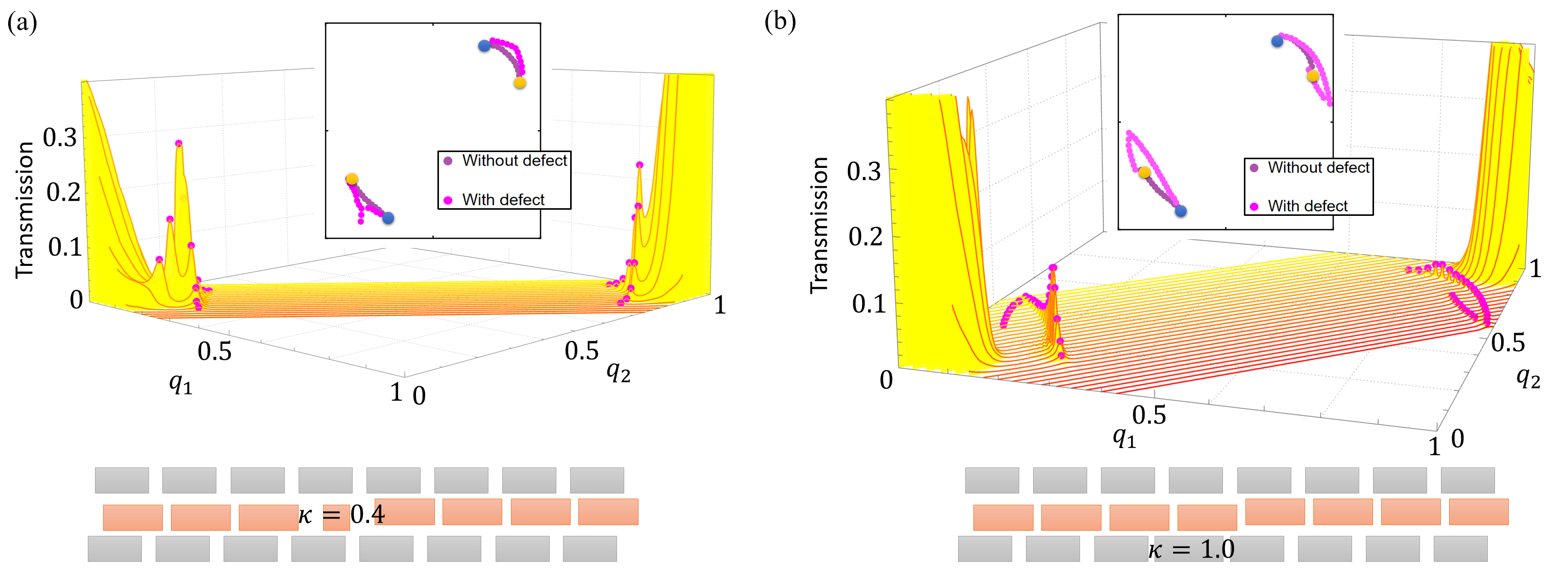}
	\caption{\label{fig:defects} The isofrequency transmission in the presence of edge defects. (a) a dielectric rod of layer 2 at the interface is shrunken. (b) a dielectric rod of layer 1 is dilated.}
\end{figure}

Besides the structural errors of fabricating the silicon gratings, we consider how the edge states are affected when a strong perturbation is present at the junction interface. In particular, we modify the size of a dielectric rod at the interface in two ways, as shown in Fig.~\ref{fig:defects}. Here, the bulk Weyl points are indicated by the blue and yellow round dots as in Fig.~4. In both cases, we still observe the reconstructed FAs connecting Weyl points of the same chirality. In Fig.~\ref{fig:defects}(a), there appears an additional mode branching off from an original arc. By examining the full spectrum, we see that this mode connects the bulk lower band with the gapless edge mode, and thus it is a trivial state. In Fig.~\ref{fig:defects}(b), the FAs are strongly deformed but importantly still stay connected with the bulk Weyl points.

\section{Computing edge states by effective model}
In this section, we present the method to compute the localized edge states of our effective Hamiltonian, i.e., a Hamiltonian linear in momentum $k$. Then, we show the interface Fermi loops (FLs) and Fermi arcs (FAs) of some different junctions.
\subsection{Method}
We consider a junction of two systems described by the bulk Hamiltonian $H_\text{L}(k,q_1,q_2)$ and $H_\text{R}(k,q_1,q_2)$, both of which are linear in $k$. Each Hamiltonian is represented by a $6\times6$ matrix and its eigenstates can be written as six independent spinors,
\begin{equation}
	\psi_{\alpha m} = \begin{pmatrix} C_{1\alpha m} & C_{2\alpha m} & C_{3\alpha m} & C_{4\alpha m} & C_{5\alpha m} & C_{6\alpha m} \end{pmatrix}^Te^{i\mathbf{k}\cdot\mathbf{r}},
\end{equation}
so that
\begin{equation}
	H_{\alpha}(k,q_1,q_2)\psi_{\alpha m}(k,q_1,q_2) = E_{\alpha m}(k,q_1,q_2)\Psi_{\alpha m}(k,q_1,q_2).
\end{equation}
Here, $\alpha=\text{L}$ or $\text{R}$ and $m=1...6$ is the band index. We consider the exponentially localized states at the interface ($x=0$) of our junction. The wave vectors of these states have complex components relative to the direction without translation symmetry, i.e., $k=\tilde{k}+i\kappa$ with $\kappa>0$ being proportional to the inverse of the localization length. In general, these complex wave numbers $k$ can be found as six energy-dependent solutions of the secular equation $\det\left|H_{\alpha}(k,q_1,q_2) - E\right|=0$. However, since our effective Hamiltonian is linear in $k$, we can write it as $H_{\alpha}(k,q_1,q_2)=h_{\alpha}(q_1,q_2)+kM_{\alpha}(q_1,q_2)$ and solve the generalized eigenequation
\begin{equation}
	[\underbrace{E_{\alpha}-h_{\alpha}(q_1,q_2)}_{\mathcal{K}(E,q_1,q_2)}]\psi_{\alpha m}(E,q_1,q_2) = kM_{\alpha}(q_1,q_2)\psi_{\alpha m}(E,q_1,q_2).\label{Eq: generalized schrodinger}
\end{equation}
Here, the band index $m$ denotes the $k$ bands with respect to energy $E$ and other momenta. This generalized eigenequation of operators $\mathcal{K}$ and $M_\alpha$ has complex eigenvalues $k$. We see that if $k$ is an eigenvalue of this equation, $k^*$ is also an eigenvalue. Consequently, in six eigenvalues of Eq.~(\ref{Eq: generalized schrodinger}) there are three values with non-negative $\kappa$ and three non-positive ones. As assumed before, we only take into account eigenvalues and eigenvectors with non-negative $\kappa$. The surface-localized state is now written as a linear combination of the three spinors
\begin{equation}
	\Psi_{\alpha}(E,q_1,q_2) = \sum_{m=1}^3 A_m\psi_{\alpha m}(E,q_1,q_2)
\end{equation}
As the wavefunction has to be continuous, we have $\Psi_{\text{L}}(E,q_1,q_2) = \Psi_{\text{R}}(E,q_1,q_2)$, or
\begin{equation}
	\sum_m A_m\psi_{\text{L}m}(E,q_1,q_2) - \sum_m B_m\psi_{\text{R}m}(E,q_1,q_2) = 0
\end{equation}
The six spinors $\psi_{\text{L}m}$ and $\psi_{\text{R}m}$ constitute a linearly independent vector space, which yields
\begin{equation}
	\det\begin{pmatrix}\psi_{\text{L}1}&\psi_{\text{L}2}&\psi_{\text{L}3}&\psi_{\text{R}1}&\psi_{\text{R}2}&\psi_{\text{R}3}\end{pmatrix} = 0.
\end{equation}
Here, $\begin{pmatrix}\psi_{\text{R}1}&\psi_{\text{R}2}&\psi_{\text{R}3}&\psi_{\text{L}1}&\psi_{\text{L}2}&\psi_{\text{L}3}\end{pmatrix}$ is a square matrix whose columns are state vectors $\psi_{\alpha m}$. This secular equation provides the condition for finding the localized edge states at $(E,q_1,q_2)$.
\subsection{Fermi loops and Fermi arcs}

Applying this method to different junctions of our system, we can obtain the corresponding interface states.
\subsubsection{Original Fermi loops and Fermi arcs}
We consider a photonic junction of a topological lattice and a trivially insulating lattice. This allows us to obtain the original surface states of the topological one, namely surface states without reconstruction. Here, for simplicity, we choose the insulating system to be our trilayer lattice when the semi-Weyl points are gapped at $q_1=q_2=0$, i.e., $U_1=U_3=1$ and $U_2=1.2$. As the synthetic momenta can be  The evanescent coupling rates will be chosen according to the topologically nontrivial system. The Hamiltonian for this trivially insulating system reads
\begin{equation}
	H_{\text{trv}}(k) = \omega_0 + \begin{pmatrix}
		v_1k & U_1 & V_1 & 0 & 0 & 0\\
		U_1 & -v_1k & 0 & V_1 & 0 & 0\\
		V_1 & 0 & v_2k & U_2 & V_2 & 0\\
		0 & V_1 & U_2 & -v_2k & 0 & V_2\\
		0 & 0 & V_2 & 0 & v_3k & U_3\\
		0 & 0 & 0 & V_2 & U_3 & -v_3k
	\end{pmatrix}.
\end{equation}
As the three dielectric gratings are always aligned, the bands are completely flat in the momentum space $(q_1,q_2)$. The original FAs and FLs of some systems of our interest are shown in Fig.~4 of the main text and Fig.~\ref{fig:sup9}. Although the curvature of these chiral surface states depends on the trivial system, their chirality remains consistent with the bulk Chern number, satisfying the bulk-boundary correspondence.

\begin{figure}
	\includegraphics[width=\linewidth]{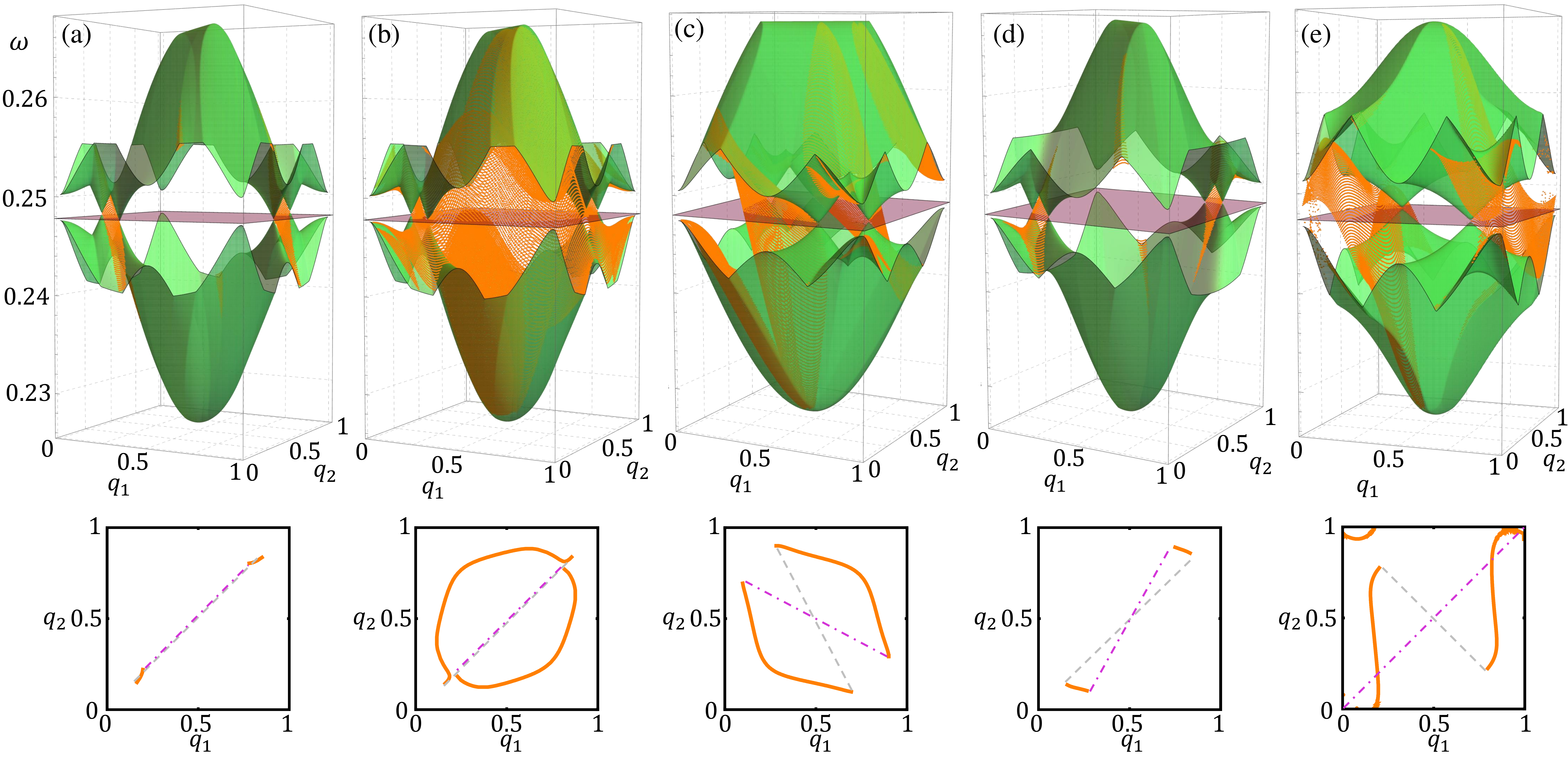}
	\caption{\label{fig:sup10} The bulk and edge bands of different photonic junctions, obtained by the effective model, and their isofrequency surface. The diffractive coupling rates of all these systems are the same with $U_1=U_3=1$ and $U_2=1.2$. (a) \textit{left:} $(V_1,V_2)=(1,1)$, \textit{right:} $(1.6,1.6)$; (b) \textit{left:} $(1,1)$, \textit{right:} $(1.6,1.6)$ with $\delta_{1/2}\rightarrow-\delta_{1/2}$; (c) \textit{left:} $(1.3,1)$ with $\delta_2\rightarrow-\delta_2$, \textit{right:} $(1,1.3)$ with $\delta_1\rightarrow-\delta_1$; (d) \textit{left:} $(1,1)$, \textit{right:} $(1,1.3)$; (e) \textit{left:} $(0.6,0.6)$, \textit{right:} $(1.6,1.6)$ with $\delta_1\rightarrow-\delta_1$. The dashed and dashed-dotted lines schematically show the original FAs and FLs.}
\end{figure}

\subsubsection{Different cases of interface state}
With the agreement between the effective model and electromagnetic simulation, we can use the effective model to study different junctions formed by WSMs and/or CIs. Here, we show five cases in Fig.~\ref{fig:sup10}. In the first two cases, the two Weyl points of both sides are aligned along the diagonal of the BZ and have different distances between them, i.e., different FA lengths. However, while the two FAs at the interface of Fig.~\ref{fig:sup10}(a) have opposite chirality, those of Fig.~\ref{fig:sup10}(b) share the same chirality. Remind that the chirality here indicates the propagating direction of edge modes. The results agree well with those obtained by Ishida \textit{et al.} \cite{Ishida2018}. The two arcs of Fig.~\ref{fig:sup10}(b) are unconnected because the Weyl points of the two sides are located at different frequencies. This can be tuned by varying the lattice constant of one side as mentioned in Sec.~\ref{Sec: 2}. Fig.~\ref{fig:sup10}(c) shows a similar case of Fig.~4(f) in the main text but for a larger angle between the two original FAs. This is a special case since the Weyl points of our bulk system can only move within two quarters of the BZ [Figs.~3(b) and 3(c)]. However, we can redefine the synthetic momenta as $\delta_i\rightarrow-\delta_i$, and the Weyl points can locate at the other two quarters owing to reflection or inversion. To get Fig.~\ref{fig:sup10}(c), we consider the junction given in Fig.~4(d) but reverse $\delta_2$ of the right system and $\delta_1$ of the left one. Fig.~\ref{fig:sup10}(d) also shows a similar case but for asymmetric original FAs. Finally, Fig.~\ref{fig:sup10}(e) shows the interface states of a WSM and a CI.

Overall, we can achieve a variety of interface states using the effective model for our photonic lattice. The limit of this system is that it cannot change the connectivity of the Weyl points while preserving the nodal configuration, which is considered in Ref.~\cite{Dwivedi2018}.

\section{Semi-Weyl point\label{Sec: 2}}
Semi-Weyl point is a band touching point whose dispersion is linear in a 2D plane and quadratic along the direction perpendicular to that plane. It is a transition phase between WSM and CI, which appears when two Weyl points of opposite chirality merge with each other. As the Weyl points are magnetic monopoles of Berry curvature, a semi-Weyl point can thus be regarded as a dipole. In this section, we study the semi-Weyl points of bands (3) and (4). Using perturbation theory, we derive an effective Hamiltonian for a semi-Weyl point and demonstrate how it can be split into two Weyl points or gapped out. We then show how the Chern number varies over the BZ in these cases. Even though the semi-Weyl points are not protected in our photonic lattice, the tendency that they transition to WSM or CI remains the same despite perturbations. Hence, we can confirm that the Weyl points split from these semi-Weyl points give rise to the Fermi-arc surface states in a photonic junction.
\subsection{Effective model}
The two semi-Weyl points of bands (3) and (4) locate on the $q_1=q_2=0$ axis. They only exist in our theoretical model when $U_1=U_2=U_3$ and their distance depends on the interlayer coupling rates. If we assume $V_1=V_2$, the two semi-Weyl points merge with each other when $V_1=V_2=U/\sqrt{2}$ and they become gapped for smaller values of $V_1$, $V_2$.

We now derive an effective two-band Hamiltonian in the vicinity of a semi-Weyl node. For simplicity, we neglect the dependence of $\omega_0$ and $v$ on the filling fraction $\kappa$ and assume that all coupling rates are the same except $U_2$. We then have $U_1=U_3=V_1=V_2=1$ and $U_2=1-\eta$. The Hamiltonian can thus be written as
\begin{equation}
	H(k,q_1,q_2) = \omega_0 + \begin{pmatrix}
		k& e^{iq_1}& 1& 0& 0& 0\\ e^{-iq_1}& -k& 0& 1& 0& 0\\ 1& 0& k& 1-\eta& 1& 0\\ 0& 1& 1-\eta& -k& 0& 1\\ 0& 0& 1& 0& k& e^{-iq_2}\\ 0& 0& 0& 1& e^{iq_2}& -k
	\end{pmatrix},
\end{equation}
where $k\leftarrow kv/\bar{U}$. From now on we also set $\omega_0=0$ besides $\bar{U}=1$ for short. When $\eta=0$, the semi-Weyl points locate at $q_1=q_2=0$, and thus we can diagonalize $H(k,0,0)$ and get
\begin{align}
	E_1=-\left|\sqrt{k^2 + 1}-\sqrt{2}\right|,&\quad E_2=\left|\sqrt{k^2 + 1}-\sqrt{2}\right|,\nonumber\\
	E_3=-\sqrt{k^2 + 1}-\sqrt{2},&\quad E_4=-\sqrt{k^2 + 1},\nonumber\\
	E_5=\sqrt{k^2 + 1},&\quad E_6=\sqrt{k^2 + 1}+\sqrt{2}.
\end{align}
The band touching points of our interest are defined at $E_1=E_2\Leftrightarrow\pm\sqrt{k^2 + 3 - \sqrt{8k^2 + 8}}=0$, which yields $\boxed{k_s=\pm1}$. We now write the Hamiltonian in the vicinity of the semi-Weyl point $(k,q_1,q_2)=(1,0,0)$ as
\begin{align}
	H_+(k,q_1,q_2,\eta) &\approx \begin{pmatrix}
		1& 1& 1& 0& 0& 0\\ 1& -1& 0& 1& 0& 0\\ 1& 0& 1& 1& 1& 0\\ 0& 1& 1& -1& 0& 1\\ 0& 0& 1& 0& 1& 1\\ 0& 0& 0& 1& 1& -1
	\end{pmatrix} + \begin{pmatrix}
		q_0& iq_1-\frac{1}{2}q_1^2& 0& 0& 0& 0\\ -iq_1-\frac{1}{2}q_1^2& q_0& 0& 0& 0& 0\\ 0& 0& q_0& -\eta& 0& 0\\ 0& 0& -\eta& q_0& 0& 0\\ 0& 0& 0& 0& q_0& -iq_2-\frac{1}{2}q_2^2\\ 0& 0& 0& 0& iq_2-\frac{1}{2}q_2^2& q_0
	\end{pmatrix} \\
	&= H_0 + H'(q_0,q_1,q_2,\eta),
\end{align}
where we have defined $k=1+q_0$. By applying a unitary transformation $\mathcal{U}_s$, we diagonalize $H_0$ and obtain a new basis set
\begin{equation}
	\mathcal{H}_+ = \mathcal{U}_s^{\dagger}H_+(k,q_1,q_2,\eta)\mathcal{U}_s = \text{diag}(0,0,-2\sqrt{2},-\sqrt{2},\sqrt{2},2\sqrt{2}) + \underbrace{\mathcal{U}_s^{\dagger}H'(q_0,q_1,q_2,\eta)\mathcal{U}_s}_{\mathcal{H}'}.
\end{equation}

\begin{figure}
	\includegraphics[width=\linewidth]{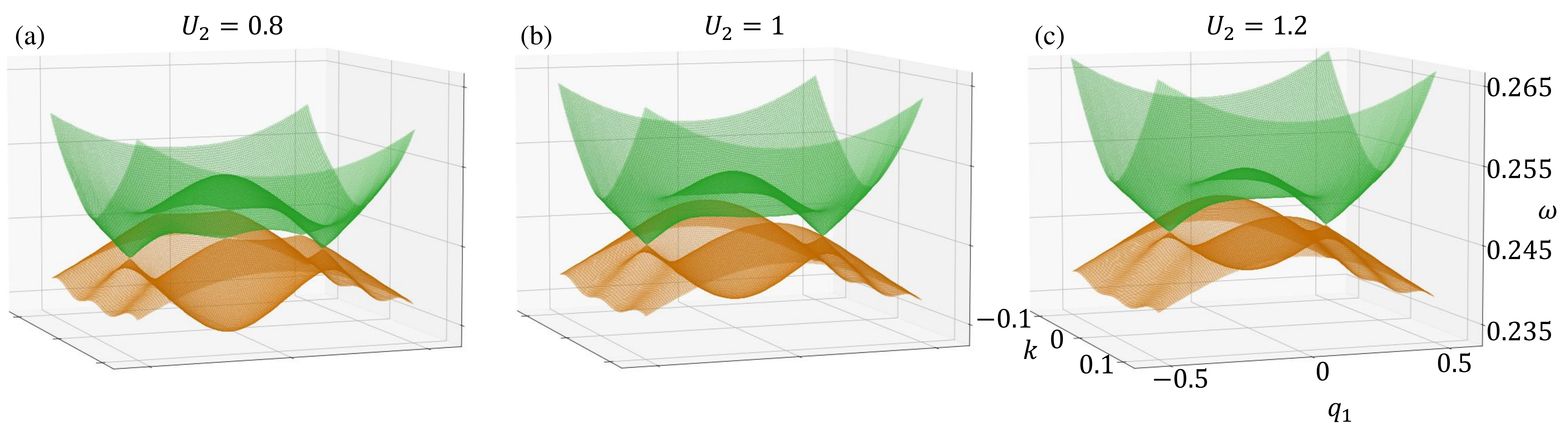}
	\caption{\label{fig:sup5} PWE simulations for bands (3) and (4) in $(q_1=q_2)$-plane when $U_1=U_3=V_1=V_2=1$. Notice that the BZ is shifted so that $q_1$ and $q_2$ run from $-\pi/\Lambda$ to $\pi/\Lambda$.}
\end{figure}

Applying L\"{o}wdin perturbation~\cite{Winkler2003,Lowdin1951} for the first two basis states gives
\begin{equation}
	\tilde{\mathcal{H}}_+(q_0,q_1,q_2) = \mathcal{H}_0 + \mathcal{H}_1 + \mathcal{H}_2 + \cdots
\end{equation}
with
\begin{align}
	\mathcal{H}_0 &= \begin{pmatrix}E_1(k=1) & 0\\ 0&E_2(k=1)\end{pmatrix}=0,\\
	\mathcal{H}_1 &= \begin{pmatrix}\psi_1\mathcal{H}'\psi_1 & \psi_1\mathcal{H}'\psi_2\\ \psi_2\mathcal{H}'\psi_1& \psi_2\mathcal{H}'\psi_2\end{pmatrix}=\begin{pmatrix}-2\eta + 2q_0& 2q_0 + i(q_1 - q_2) - \frac{1}{2}(q_1^2+q_2^2)\\2q_0 - i(q_1 - q_2) - \frac{1}{2}(q_1^2+q_2^2)& 2\eta - 2q_0\end{pmatrix},\\
	\mathcal{H}_2 &= \begin{pmatrix}h_{11} & h_{12}\\ h_{21} & h_{22}\end{pmatrix},\text{ for } h_{mn} = \frac{1}{2}\sum_l\mathcal{H}_{ml}'\mathcal{H}_{ln}'\left[\frac{1}{E_m-E_l}+\frac{1}{E_n-E_l}\right]\cdots.
\end{align}
If we take into account only the first-order correction, we arrive at an effective Hamiltonian
\begin{align}
	\tilde{\mathcal{H}}_+ &\approx \begin{pmatrix} -2\eta + 2q_0& 2q_0 + i(q_1 - q_2) - \frac{1}{2}(q_1^2+q_2^2)\\2q_0 - i(q_1 - q_2) - \frac{1}{2}(q_1^2+q_2^2)& 2\eta - 2q_0\end{pmatrix} \nonumber\\
	&= \left[2q_0-\frac{1}{2}(q_1^2+q_2^2)\right]\sigma_x - (q_1-q_2)\sigma_y + 2(q_0-\eta)\sigma_z.
\end{align}
The energy eigenvalues are
\begin{equation}
	E_{1/2}^{(+)}(q_0,q_1,q_2) = \pm\sqrt{4(q_0-\eta)^2 + 4\left[q_0-\frac{1}{4}(q_1^2+q_2^2)\right]^2 + (q_1-q_2)^2},
\end{equation}
from which we see that the spectrum is gapless when $\eta=0$, disperses linearly along $q_0$ and $q_x=(q_1-q_2)/\sqrt{2}$ and quadratically along $q_y=(q_1+q_2)/\sqrt{2}$. It has a gap for $\eta<0$ and two Weyl points at $(\eta,-\sqrt{2\eta},-\sqrt{2\eta})$ and $(\eta,\sqrt{2\eta},\sqrt{2\eta})$ for $\eta>0$. The two Weyl points split along the quadratic direction ($q_y$).

Following the same procedure, we also obtain the effective two-band Hamiltonian near $(-1,0,0)$, which reads
\begin{align}
	\tilde{\mathcal{H}}_- &\approx \begin{pmatrix}2\eta + 2q_0& -2q_0 + i(q_1 - q_2) - \frac{1}{2}(q_1^2+q_2^2)\\-2q_0 - i(q_1 - q_2) - \frac{1}{2}(q_1^2+q_2^2)& -2\eta - 2q_0\end{pmatrix} \nonumber\\
	&= -\left[2q_0+\frac{1}{2}(q_1^2+q_2^2)\right]\sigma_x - (q_1-q_2)\sigma_y + 2(q_0+\eta)\sigma_z,
\end{align}
with $k=-1+q_0$. The corresponding energy eigenvalues are
\begin{equation}
	E_{1/2}^{(-)}(q_0,q_1,q_2) = \pm\sqrt{4(q_0+\eta)^2 + 4\left[q_0+\frac{1}{4}(q_1^2+q_2^2)\right]^2 + (q_1-q_2)^2}.
\end{equation}
Similar to the previous case, the spectrum has a gap for $\eta<0$, one semi-Weyl band touching point at $(0,0,0)$ for $\eta=0$, and two Weyl points at $(-\eta,-\sqrt{2\eta},-\sqrt{2\eta})$ and $(-\eta,\sqrt{2\eta},\sqrt{2\eta})$ for $\eta>0$. These analytical expressions agree well with the numerical diagonalization of the effective six-band Hamiltonian and qualitatively with the PWE simulation [Fig.~\ref{fig:sup5}] besides the fact that semi-Weyl points are not stable.
\subsection{Chern number}
\begin{figure}
	\includegraphics[width=\linewidth]{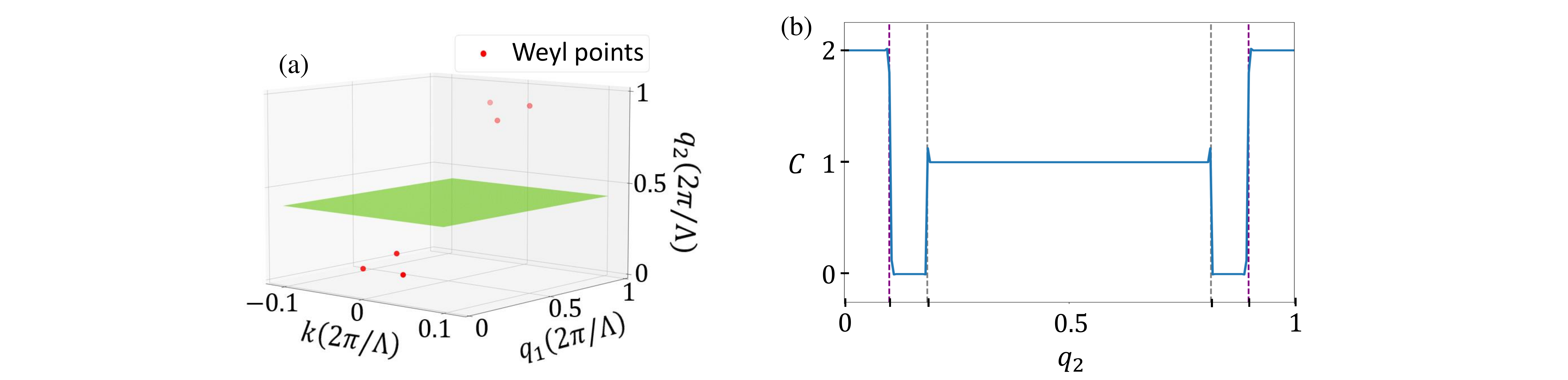}
	\caption{\label{fig:sup6} (a) The BZ with six Weyl points when $U_1=U_3=V_1=V_2=1$ and $U_2=0.8$. The green plane indicates the cross sections of the BZ perpendicular to $q_2$ where the Chern number is computed. (b) The dependence of Chern number on $q_2$.}
\end{figure}

When the semi-Weyl points split into pairs of Weyl points, the topology of the BZ changes. To illustrate the distribution of the Chern number, we consider a cross-section of the BZ that is perpendicular to $q_2$, compute the Chern number in this plane and examine how it varies with respect to $q_2$. In Fig.~\ref{fig:sup6}(a), we show the six Weyl points in the first BZ and the cross-section used for computing the Chern number. The result is shown in Fig.~\ref{fig:sup6}(b). Here, we note that the oscillations near the band touching points result from the fact that the integration mesh of $(k,q_1)$-plane is not dense enough. It decreases when we increase the mesh density. Importantly, as expected, we see that the Chern number is nonzero in regions between the Weyl points. The Chern number is 2 between the Weyl points split from the semi-Weyl points since we have two pairs of Weyl points sharing the same nodal configurations and FA connectivity.

\subsection{Heterojunction}
\begin{figure}
	\includegraphics[width=\linewidth]{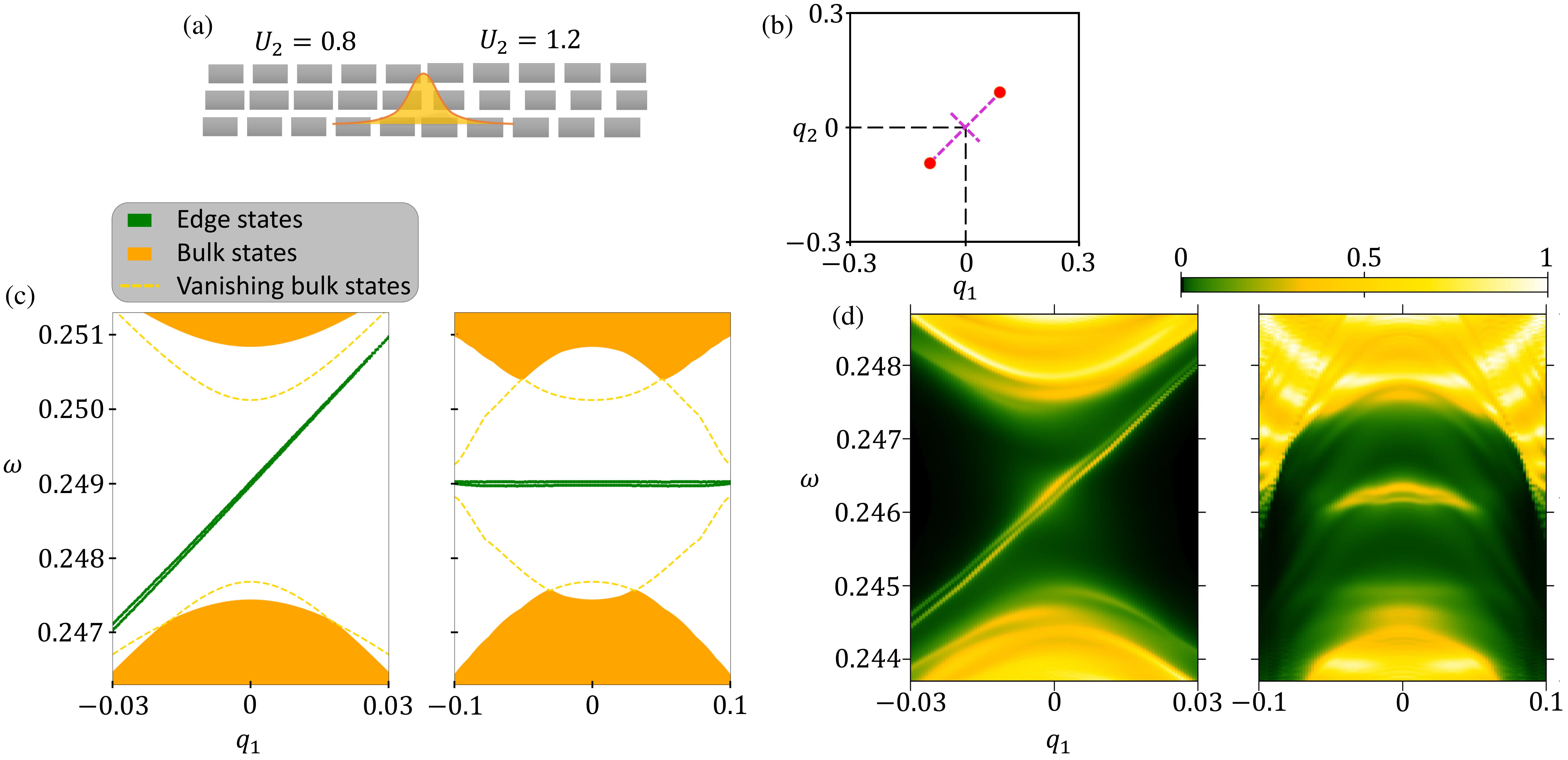}
	\caption{\label{fig:sup7} (a) Sketch of the photonic junction. The two sides have $U_1=U_3=V_1=V_2=1$. The left side has $U_2=0.8$ ($\kappa_2=0.836$) whereas the right side has $U_2=1.2$ ($\kappa_2=0.764$). The length scale of the left side is scaled by a factor of $1.0126$. (b) The projections of Weyl points on the $(q_1,q_2)$-plane. The two dashed lines are used for visualizing the energy bands obtained by (c) the effective model and (d) the FDTD simulation. Here, bands along the dashed line connecting two Weyl points are shown on the right, and those along the dashed line perpendicular to that are shown on the left.}
\end{figure}

With two Weyl points split from each semi-Weyl point, we can construct a photonic junction to observe their chiral surface states. In particular, the junction consists of two sides as shown in Fig.~\ref{fig:sup7}(a) where the right system simulates a trivial insulator and the left one simulates a Weyl semimetal. The two systems are almost identical besides having different coupling rates $U_2$, namely different filling fractions $\kappa_2$. We will show the existence of the chiral FAs by using three methods: the two-band effective model, the six-band effective model, and the FDTD simulation.

First, we use the two-band Hamiltonian to describe the chiral surface states. We apply a unitary transformation $\tilde{U}=\dfrac{1}{\sqrt{2}}(1-i\sigma_x)$ to $\tilde{\mathcal{H}}_+$ and $\tilde{\mathcal{H}}_-$ that transforms $\sigma_y$ into $-\sigma_z$ and $\sigma_z$ into $\sigma_y$, which gives
\begin{align}
	\tilde{\mathcal{H}}_+ &= \left[2q_0-\frac{1}{2}(q_1^2+q_2^2)\right]\sigma_x + (q_1-q_2)\sigma_z + 2(q_0-\eta)\sigma_y,\\
	&= \begin{pmatrix} q_1-q_2& 2(1 - i)q_0 + 2i\eta - \frac{1}{2}(q_1^2+q_2^2)\\2(1 + i)q_0 - 2i\eta - \frac{1}{2}(q_1^2+q_2^2)& -q_1+q_2\end{pmatrix},
\end{align}
\begin{align}
	\tilde{\mathcal{H}}_- &= -\left[2q_0+\frac{1}{2}(q_1^2+q_2^2)\right]\sigma_x + (q_1-q_2)\sigma_z + 2(q_0+\eta)\sigma_y,\\
	&= \begin{pmatrix}q_1-q_2& -2(1+i)q_0 - 2i\eta - \frac{1}{2}(q_1^2+q_2^2)\\-2(1-i)q_0 + 2i\eta - \frac{1}{2}(q_1^2+q_2^2)& -q_1+q_2\end{pmatrix}.
\end{align}
We assume an interface at $x=0$ between two systems: $\eta(x)=-\eta_0<0$ for $x<0$ and $\eta(x)=\eta_0>0$ for $x>0$. Since the translation symmetry is broken, we write the operators as
\begin{align}
	\tilde{\mathcal{H}}_+ &= \begin{pmatrix} q_1-q_2& -2(1 + i)\partial_x + 2i\eta - \frac{1}{2}(q_1^2+q_2^2)\\2(1 - i)\partial_x - 2i\eta - \frac{1}{2}(q_1^2+q_2^2)& -q_1+q_2\end{pmatrix},\\
	\tilde{\mathcal{H}}_- &= \begin{pmatrix}q_1-q_2& -2(1-i)\partial_x - 2i\eta - \frac{1}{2}(q_1^2+q_2^2)\\2(1+i)\partial_x + 2i\eta - \frac{1}{2}(q_1^2+q_2^2)& -q_1+q_2\end{pmatrix}.
\end{align}
Following the same procedure for solving edge states of the Haldane model \cite{Bernevig2013}, we obtain the bound states at the interface
\begin{align}
	\psi_+(x) \sim \text{exp}\left(\frac{q_1^2+q_2^2-4\eta(x)}{8}x + i\frac{q_1^2+q_2^2+4\eta(x)}{8}x\right)\begin{pmatrix}1\\0\end{pmatrix},\quad\text{for }\tilde{\mathcal{H}}_+\psi_+(x) = (q_1-q_2)\psi_+(x),\\
	\psi_-(x) \sim \text{exp}\left(\frac{q_1^2+q_2^2-4\eta(x)}{8}x + i\frac{q_1^2+q_2^2-4\eta(x)}{8}x\right)\begin{pmatrix}1\\0\end{pmatrix},\quad\text{for }\tilde{\mathcal{H}}_-\psi_-(x) = (q_1-q_2)\psi_-(x).
\end{align}
We get two FA surface states chiral along the $(q_1-q_2)$ direction. They are constrained by the condition $q_1^2+q_2^2<4\eta_0$, showing that the arcs connect between the two Weyl cones.

We verify these chiral edge states using the six-band model and FDTD simulation. The method for solving the six-band Hamiltonian is shown in the next section. A technical problem of this junction is that the trivial gap and the Weyl cones are not aligned in energy, i.e., the Weyl cones of the left system is merged to the bulk states of the right one. This problem can be remedied by tuning the lattice constant or dielectric constant of one side. Here, we increase the lattice constant of the right system by a factor $1.0126$, which also leads to scaling the filling fraction and layer thickness.

Fig.~\ref{fig:sup7}(b) shows the projections of Weyl points onto the $(q_1,q_2)$-plane. The momenta $q_1$ and $q_2$ are limited so that the Weyl points at $k=0$ are not shown. The two semi-Weyl points are split into two pairs of Weyl points, but those with the same chirality share the same position in the $(q_1,q_2)$-plane. Hence, there are only two projections, each of which belongs to two bulk Weyl points. The energy bands obtained by the effective model are shown in Fig.~\ref{fig:sup7}(c). We can see the chiral band with an open Fermi surface, which corresponds to the FA surface state. The chiral band disperses along the $(q_1-q_2)$ direction, consistent with results obtained by the two-band model. Notice that, in order to compare this result with the transmission spectrum, the bulk states belong to the intersection of two sets of states, each set contains the bulk states of one side of our junction. If a bulk mode on the left side lies in a gap of the right side, it is most likely to be unable to propagate through the right side, and vice versa. The dashed lines in Fig.~\ref{fig:sup7}(c) show the bulk states that vanish due to this reason.

The transmission spectra obtained by FDTD simulation also confirm the appearance of the chiral FA in our photonic junction. The deviation between Figs.~\ref{fig:sup7}(c) and \ref{fig:sup7}(d) results from the mismatch in energy of the two sides of the junction. The two chiral edge states can be seen clearly, showing the existence of two pairs of Weyl points.

\section{Nodal line spectra}
\begin{figure}
	\includegraphics[width=\linewidth]{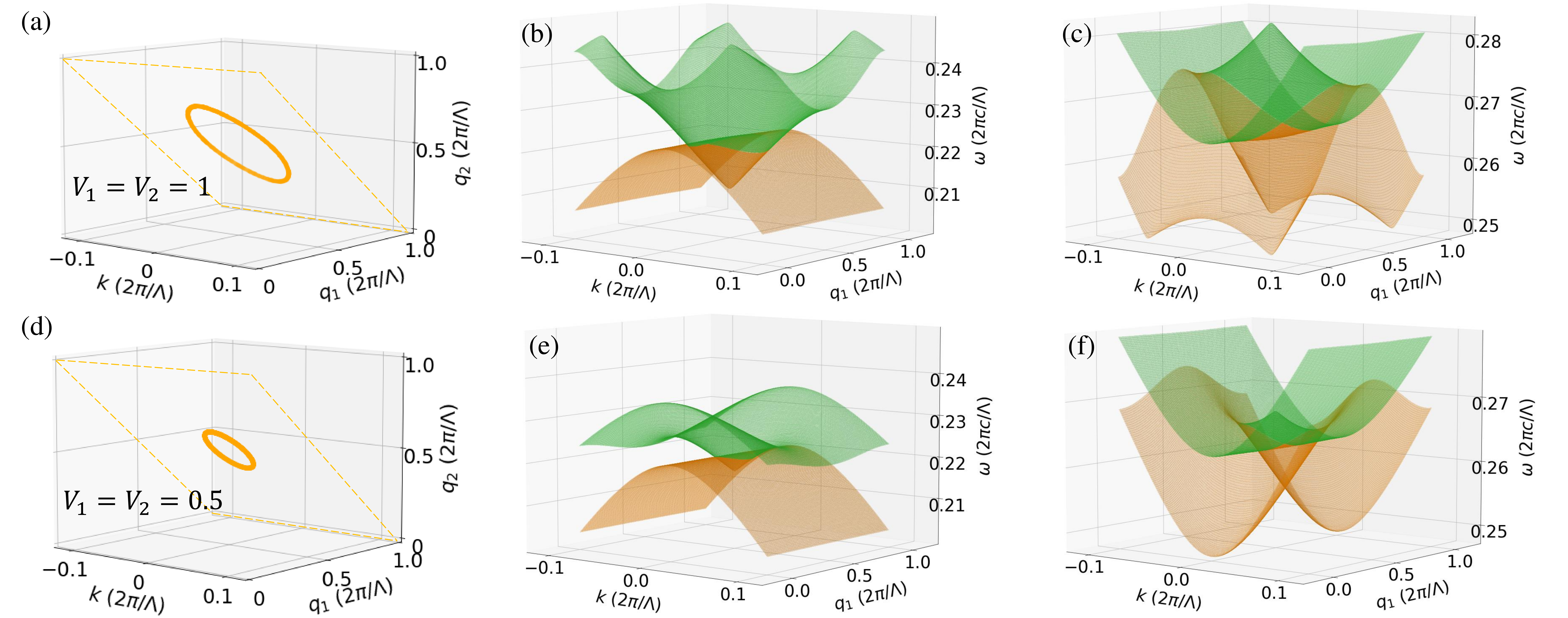}
	\caption{\label{fig:sup14} Nodal line spectra. (a)-(c) $V_1=V_2=1$. (d)-(f) $V_1=V_2=0.5$. (a) and (d) show the nodal obtained by the effective model, while the others are spectra obtained by PWE simulations in the $(q_1+q_2=2\pi/\Lambda)$-plane. (b) and (e) show bands (2) and (3). (c) and (f) show bands (4) and (5).}
\end{figure}

Finally, we show some spectra of the nodal line obtained by PWE simulation following the prediction by the effective model. We consider two cases where the evanescent coupling rates are $V_1=V_2=1$ and $V_1=V_2=0.5$. In both cases, the diffractive coupling rates remain $U_1=U_2=U_3=1$. The first case is shown in Figs.~\ref{fig:sup14}(a)-(c) and the second one correspond to Figs.~\ref{fig:sup14}(d)-(f). The nodal line can be formed by bands (2) and (3) [Figs.~\ref{fig:sup14}(b) and \ref{fig:sup14}(e)], or bands (4) and (5) [Figs.~\ref{fig:sup14}(c) and \ref{fig:sup14}(f)]. The PWE simulations show that the nodal lines shrink when the interlayer coupling decreases, as predicted by the effective model. These nodal lines can be gapped out by breaking the mirror symmetry in the $z$-direction, e.g., by letting $U_1\neq U_2$ or $V_1\neq V_2$.

\end{document}